\documentclass[]{aa}  

\usepackage{graphicx}
\usepackage{txfonts}
\usepackage{natbib,twoopt}
\makeatletter
\newcommand{\AJ}[1]{{{#1}}}
\newcommand\omicron{o}
\newcommand{\Msun}{M$_{\odot}$}
\newcommand{\kms}{km~s$^{-1}$}
\newcommandtwoopt{\citeads}[3][][]{\href{http://adsabs.harvard.edu/abs/#3}%
{\def\hyper@linkstart##1##2{}%
\let\hyper@linkend\@empty\citealp[#1][#2]{#3}}}
\newcommandtwoopt{\citepads}[3][][]{\href{http://adsabs.harvard.edu/abs/#3}%
{\def\hyper@linkstart##1##2{}%
\let\hyper@linkend\@empty\citep[#1][#2]{#3}}}
\newcommandtwoopt{\citetads}[3][][]{\href{http://adsabs.harvard.edu/abs/#3}%
{\def\hyper@linkstart##1##2{}%
\let\hyper@linkend\@empty\citet[#1][#2]{#3}}}
\newcommandtwoopt{\citeyearads}[3][][]%
{\href{http://adsabs.harvard.edu/abs/#3}
{\def\hyper@linkstart##1##2{}%
\let\hyper@linkend\@empty\citeyear[#1][#2]{#3}}}
\makeatother

\hypersetup{draft}
\begin{document}

  \title{
Barium and related stars, and their white-dwarf companions  
    \thanks{Based on observations made with the Mercator Telescope, operated on the island of La Palma by the Flemish Community, at the Spanish Observatorio del Roque de los Muchachos of the Instituto de Astrof\'\i sica de Canarias.
 }$^{,}$\thanks{\AJ{Tables 2 and 3 are only available in electronic form at the CDS via anonymous ftp to cdsarc.u-strasbg.fr (130.79.128.5) or via http://cdsweb.u-strasbg.fr/cgi-bin/qcat?J/A+A/ }} }
 \subtitle{I. Giant stars}
\titlerunning{Orbital properties of barium and S stars}


\author{
A. Jorissen \inst{1}
\and
H.M.J. Boffin \inst{2}
   \and
D. Karinkuzhi \inst{1,3}
\and
S. Van Eck \inst{1}
\and
A. Escorza \inst{1,4}
\and
S. Shetye \inst{1,4}
\and
H. Van Winckel \inst{4}
  }

\institute{
Institut d'Astronomie et d'Astrophysique, Universit\'e Libre de Bruxelles, Campus Plaine C.P. 226, Boulevard du Triomphe, B-1050 Bruxelles, Belgium
         \email{ajorisse,svaneck@ulb.ac.be}
\and
ESO, K. Schwarzschild Stra\ss e 2, Garching bei M\"unchen, Germany\\
              \email{hboffin@eso.org}
\and
Department of Physics,  Bangalore University, Jnana Bharathi Campus, Bangalore, India 560056
\and
Institute of Astronomy, KU Leuven, Celestijnenlaan  200D, 3001 Leuven, Belgium
}

\date{Received X; accepted Y}

 \abstract{Barium and S stars without technetium are red giants suspected of being all members of binary systems.
}
{
 This paper provides both long-term and revised, more accurate orbits for barium and S stars adding to previously published ones. The sample of barium stars with strong anomalies 
 comprise all such stars present in the L\"{u} et al. catalogue.
}
{
Orbital elements are derived from radial velocities collected from a long-term radial-velocity monitoring performed with the HERMES spectrograph mounted on the Mercator 1.2 m telescope. These new measurements were combined with older, CORAVEL measurements. With the aim of investigating possible correlations between orbital properties and abundances, we collected as well an as homogeneous as possible set of abundances for barium stars with orbital elements. 
}
{
We find orbital motion for all barium and extrinsic S stars monitored. We obtain the longest period known so far for a spectroscopic binary involving an S star, namely 57~Peg with a period of the order of  100 -- 500~yr.
We 
present the mass distribution for the barium stars, which ranges from 1 to 3~M$_\odot$, with a tail extending up to 5~M$_\odot$ in the case of mild barium stars. This high-mass tail comprises mostly high-metallicity objects ([Fe/H]~$ \ge -0.1$). 
Mass functions are compatible with WD companions whose masses range from 0.5 to 1~M$_\odot$.
Strong barium stars have a tendency to be found in  systems with shorter periods than mild barium stars, although this correlation is rather lose, metallicity and WD mass playing a role as well. Using the initial -- final mass relationship established for field WDs, we derived the distribution of the mass ratio $q' = M_{\rm AGB, ini}/M_{\rm Ba}$ (where $M_{\rm AGB, ini}$ is the WD progenitor initial mass, i.e., the mass of the system former primary component) which is a proxy for the initial mass ratio (the more so, the less mass the barium star has accreted). It appears that the distribution of $q'$ is highly non uniform, and significantly different for mild and strong barium stars, the latter being characterized by values mostly in excess of 1.4, whereas mild barium stars occupy the range 1 -- 1.4.
}
{
The 
orbital properties presented in this paper pave the way for a comparison with binary-evolution models.
}


   \maketitle
%

\section{Introduction}

Barium stars \citep{1951ApJ...114..473B} are a class of G-K red-giant stars with strong spectral lines of barium and other elements produced by the slow neutron-capture process \citep[s-process; e.g.,][]{2011RvMP...83..157K}. Similar spectral peculiarities are also found in main sequence stars known as barium dwarfs, which cover spectral types all the way from F to K \citep{1994A&A...281..775N}. 
Another related family comprises S stars \citep{1954ApJ...120..484K}, which are giants cooler than barium stars, exhibiting ZrO bands in their spectra. As shown by \citet{1988ApJ...333..219S} and \citet{1993A&A...271..463J} for example, two kinds of S stars arise: Tc-rich (also known as intrinsic) and no-Tc (also known as extrinsic) S stars, depending on the presence or absence of Tc lines, an element with no stable isotopes. Extrinsic S stars are the cooler analogs of barium stars.

These families of stars exhibiting strong lines of s-process elements have been intensively studied in the past 
\citep[e.g.,][]{1957ApJ...126..357B,1965MNRAS.129..263W,1980ApJ...238L..35M,1988A&A...205..155B,1990ApJ...352..709M,1992A&A...260..115J,1994A&A...281..775N,
1998A&A...332..877J,2000IAUS..177..269N,2016MNRAS.459.4299D,2016A&A...586A.151M}, being
benchmarks of post mass-transfer binaries involving
low- and intermediate-mass stars. They provide strong constraints
on the mass-transfer phase they experienced when the former primary,
now a white dwarf, was an asymptotic giant branch (AGB) star and transferred material
enriched in heavy elements produced by the s-process of nucleosynthesis \citep[e.g.,][]{2011RvMP...83..157K}, among which is barium. 
The polluted companion indeed kept this chemical
signature up to now, long after the mass transfer ceased, and exhibits strong absorption lines of ionised barium in its spectrum. 
This binary scenario was convincingly confirmed by the observation
that statistically all barium stars reside in binary systems 
\citep{1980ApJ...238L..35M,1983ApJ...268..264M,1988A&A...198..187J,1990ApJ...352..709M,1998A&A...332..877J}.

Previous binary-evolution models have shown how difficult it is to account for the
orbital properties of these objects \citep{2003ASPC..303..290P,2003agbs.conf..461J,2008A&A...480..797B}.
These models need improved prescriptions for the mass-transfer process \citep{Frankowski2007,2010A&A...523A..10I,2011ASPC..445..447D}.
These studies have shown how important it is to derive the orbital periods
and eccentricities of post-mass-transfer systems such as barium stars in order to constrain evolutionary models. CH and carbon-enriched metal-poor (CEMP) stars are post-mass-transfer objects as well, albeit of low metallicity, and new and updated orbits for these classes were presented in a recent paper \citep{2016A&A...586A.158J}. Post-AGB stars with a near-infrared excess indicative of a dusty disk form another possibly related family of post-mass-transfer objects \citep[e.g.,][]{2009A&A...505.1221V,2019Oomen}. 

Orbital elements provide constraints on evolutionary models through the period--eccentricity ($P - e$) diagram and the mass-function distribution, which is sensitive to the mass of the companion. For post-mass-transfer systems (like barium, CH, and CEMP-s systems enriched in heavy elements synthesised by the s-process), the companion should be a CO white dwarf   \citep{2016A&A...586A.151M}.
This paper presents the orbits for all known giant barium stars with strong chemical anomalies
(i.e., all those classified as Ba3, Ba4, or Ba5 in the 1983 edition of the L\"{u} et al. catalog), plus an extended sample of mild barium stars, along with their cooler analogs, the extrinsic S stars lacking the unstable element technetium.
A detailed analysis of the mass functions, the mass-ratio and mass distributions, the $P - e$ diagram, and their relationship with chemical pollution  concludes this paper. A twin paper \citep{2019Escorza} addresses the same questions for dwarf barium stars.

\section{Samples of barium and S stars without Tc}

The present  study is a follow-up of the monitoring campaign of barium and S stars initiated in 1984 with the CORAVEL spectrograph \citep{1979VA.....23..279B}, the results of which were presented in 
\citet{1988A&A...198..187J,1992A&A...260..115J}, \citet{1998A&A...332..877J}, and \citet{1998A&AS..131...25U,1998A&AS..131...43U}.

The CORAVEL monitoring was not able to derive all the orbits either because several turned out to be  much  longer than its time span, or because its precision (about 0.3~\kms) was not good enough to detect the orbits with the smallest semi-amplitudes (like 0.6~\kms\  for HD~183915 and HD~189581, as we report in Sect.~\ref{Sect:orbits}). These shortcomings  motivated the pursuit of this former monitoring campaign after several years of interruption, with a much
more accurate spectrograph (HERMES, as described in Sect.~\ref{Sect:HERMES}) than the old CORAVEL. The new monitoring, described in e.g., \citet{2010MmSAI..81.1022V} and \citet{2013EAS....64..163G}, could therefore reveal binary systems with much lower velocity amplitudes, not accessible to CORAVEL.

The sample comprises all 37 barium stars with strong chemical anomalies \citep[dubbed Ba3, Ba4, or Ba5 in Warner scale;][]{1965MNRAS.129..263W} from the list of  \citet{1983ApJS...52..169L}, as well as 40 among the mild\footnote{See Table~\ref{Tab:abundances} in Sect.~\ref{Sect:abundances} for a rough calibration of the qualifications mild and strong in terms of quantitative s-process overabundances; there we show that [La/Fe] and [Ce/Fe] values of 1~dex fairly represent the transition between mild and strong barium stars. Conversely, no mild barium stars are found with [Ce/Fe] values below 0.2~dex.}  barium stars of that list.
Although the latter sample is by no means complete, it provides a good comparison to the (complete) sample of strong barium stars. The binary status of the targets mentioned in the list below refers to the situation prevailing at the start of the HERMES monitoring.

The sample of barium stars monitored by HERMES is comprised of the following.
\begin{itemize}
\item 2 strong barium stars (HD~123949, HD~211954) with long and uncertain orbital periods;
\item 1 strong barium star with no evidence for binary motion (HD 65854);
\item 11 mild barium stars with long, uncertain periods (HD~22589, HD~53199, HD~196673), or with a lower limit on the period (HD~40430, HD~51959, HD~98839, HD~101079, HD~104979, HD~134698, HD~165141, BD~$-10^{\circ}$4311);
\item 3 suspected binaries (HD~18182, HD~183915, HD~218356) among mild barium stars, and 3 mild barium stars with no evidence for binary motion (HD~50843, HD~95345, HD~119185). 
\end{itemize}

The S-star sample monitored by HERMES is constructed as follows.
\begin{itemize}
\item 6 stars (HD 30959 = $o^1$~Ori, HD~184185, HD~218634, HDE~288833, BD+31$^\circ$4391, and  BD+79$^\circ$156\footnote{A recent re-analysis of that star by Shetye et al. (in preparation) concludes that it shares properties of extrinsic (Nb-rich; for the correlation extrinsic / Nb-rich, see \citealt{2018A&A...MMMA.NNNK}) and intrinsic  (Tc-rich) S stars.}) with a lower limit on the orbital period from Table~3a of \citet{1998A&A...332..877J}; 
\item 2 stars with no Tc lines and no evidence for binarity (BD~$-21^\circ$2601 and HD~189581) from Table~3c of \citet{1998A&A...332..877J}; 
\item 4 poorly studied symbiotic S stars not present in the original sample of \citet{1998A&A...332..877J}: 
Hen 4-18, V420 Hya, and ER Del from \citet{2002A&A...396..599V} and HR~363 from \citet{1996A&A...306..467J}. 
\end{itemize}

To these twelve S stars monitored by HERMES, 22 supplementary systems with orbital elements already obtained by CORAVEL \citep[as listed in Table~3a of][]{1998A&A...332..877J} must be added. 
In total, the sample of  S stars monitored thus comprises 34 objects. 


\section{Radial-velocity monitoring with the HERMES spectrograph}
\label{Sect:HERMES} 

The radial-velocity (RV) monitoring was performed with the  HERMES spectrograph attached to the 1.2m Mercator telescope 
from the Katholieke Universiteit Leuven, installed at the Roque 
de los Muchachos Observatory (La Palma, Spain). The spectrograph began regular science observations in April 2009, and is fully described 
in \citet{2011A&A...526A..69R}.  The fiber-fed HERMES spectrograph is designed to 
be optimized both in stability as well as in efficiency. It samples the whole optical range from 
380 to 900~nm in one shot, with a spectral resolution of about $86\,000$ for the high-resolution science fiber. This fiber has a 2.5 arcsec aperture on the 
sky and the high resolution is reached by mimicking a narrow slit using a two-sliced image 
slicer. 

The MERCATOR-HERMES combination
is precious because it guarantees regular telescope time. This is needed for our monitoring programme and the operational agreement
reached by all consortium partners (KULeuven, Universit\'e libre de Bruxelles, Royal Observatory of Belgium, Landessternwarnte Tautenburg) is optimized to allow efficient long-term monitoring, which is indispensable for this programme. 
The long-term monitoring of barium and S stars is performed within the framework of this HERMES consortium, with some further data points acquired during KULeuven observing runs \citep{2010MmSAI..81.1022V,2013EAS....64..163G}. In total, about 200 nights per year are devoted to this monitoring campaign, 
and the observation sampling is adapted to the known variation timescale. 

A Python-based pipeline extracts a wavelength-calibrated, cosmic-ray cleaned spectrum. A separate routine is used for measuring RVs, by means of a cross-correlation with a spectral mask constructed on an Arcturus spectrum.  A restricted region covering the range 478.11 -- 653.56~nm (orders 55 -- 74) and containing 1543 useful spectral lines was used to derive the RV, in order to avoid telluric lines on the red end, and often poorly exposed and crowded spectrum on the blue end. A spectrum with a signal-to-noise ratio of 20 is usually sufficient to obtain a cross-correlation function (CCF) with a well-defined minimum.  An example of CCF  is shown in Fig.~1 of  \citet{2016A&A...586A.158J}.

A Gaussian fit is performed on the CCF, and delivers an internal precision of less than 10~m~s$^{-1}$ on the position of the center (depending on the CCF shape). 
The absolute precision of a single RV measurement is $\sim200$~m/s, limited by the pressure fluctuations during the night in the spectrograph room
\citep[see Fig.~9 of][]{2011A&A...526A..69R}. However, this drift has no effect if the arc spectrum used for wavelength calibration is taken consecutive to the science exposure. 
The long-term accuracy (i.e., over several years) may be estimated from the stability of the RV standard stars monitored along with the science targets. These standard stars are taken from the list of \citet{1999ASPC..185..367U}, available at {\tt http://obswww.unige.ch/$\sim$udry/std/std.html}. The distribution of the standard-star velocity standard deviations peaks at $\sigma(Vr) = 55$~m~s$^{-1}$ \citep[as shown in Fig.~2 of ][]{2016A&A...586A.158J}, which may thus be adopted as the typical uncertainty on the radial velocities over the long term.

The difference between the standard-star-catalogue velocity and the measured value is on average 4~m~s$^{-1}$ with a standard deviation of 109~m~s$^{-1}$.
This difference of 4~m~s$^{-1}$ indicates that there is no zero-point offset between HERMES and \citet{1999ASPC..185..367U} list of  standard velocity stars. The standard radial-velocity stars from \citet{1999ASPC..185..367U} are tied to the ELODIE velocity system (see \citealt{1999ASPC..185..367U} for more details), but the CORAVEL radial velocities used here are on the old CORAVEL system, before its conversion to the ELODIE system. Hence, a zero-point offset needs to be applied to these old measurements in order to make them compatible with the ELODIE/HERMES system.  However, this zero-point offset is not easy to predict accurately because it depends on stellar velocity and color. Consequently, its value was, whenever possible, derived a posteriori by ensuring minimal orbital residuals $O-C$ (as displayed by the bottom panels of Figs.~\ref{Fig:95345} and \ref{Fig:Orbit_18182}--\ref{Fig:Orbit_ERDel}). The applied offset is given in the caption of these figures displaying the orbital solutions.

\section{Results of radial-velocity monitoring}

\subsection{Binary frequency}
\label{Sect:frequency}

This section reviews the binary frequency  for barium and (extrinsic) S stars. 
It updates our previous review \citep{1998A&A...332..877J} with the new results from the HERMES monitoring.
Our 1998 review  concluded that 35 out of 37 barium stars with strong chemical anomalies, 34 out of 40 mild barium stars (plus an additional 3 stars with binary suspicion), and 25 out of 27 Tc-poor S stars  
showed evidence of being binary systems.   

The present situation is summarized in Table~\ref{Tab:formerly}, which also includes stars monitored by McClure at the Dominion Astrophysical Observatory  \citep{1980ApJ...238L..35M,1983ApJ...268..264M,1990ApJ...352..709M}.

For S stars, the two stars  (HD~189581 and BD~$-21^{\circ}$2601) previously lacking evidence for orbital motion now reveal their binary nature, thanks to the more accurate HERMES data (see Fig.~\ref{Fig:21.2601} for BD~$-21^{\circ}$2601 and Fig.~\ref{Fig:Orbit_189581} for HD~189581).

\begin{table*}
\caption[]{\label{Tab:formerly}
Updated binary frequency among barium and S star samples from \citet{1998A&A...332..877J}. The S stars with radial-velocity jitter were not included.  SB19 and SB98 stand for the number of spectroscopic binaries known in 2019 (present paper)  and in 1998 (before HERMES), respectively, SBO denotes spectroscopic binaries with either good or preliminary orbits, SB stands for  spectroscopic binaries with no orbit available yet, and SB? stands for suspected spectroscopic binary. 
}
\begin{tabular}{ll|c|c|ccc|ccc|c}
\hline
Class     & $N$ & SB19 & SB98 & \multicolumn{3}{c}{SBO}  & SB  & SB? & no evidence SB & New HERMES SBO\\
\cline{5-7}
  & & & & total &good & preliminary\\
\hline
Strong barium (Ba 3,4,5) & 37  & 37 & 35 & 36 & 36 & 0 & 1$^a$ & 0 & 0 & 2\\ 
Mild barium (Ba 1,2) &  40  & 40 & 34 & 37 & 32  &5  & 2$^b$ & 1$^c$ & 0 & 15\\ 
S (no Tc) & 34   & 34 & 25/27 & 32 & 29 & 3  & 2$^d$ & 0 & 0 & 14\\
\hline\\
\end{tabular}

Notes: (a) HD 19014 (b) HD 50843, HD 65854 (c) HD 95345  (d) BD~$-21^{\circ}$2601, T Sgr\\

\end{table*}

All barium stars monitored with HERMES 
(but one, HD~95345, as discussed below) now show clear signatures of binarity, although orbits are not yet available for all of them. 
For instance, HD~50843 and HD~65854 are clearly long-period binaries of small amplitude (Figs.~\ref{Fig:50843} and \ref{Fig:65854}), irrespective of  the uncertain zero-point offset, since the HERMES data alone reveal a clear drift. There is not enough data  yet to look for an orbital solution however. 

The situation is not as clear for HD~95345, since there is a short-period, very low-amplitude ($K = 78\pm11$~m~s$^{-1}$, thus not significantly above the instrumental error; see Sect.~\ref{Sect:HERMES}) orbit possibly fitting the HERMES data points (bottom panel of Fig.~\ref{Fig:95345} and Table~\ref{Tab:orbits}).  Since the standard dispersion of the $O - C$ residuals amounts to 73~m~s$^{-1}$, almost identical to the semi-amplitude $K$ of the orbit, the significance of this orbit should be considered as very low.
In the absence of  any long-term drift, an offset of  0.6~\kms\  is needed to bring the old CORAVEL measurements in agreement with the new HERMES ones  (this offset has been applied in the upper panel of Fig.~\ref{Fig:95345}). Since this zero-point offset is of the same order as \AJ{that} applied to other barium stars, there is therefore no indication in favor of the duplicity of this star, which could nevertheless be a binary system seen very close to face-on.

HD 19014 is a star monitored by the southern CORAVEL \citep{1998A&AS..131...25U} but not by HERMES (because it is located too far south). Evidence for binarity is nevertheless provided  by the comparison of the old CORAVEL velocities, yielding an average velocity of $13.3\pm0.11$~\kms\  \citep[Table 2b of][]{1998A&A...332..877J}, with the Gaia DR2 velocity of  $15.98\pm0.17$~\kms\  \citep{2018arXiv180409365G}. There is thus a difference of 2.7~\kms\  between the two data sets. 
To assess whether this difference is a signature of duplicity or is caused by an offset between CORAVEL and Gaia DR2, we compared the average velocities obtained in these two monitoring campaigns for the supposedly constant star HD~95345 described above. Gaia DR2 yields $6.2\pm0.2$~\kms, in perfect agreement with the HERMES and CORAVEL results 
(after applying the +0.6~\kms\  offset to the latter; see top panel of Fig.~\ref{Fig:95345}). This comparison confirms that the CORAVEL/Gaia-DR2 offset is not expected to be larger than a few tenths of a kilometer per second. Consequently, the difference of 2.7~\kms\  obtained 
between the measurements of CORAVEL and Gaia DR2 is very unlikely to be of instrumental origin, and probably indicates that  HD~19014 belongs to a binary system.

It is interesting to extend this comparison to the stars  
HD~50843 and HD~65854, for which HERMES data reveal a low-amplitude long-term drift. For them, Gaia DR2 yields $13.95\pm0.15$ and $0.82\pm0.17$~\kms, respectively, as compared to 
the HERMES values of  $\sim 13.7$  and $\sim 0.8$~\kms\  (Figs.~\ref{Fig:50843} and   \ref{Fig:65854}).  Also in these two stars, HERMES and Gaia DR2 agree within a few tenths of a kilometer per second, despite the long-term drifts.

\subsection{Individual radial velocities}

The individual radial velocities, referring to the barycenter of the solar system (from {\sc IRAF} {\it astutils} routine using the Stumpff 1980\nocite{1980A&AS...41....1S} ephemeris), are presented in Table~\ref{Tab:VrBa} (for barium stars) and Table \ref{Tab:VrS} (for S stars). 
The CORAVEL radial velocities \citep[described in the papers by][]{1998A&AS..131...25U,1998A&AS..131...43U} used to compute the orbital solutions are repeated here. 

The data before JD~$2\,455\,000$ are from the CORAVEL monitoring \citep{1988A&A...198..187J,1998A&A...332..877J},
and the more recent data are from HERMES \citep{2010MmSAI..81.1022V,2013EAS....64..163G}. No zero-point correction has been applied to the data listed in Tables~\ref{Tab:VrBa} and \ref{Tab:VrS}, \AJ{although the recommended value is listed in column 6}.
\vspace{1mm}\\

As discussed in Sect.~\ref{Sect:HERMES}, the HERMES radial velocities are tied to the 
ELODIE system, defined by the RV standard stars  of  \citet{1999ASPC..185..367U}, while the CORAVEL data are still  on the old (pre-1999) CORAVEL system.

\setlength{\tabcolsep}{3pt}
\begin{table}
\caption[]{
\label{Tab:VrBa}
Individual radial velocities for barium stars \AJ{with no offset applied. The recommended offset is listed in the last column.} The full table is only available as online material.}
\begin{tabular}{rrrrcc}
\hline
 \multicolumn{1}{c}{Star} & \multicolumn{1}{c}{JD} & \multicolumn{1}{c}{$Vr$} & \multicolumn{1}{c}{$\epsilon(Vr)$} & \multicolumn{1}{c}{Instrument} & \multicolumn{1}{c}{Offset}\\
&   & \multicolumn{1}{c}{(km~s$^{-1}$)} &  \multicolumn{1}{c}{(km~s$^{-1}$)}
&&
\multicolumn{1}{c}{(km~s$^{-1}$)}\\
\hline\\
HD 18182  &   2446819.2676  & 24.830& 0.320& CORAVEL&         0.6\\ 
HD 18182  &   2447036.6647  & 24.770& 0.370& CORAVEL&         0.6\\ 
HD 18182  &   2447455.5471  & 25.150& 0.370& CORAVEL&         0.6\\ 
HD 18182  &   2447540.3287  & 26.030& 0.310& CORAVEL&         0.6\\ 
HD 18182  &   2447838.4685  & 25.770& 0.320& CORAVEL&         0.6\\ 
HD 18182  &   2447911.2653  & 25.920& 0.320& CORAVEL&         0.6\\ 
HD 18182  &   2448229.4530  & 25.740& 0.300& CORAVEL&         0.6\\ 
HD 18182  &   2448569.5505  & 25.810& 0.330& CORAVEL&         0.6\\ 
HD 18182  &   2448586.3953  & 25.330& 0.340& CORAVEL&         0.6\\ 
HD 18182  &   2448648.2547  & 25.850& 0.310& CORAVEL&         0.6\\ 
HD 18182  &   2448917.5595  & 25.660& 0.360& CORAVEL&         0.6\\ 
HD 18182  &   2449308.4279  & 25.790& 0.340& CORAVEL&         0.6\\ 
HD 18182  &   2449374.2786  & 25.910& 0.300& CORAVEL&         0.6\\ 
HD 18182  &   2449606.6341  & 25.350& 0.350& CORAVEL&         0.6\\ 
HD 18182  &   2449955.6304  & 25.760& 0.410& CORAVEL&         0.6\\ 
HD 18182  &   2450039.4590  & 25.270& 0.310& CORAVEL&         0.6\\ 
HD 18182  &   2450071.3952  & 25.150& 0.360& CORAVEL&         0.6\\ 
HD 18182  &   2450342.5884  & 25.330& 0.330& CORAVEL&         0.6\\ 
HD 18182  &   2450353.6565  & 25.450& 0.380& CORAVEL&         0.6\\ 
HD 18182  &   2450354.5067  & 24.790& 0.320& CORAVEL&         0.6\\ 
HD 18182  &   2450379.5194  & 25.190& 0.300& CORAVEL&         0.6\\ 
HD 18182  &   2455037.7017  & 25.845& 0.045& HERMES &         0.0\\ 
HD 18182  &   2455085.7070  & 25.850& 0.044& HERMES &         0.0\\ 
HD 18182  &   2455085.7231  & 25.852& 0.044& HERMES &         0.0\\ 
HD 18182  &   2455106.6543  & 25.919& 0.043& HERMES &         0.0\\ 
HD 18182  &   2455160.4400  & 25.688& 0.043& HERMES &         0.0\\ 
... \\
\hline
\end{tabular}
\end{table}

\begin{table}
\caption[]{
\label{Tab:VrS}
As in Table~\ref{Tab:VrBa} but for S stars. The full table is only available as online material.}
\begin{tabular}{rrrrcc}
\hline
 \multicolumn{1}{c}{Star} & \multicolumn{1}{c}{JD} & \multicolumn{1}{c}{$Vr$} & \multicolumn{1}{c}{$\epsilon(Vr)$} & \multicolumn{1}{c}{Instrument} & \multicolumn{1}{c}{Offset}\\
&   & \multicolumn{1}{c}{(km~s$^{-1}$)} &  \multicolumn{1}{c}{(km~s$^{-1}$)}
&&
\multicolumn{1}{c}{(km~s$^{-1}$)}\\
\hline\\
CD-28 3719 &  2448643.6500 &  72.180& 0.620& CORAVEL&         0.0\\ 
CD-28 3719 &  2449056.5220 &  71.310& 0.430& CORAVEL&         0.0\\ 
CD-28 3719 &  2449404.6210 &  72.680& 0.470& CORAVEL&         0.0\\ 
CD-28 3719 &  2449801.5420 &  73.450& 0.430& CORAVEL&         0.0\\ 
CD-28 3719 &  2450468.7230 &  59.660& 0.410& CORAVEL&         0.0\\ 
CD-28 3719 &  2450471.6830 &  60.760& 0.420& CORAVEL&         0.0\\ 
CD-28 3719 &  2455222.5241 &  57.738& 0.070& HERMES &         0.0\\ 
CD-28 3719 &  2455619.4348 &  57.979& 0.070& HERMES &         0.0\\ 
CD-28 3719 &  2455660.3596 &  62.429& 0.067& HERMES &         0.0\\ 
CD-28 3719 &  2455859.7530 &  68.442& 0.069& HERMES &         0.0\\ 
CD-28 3719 &  2455932.5593 &  59.682& 0.074& HERMES &         0.0\\ 
CD-28 3719 &  2455935.5595 &  59.184& 0.070& HERMES &         0.0\\ 
CD-28 3719 &  2455958.4853 &  58.000& 0.073& HERMES &         0.0\\ 
CD-28 3719 &  2455992.3629 &  56.971& 0.064& HERMES &         0.0\\ 
CD-28 3719 &  2456015.3915 &  58.447& 0.079& HERMES &         0.0\\ 
CD-28 3719 &  2456316.4978 &  61.129& 0.064& HERMES &         0.0\\ 
CD-28 3719 &  2456317.5105 &  60.907& 0.071& HERMES &         0.0\\ 
CD-28 3719 &  2456321.4947 &  59.900& 0.074& HERMES &         0.0\\ 
CD-28 3719 &  2456349.4299 &  57.758& 0.062& HERMES &         0.0\\ 
CD-28 3719 &  2456563.7395 &  71.906& 0.080& HERMES &         0.0\\ 
CD-28 3719 &  2456598.7162 &  72.522& 0.077& HERMES &         0.0\\ 
CD-28 3719 &  2456608.7550 &  72.060& 0.069& HERMES &         0.0\\ 
CD-28 3719 &  2456662.5537 &  67.903& 0.086& HERMES &         0.0\\ 
CD-28 3719 &  2456668.5507 &  67.078& 0.079& HERMES &         0.0\\ 
CD-28 3719 &  2456700.4912 &  63.563& 0.059& HERMES &         0.0\\ 
CD-28 3719 &  2457461.3750 &  66.854& 0.072& HERMES &         0.0\\ 
...\\
\hline

\end{tabular}
\end{table}

\subsection{Orbits}
\label{Sect:orbits}

The orbital elements of the newly derived orbits are listed in Table~\ref{Tab:orbits}. Some among these are not yet 
well constrained. For barium stars, these are  HD~18182, HD~104979, HD~119185, HD~134698, and HD~199394. Among S stars, HD~184185, HD~218634 (57~Peg), and HDE~288833 have poorly constrained orbital elements. All figures with the orbital solution superimposed on the radial-velocity data are presented in Appendix~\ref{Sect:Appendix}. 
The orbital elements derived earlier may be found in \cite{1998A&A...332..877J} and \citet{1998A&AS..131...25U,1998A&AS..131...43U}.

Before proceeding to the analysis of this orbital material in Sects.~\ref{Sect:P-e} and \ref{Sect:mass}, 
we hereafter comment on individual stars.
\vspace{1mm}\\

\setlength{\tabcolsep}{3pt}
\begin{table*}
\caption{New, revised, and preliminary orbital elements (the latter data are followed by ":").
\label{Tab:orbits}
}
{\tiny \begin{tabular}{lllllllllll}
 \hline
HD/DM & Period  & $e$ & $V_\gamma$ & $T_0$ & $K$ & $\omega$ & $a_1 \mathrm{sin}\; i$ & $f(m)$ & $\sigma$(O-C) &$N$\\
     & (d)     &     &  (\kms)    &       & (\kms) & ($^\circ$) & (Gm) & (M$_\odot$) & (\kms)\\ 
\hline
\medskip\\
\noalign{\bf Mild Ba stars}
\medskip\\
18182 & 8059: & 0.3: \\
40430 & $5609\pm55$ & $0.22\pm0.01$ & $-23.27\pm0.04$ & $2463116 \pm 80$ & $1.67\pm0.06$ & $88\pm5$ & $126\pm7$ & $0.0025\pm0.0004$ & 0.31 & 36  \\
51959 & $9718\pm157$ & $0.53\pm0.04$ & $38.21\pm0.04$ & $2458537 \pm 113$ & $0.92\pm0.06$ & $50\pm5$ & $104\pm11$ & $0.00047\pm0.00014$ & 0.23 & 64 \\
53199 & $8314 \pm 99$ & $0.24 \pm 0.01$ & $23.6 \pm 0.1$ & $2448483 \pm 117$ & $3.3 \pm 0.1$ & $63 \pm 3$ & $364 \pm 14$ & $0.028 \pm 0.003$ & $0.09$ & $51$ \\
95345 & 485? & 0.3? & & & $0.08\pm0.01$? \\
98839$^b$ & $16471\pm113$ & $0.560\pm0.005$ & $0.13\pm0.01$ & $2451547\pm17$ & $3.86\pm0.03$ & $288.9\pm0.7$ & $724\pm13$ & $0.056\pm0.002$ & 0.41 & 143 \\  
101079 & $1565.8\pm 1.7$ & $0.175\pm 0.005$ & $-2.000 \pm 0.007$ & $2458486 \pm 6$ & $2.48 \pm 0.02$ & $139.9 \pm 1.2$ & $52.6 \pm 0.4$ & $0.00236 \pm 0.00005$ & 0.13& $55$ \\
104979 & 19295:  & 0.1:  \\
119185 & 22065: & 0.6:  \\
134698 & 10005:  & 0.95:  \\
183915 & $4382 \pm 21$ & $0.27 \pm 0.02$ & $-49.83 \pm 0.01$ & $2462214 \pm 80$ & $0.56\pm 0.01$ & $130 \pm 6$ & $32.6\pm 0.8$ & $(7.2 \pm 0.5)\time10^{-5}$ & 0.36 & 98\\
196673 & $7780 \pm 117$ & $0.59 \pm 0.02$ & $-24.2 \pm 0.1$ & $2451698 \pm 128$ & $3.7 \pm 0.1$ & $116 \pm 2 $ & $314 \pm 20$ & $0.020 \pm 0.003$ & 0.40 & $75$ \\
199394 & 5232:$^d$ & 0.11: &  \\
\medskip\\
\noalign{\bf Strong Ba stars}
\medskip\\
123949 & $8523\pm8$ & $0.9162\pm0.0003$ & $-9.56\pm0.01$ & $2466294\pm8$ & $9.33\pm0.02$ & $96.5\pm0.2$ & $438\pm2$ & $0.0462\pm0.0005$ & 0.20 & 86 \\
211954 & $10889\pm113$ & $0.24\pm0.05$ &  $-6.0\pm0.5$ &   $2461595\pm 803$ & $4.1\pm0.8$ &  $357\pm17$  & $601\pm137$ &  $0.07\pm0.06$  & 0.25 & 23\\
\medskip\\
\noalign{\bf S stars}
\medskip\\
7351 & $4596\pm7$ & $0.18\pm0.01$ & $1.67\pm0.02$ & $2444703\pm20$ & $5.38\pm0.03$ & $105.7\pm1.4$ & $334.4\pm2.5$ & $0.070\pm0.001$ & 0.64 & 76 \\
170970 & $4651\pm10$ & $0.19\pm0.01$ & $-35.68\pm0.03$ & $2457482\pm40$ & $3.60\pm0.03$ & $234\pm3$ & $226\pm3$ & $0.0213\pm0.0007$ & 0.31& 50\\
184185 &15723: & 0: \\
189581 & $618 \pm 1$ & $<0.02$ & $-17.12\pm 0.01$ & $2457037^e$ & $0.59 \pm 0.02$ & - & $5.0 \pm 0.2$ & $(1.3\pm0.1)\times10^{-5}$ & $0.36$ & $52$ \\
215336 & $1143.6 \pm 0.7$ & $0.040\pm0.009$ & $-2.28 \pm 0.05$ & $2454855\pm 30$& $6.90\pm0.02$ & $188\pm9$ & $108.5 \pm 0.1$ & $0.03887 \pm 0.00006$ & $0.25$ & $35$ \\
288833 &28557: & 0.35: \\
218634$^c$ & 194313: & 0.8: \\
BD +79$^{\circ}$156 & $10931\pm41$ & $0.461\pm0.005$ & $-31.87\pm0.05$ & $2468040\pm42$ & $3.16\pm0.05$ & $184\pm1$ & $428\pm8$ & $0.025\pm0.001$ & 0.48 & 68\\
CD $-28^\circ3719$ & $397.5 \pm 0.1$ & $0.042\pm0.002$ & $65.18 \pm 0.02$ & $2457148\pm4$ & $7.84 \pm 0.01$ & $152\pm3$ & $42.83 \pm 0.07$ & $0.0198 \pm 0.0001$ & $0.75$ & 27 \\
BD~$+31^{\circ}$4391 & $6748\pm35$ & $0.16\pm0.02$ & $24.0\pm0.1$ & $2457525\pm91$ & $3.0\pm0.1$ & $149\pm6$ & $280\pm9$ & $0.019\pm0.02$ & 0.47 & 62\\
ER Del & $2081 \pm 2$ & $0.281 \pm 0.003$ & $-48.80 \pm 0.01$ & $2454427 \pm 4$ & $7.12 \pm 0.03$ & $116.8 \pm 0.6$ & $195.7 \pm 0.3$ & $0.0689 \pm 0.0002$ & 1.2 & 41 \\
V420 Hya & $751.4 \pm 0.2$ & $0.099 \pm 0.004$ & $-5.91 \pm 0.05$ & $2449838 \pm 6$ & $10.49 \pm 0.03$ & $271\pm 3$ & $107.9 \pm 0.4$ & $0.089 \pm 0.001$ & $1.19$ & 69 \\
Hen 4-147 & $346.62 \pm 0.04$ & $0.112 \pm 0.006$ & $-5.2 \pm 0.1$ & $2454195 \pm 4$ & $12.3\pm0.2$ & $358 \pm 4$ & $58.2\pm0.8$ & $0.065 \pm 0.003$ & $0.26$ & $45$ \\
$\omicron^1$ Ori & $574.7\pm1.5$ & $0.22\pm0.02$ & $-9.79\pm0.01$ & $2457525\pm6$ & $0.79\pm0.02$ & $196\pm4$ & $6.1\pm0.2$ & $(2.7\pm0.2)\times 10^{-5}$ & 0.44 & $60$\\
\medskip\\
\hline
\end{tabular}

$^a$ Epoch of maximum velocity (circular orbit)\\
$^b$ HD 98839 = 56 UMa \\
$^c$  HD 218634 = 57 Peg\\
$^d$  A solution with $P =$ 10481~d and $e =$ 0.36 is also possible. However, considering the unusual mass function of
$ 0.128\pm0.007$~M$_{\odot}$, this solution is less likely than the shorter solution which has a more common mass function of $0.030\pm0.001$.\\
$^e$ Epoch of maximum velocity. 
}

\end{table*}

\section{Stars of special interest}

\subsection{HD 22589, HD 120620, HD 216219, and BD -10$^\circ$4311}

Although for backward compatibility we kept HD~22589, HD~120620, HD~216219, and BD -10$^\circ$4311 in the binary statistics of our original sample of (giant) barium stars (Table~\ref{Tab:formerly}), the analysis of the Gaia Hertzsprung-Russell diagram \citep{2017A&A...608A.100E} reveals that these stars are dwarf barium stars instead \citep[see Fig.~7 of][]{2019Escorza}. Further discussion of these stars is therefore presented in the companion paper about dwarf barium stars \citep{2019Escorza}.

\begin{figure*}
\includegraphics[width=19cm]{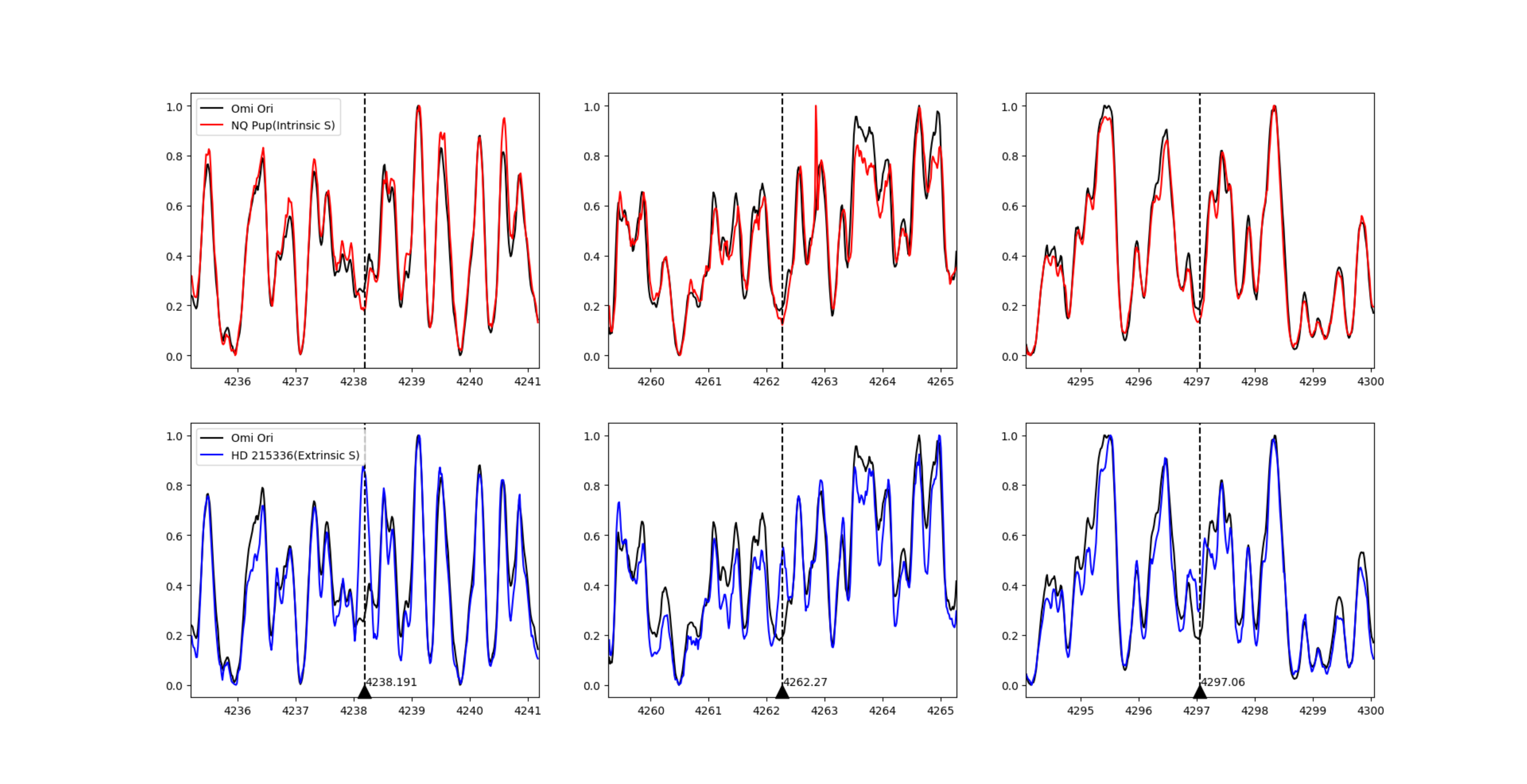}
\includegraphics[width=19cm]{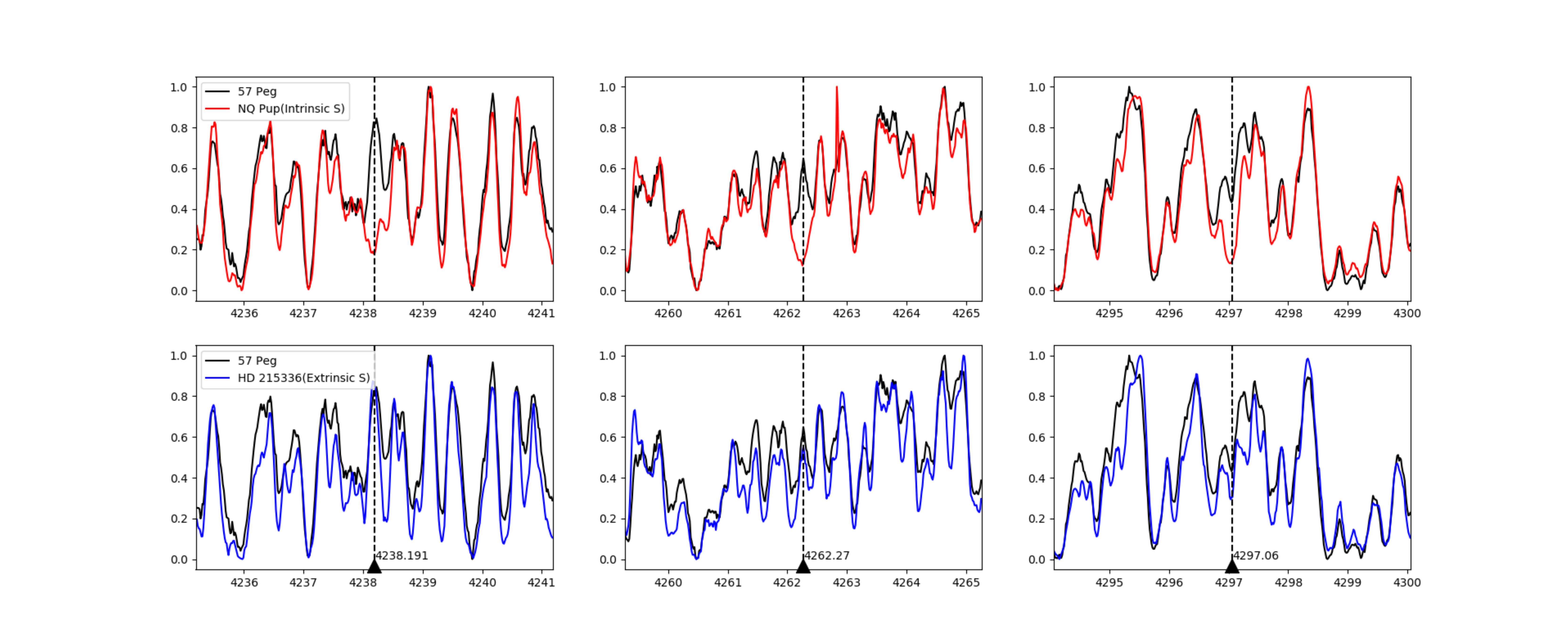}
\caption{\label{Fig:57Peg_Tc} 
The spectral regions around the Tc~I $\lambda$~423.82~nm (left column), 426.23 (middle column), and 429.71 lines (right column). The star   $\omicron$1~Ori (black line) is compared to an intrinsic S star (NQ~Pup; red line, first row) and  to an extrinsic S star (HD~215336; blue line, second row), from HERMES spectra. Third and fourth rows: As in first and second rows, but for 57~Peg compared to the same intrinsic and extrinsic S stars.}
\end{figure*}

\begin{table}
\setlength{\tabcolsep}{10pt}
    \caption{Abundances in the S stars $\omicron$1~Ori (top panel) and 57~Peg  (bottom panel). The column labeled $N$ gives the number of lines used to derive the corresponding abundances.            The uncertainty listed in the $\log \epsilon_X$ column corresponds to the line-to-line scatter, whereas the one listed in column [X/Fe] is the root mean square of the former value and the uncertainty propagating from the  model-atmosphere uncertainties, as estimated for V915~Aql by Shetye et al. (2018; their Table~8). V915~Aql has atmospheric parameters similar to those of $\omicron$1~Ori and 57~Peg.                         
}
    \label{Tab:abundances_omiOri}
    \begin{tabular*}{0.5\textwidth}{rrlrrrrrrrrrrr}
\hline\\
\noalign{\bf \hfill $\omicron$1 Ori \hfill\mbox{}}\\
\noalign{\hfill MARCS model:\hfill\mbox{}}\\
\noalign{$T_{\rm eff} = 3500$~K, $\log g = 0.0$, [Fe/H] = -0.50, C/O = 0.50, [s/Fe]= 0.00}
        \medskip\\      
        \hline\\
$Z$     & X &   $\log \epsilon_X $ &    [X/H] &         \multicolumn{1}{c}{[X/Fe]} & $N$\medskip\\
\hline\\
6       &C      &$8.06$&                -0.37&  \multicolumn{1}{l}{$\phantom{-}0.13$}\\
7       &N      &$7.4$&         -0.4&   $0.07\pm0.63$\\
26      &Fe     &$7.0\pm0.13$&  -0.5 & $\pm0.16$& 12\\  
39      &Y      &$1.8\pm0.00$&  -0.41&  $0.09\pm0.00$& 2\\
40      &Zr     &$2.45\pm0.07$& -0.13&  $0.37\pm0.07$& 2\\
41      &Nb     &$1.02\pm0.17$& -0.44&  $0.06\pm0.17$& 4\\
56      &Ba     &$1.8\pm0$&     -0.4&   $0.12\pm0.10$ & 1\\
60      &Nd     &$1.0\pm0$&     -0.4&   $0.08\pm0.20$& 2\\
\hline\\
\noalign{\bf \hfill 57 Peg \hfill\mbox{}}\\
\noalign{\hfill MARCS model:\hfill\mbox{}}\\
\noalign{$T_{\rm eff} = 3400$~K, $\log g = 1.0$, [Fe/H] = 0.00, C/O = 0.50, [s/Fe] = 0.00}
\medskip\\
\hline\\
$Z$     & X &   $\log \epsilon_X $ &    [X/H] &         \multicolumn{1}{c}{[X/Fe]} & $N$\medskip\\
\hline\\
6       &C      &$8.16$         & -0.27 &\multicolumn{1}{l}{$-0.07$}&\\
7       &N      &$8.60$         &  0.77 &$\phantom{-}0.97\pm0.63$\\
26      &Fe     &$7.25\pm0.14$&-0.25& $\pm0.17$    & 13\\
39      &Y      &$1.95\pm0.07$&-0.26&$0.04\pm0.07$ & 2\\
40      &Zr     &$2.45\pm0.21$& -0.13&  $0.12\pm0.21$ & 2\\
56      &Ba     &$2.0\pm0$&     -0.18&  $0.07\pm0.10$ & 1 \\
60      &Nd     &$1.4\pm0$&     -0.02&  $0.23\pm0.20$ & 2\\
62      &Sm     &$0.85\pm0.21$& -0.11&  $0.14\pm0.21$ & 2\\
\hline  
            \end{tabular*}
\end{table}

\subsection{The star $\omicron$1 Ori (= HD 30959)} 

The star $\omicron$1~Ori is peculiar in many respects. First, it is one of the few S stars with a direct detection of the WD companion from the {\it International Ultraviolet Explorer}  \citep[IUE;][]{1988ApJ...327..214A,1993ApJ...402..667J}\footnote{The title of the paper by \citet{1988ApJ...327..214A} reads 
{\it A white dwarf companion to the main-sequence star 4 Omicron1 Orionis and the binary hypothesis for the origin of peculiar red giants}. We  met H. Johnson soon after his paper was published by  {\it The Astrophysical Journal}, and he confessed that the language editor had changed the MS letters standing for the spectral type, as  originally present in the title, into `main sequence', a change which of course turned the title into astrophysical nonsense. To avoid this ambiguity, we shall use the terminology M/S to denote a star intermediate between M- and S-spectral types.}.
However, $\omicron$1~Ori is also Tc-rich \citep [][and bottom panel of Fig.~\ref{Fig:57Peg_Tc}]{1988ApJ...333..219S}, and as concluded by 
\citet{1988ApJ...327..214A},  is clearly something of an anomaly in that it shows Tc
lines \AJ{and at the same time} also hosts a WD companion, and could therefore be considered as an extrinsic S star. Simple considerations about the implied time scales (as explained below) make it more likely however that $\omicron$1~Ori is a unique example of an intrinsic--extrinsic S star. In other words, this star must have recently entered the thermally pulsing AGB phase responsible for the Tc production, adding to the possible former s-process pollution
from the now extinct AGB companion.   

Since the half-life of $^{99}$Tc, the isotope of Tc involved in the s-process, is
$2.11 \times 10^5$~yr, the presence of Tc on the stellar surface indicates
that less than $10^6$~yr (i.e., a few half-lives) have elapsed
since the last episode of Tc deposition on the surface. This constraint must be compared to the
cooling time of $10^8$~yr for a white dwarf with $T_{\rm eff} = 22\,000$~K \citep{2013A&A...555A..96S}, as observed for $\omicron$1~Ori.
As these time scales are mutually incompatible,  there is no possibility that
the Tc now present on the M/S star (star intermediate between M- and S-spectral types) was transferred from the companion while it was still an AGB star.
The location of $\omicron$1~Ori in the Hertzsprung-Russell diagram, just at the onset of the thermally pulsing AGB\footnote{See Shetye et al. (2018, 2019) \nocite{Shetye2018,Shetye2019}  for other examples of Tc-rich S stars located just at the onset of the TP-AGB.} \citep[see Fig.~6 of][confirmed by Gaia DR2, since the Hipparcos and Gaia DR2 parallaxes are mutually consistent: $\varpi = 6.0\pm0.9$ and $6.2\pm0.4$, respectively]{1998A&A...329..971V}, confirms the intrinsic 
nature of  $\omicron$1~Ori, that is, it is an S star on the TP-AGB capable of producing Tc in its interior and bringing it to the surface. 
The question remains however as to the origin (intrinsic or extrinsic) of the s-process enhancement which
confers $\omicron$1~Ori its distinctive status as an M/S star. 
An important clue in that respect comes from the Nb/Zr chronometer \citep{2015Natur.517..174N,2018A&A...MMMA.NNNK}, and from the orbital elements. In particular, is the system close enough to make the s-process pollution through mass transfer efficient?

\begin{figure*}
    \vspace*{-2cm}
    \includegraphics[width=15cm,height=20cm,angle=-90]{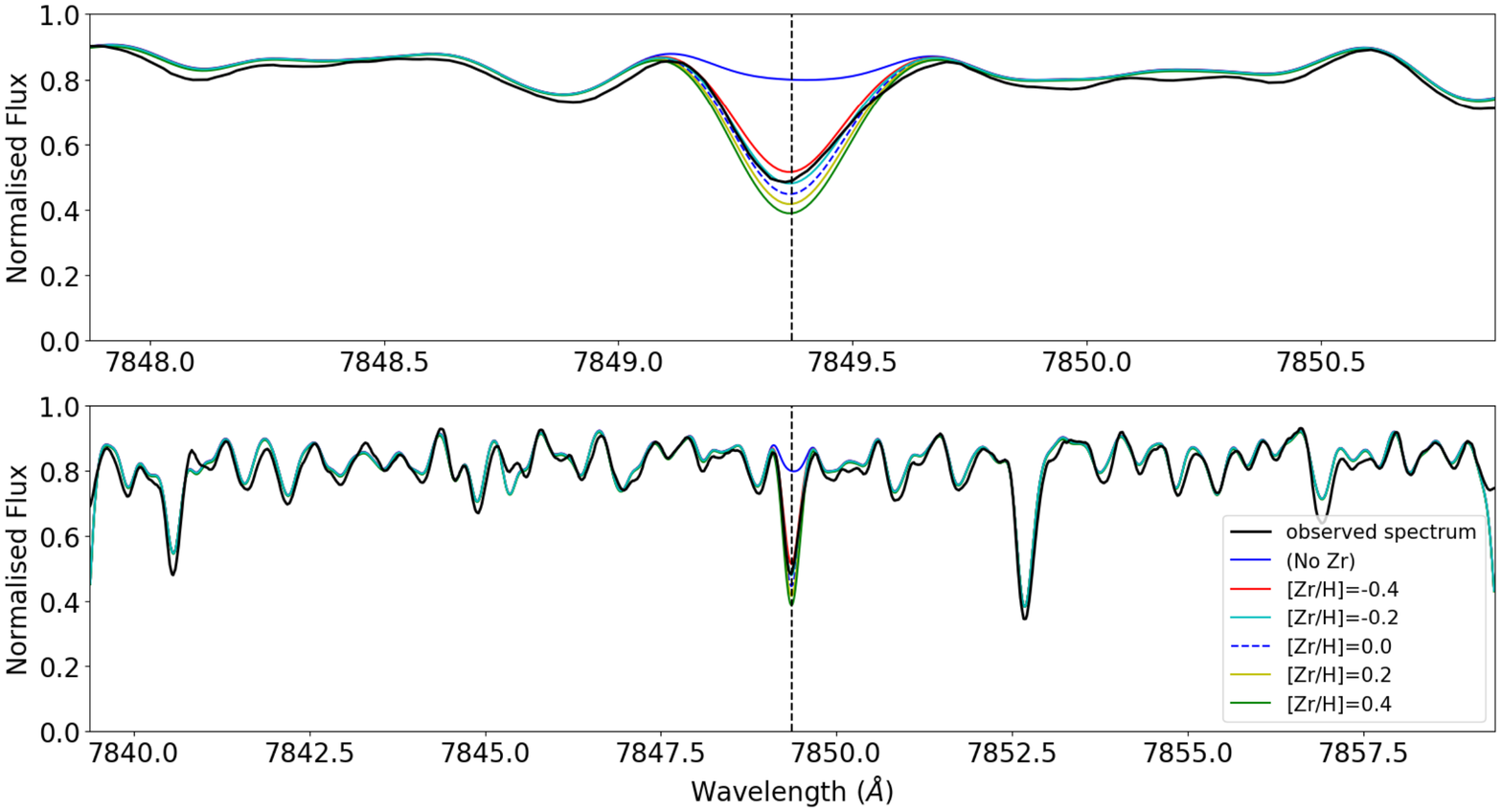}
        \vspace*{-2cm}
    \caption{
    Illustration of the quality of the match between observed and synthetic spectra obtained for the extrinsic S star $\omicron$1~Ori around the \ion{Zr}{I} line at 7849.37~\AA. The upper panel presents a $\pm~3~\AA$ zoom.}
    \label{omiOri_Zr1}
\end{figure*}

To address the first question (intrinsic vs. extrinsic s-process), a basic abundance analysis of $\omicron$1~Ori was performed, following the guidelines presented in \citet{Shetye2018}, using the same iron and s-process lines, and the same procedure to select the model parameters among the large MARCS grid of S-star model atmospheres \citep{2017A&A...601A..10V}.
The adopted model parameters for $\omicron$1~Ori are listed in Table~\ref{Tab:abundances_omiOri}, and these parameters have been validated by the good match between observed and synthetic spectra around CH, Fe, and Zr lines. 
Figure~\ref{omiOri_Zr1} 
for instance illustrates this good match around a \ion{Zr}{I} 
line.

Table~\ref{Tab:abundances_omiOri} presents the abundances derived in $\omicron$1~Ori for elements C, N, Fe, Y, Zr, Nb, Ba, and Nd. The specific MARCS model atmosphere selected for $\omicron$1~Ori is validated {\it a posteriori} by the agreement between the [Fe/H] and [s/Fe] values for the adopted MARCS model and those derived from the detailed abundance analysis (more precisely, they differ by less than one step in the model grid for both [Fe/H] and [s/Fe]).
Comparing the [Nb/Fe] and [Zr/Fe] abundances in $\omicron$1~Ori with those observed in extrinsic and intrinsic S stars (\citealt{2015Natur.517..174N}; also Fig.~14 of \citealt{2018A&A...MMMA.NNNK}; Fig.~15 of \citealt{Shetye2018}) very clearly points towards an intrinsic origin of the $\omicron$1~Ori s-process abundances. The small [Nb/Fe] value indeed indicates that $^{93}$Zr has not yet had time to decay into the only stable Nb isotope, $^{93}$Nb.  Overall, the overabundance in s-process elements in $\omicron$1~Ori is very moderate, with even some negative [X/Fe] values.

\begin{figure}[]
\vspace{-3.5cm}
\includegraphics[width=9cm]{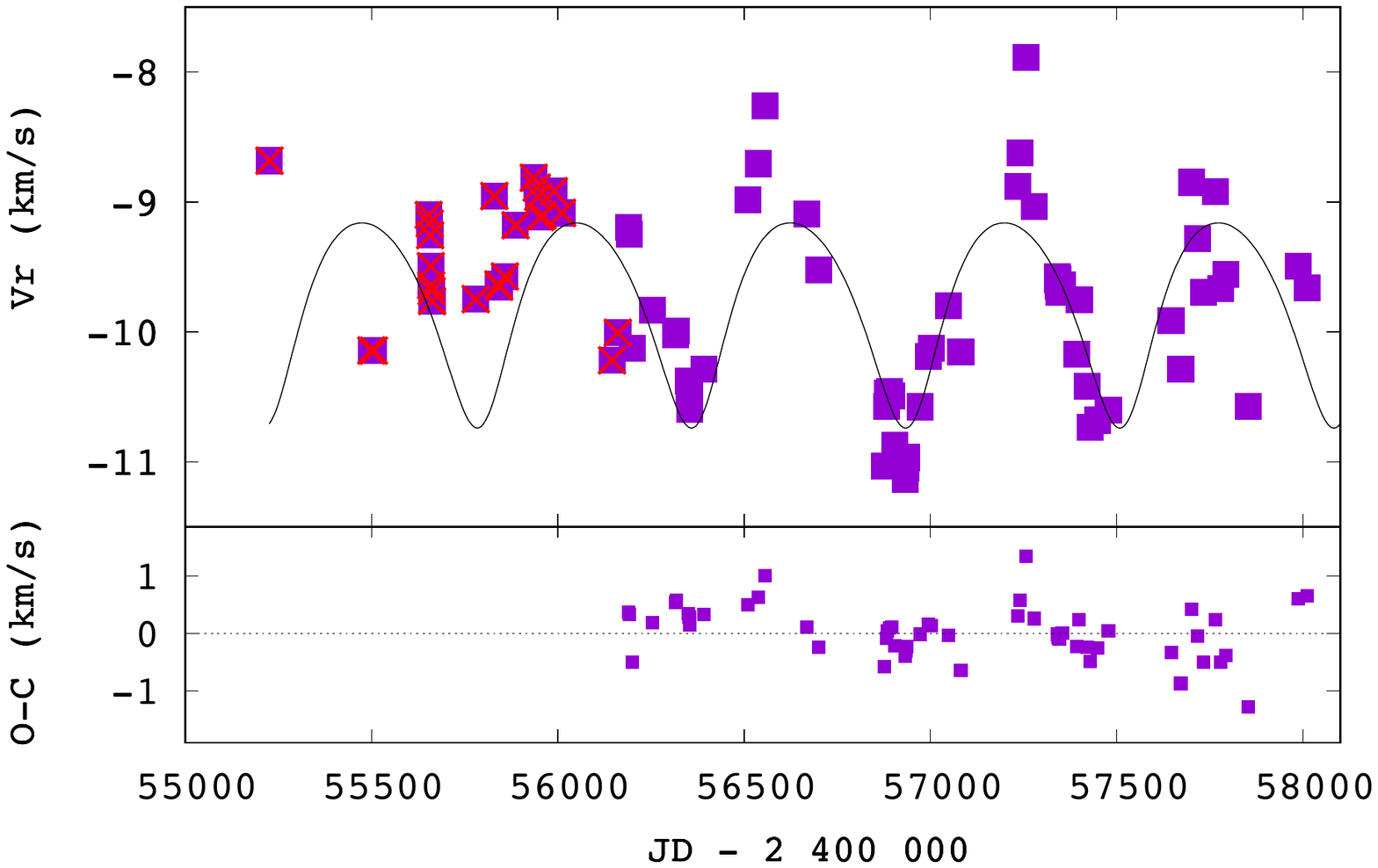}
\vspace{-4cm}
\caption[]{The tentative orbit of $\omicron$1~Ori, based on the magenta square points.
\label{Fig:omiOri}
}
\end{figure}

As far as the latter question (regarding the efficiency of s-process pollution through mass transfer) is concerned, the answer may come from \AJ{the} knowledge of $\omicron$1~Ori orbital elements. It has been very difficult however to extract an orbital signal from the long-term radial-velocity monitoring, because the amplitude of variations is small (2 to 3~\kms) and there seems to be some velocity jitter (Fig.~\ref{Fig:omiOri}).
This jitter could be associated with the envelope semi-regular pulsations with periods of 30.8 and 70.7~d, and amplitudes of 0.047 and 0.046~mag, respectively \citep{2009MNRAS.400.1945T}. Using Eq.~5 of \citet{1995A&A...293...87K} rewritten as Eq.~6 of \citet{1997A&A...324..578J} to relate photometric and radial-velocity jitter, a radial-velocity amplitude of 0.75~\kms\ is predicted to be associated with a photometric visual amplitude of 0.047~mag (adopting $T_{\rm eff} = 3500$~K for $\omicron$1~Ori; Table~\ref{Tab:abundances_omiOri}), in reasonable agreement with the data (bottom panel of Fig.~\ref{Fig:omiOri}).

An orbital solution may be obtained only after discarding data points obtained prior to 2012.7 (red crosses on Fig.~\ref{Fig:omiOri}). Even after this however, the $O-C$ residuals remain large ($\sigma(O-C) = 0.44$~\kms; bottom panel of Fig.~\ref{Fig:omiOri}). The mass function is indeed the second smallest of those reported in Table~\ref{Tab:orbits}, at $(2.7\pm0.2)\times10^{-5}$~M$_\odot$. As we show below, this small mass function is very likely due to a small inclination angle on the plane of the sky.

\citet{1988ApJ...327..214A} fit the IUE spectrum of $\omicron$1~Ori~B with a WD model of 
$\log g = 8$, which according to \citet{1997MNRAS.287..381B} corresponds to a mass of 0.65~M$_\odot$ and a radius of 0.014~R$_\odot$ for the WD (however, \citealt{1988ApJ...327..214A} do not exclude that the gravity might be slightly larger, with $\log g = 8.5$ then resulting in a mass of 0.96~M$_\odot$).
Moreover, 
\citet{2013MNRAS.434..437C} have performed a detailed analysis of  $\omicron$1~Ori using AMBER/VLTI data. They obtained an angular diameter of $9.78\pm0.10$~mas for that star, which, combined with the Gaia DR2  \citep[DR2;][]{2016A&A...595A...1G,2018arXiv180409365G} parallax of $6.2\pm0.4$~mas, yields a radius of 
$170\pm14$~R$_\odot$. Combining this radius with the effective temperature of 3500~K gives a luminosity of 3900~L$_\odot$.
Given the [Fe/H]~$ = -0.5$ metallicity of $\omicron$1~Ori (Table~\ref{Tab:abundances_omiOri}), the above parameters locate the star on the evolutionary track of a 2 -- 2.5~M$_\odot$ star according to Fig.~16 of
\citet{Shetye2018}.
Inserting these masses and their uncertainties into the orbital mass function, we obtain an inclination on the order of only $\sim 3.7^\circ - 5.7^\circ$ on the plane of the sky.

Combining the above mass estimates with the orbital period of 575~d (Table~\ref{Tab:orbits}), one finds a relative semi-major axis in the range 1.9 -- 2.0~au or 404~R$_\odot$. The corresponding Roche radius around the giant component is then on the order of 184 -- 235~R$_\odot$, corresponding to a filling factor on the order of 72 -- 92\% for the observed radius of  $170\pm14$~R$_\odot$. Orbital and diameter data thus indicate that $\omicron$1~Ori is a detached system, possibly with a large filling factor.  
Ellipsoidal variations are not expected though, since the system appears to be seen almost face-on. However, in this case, a noncircular stellar disk could be detected by interferometry, using three non-aligned telescopes. The closure-phase parameter (CSP) may be used to measure the deviation from centrosymmetry of the stellar surface brightness distribution, as done by \citet{2015MNRAS.446.3277C}. The CSP relies on the triple product of the complex visibilities recorded by the three telescopes; its exact definition is beyond the scope of this paper, and we refer the interested reader to the paper by \citet{2014MNRAS.443.3550C}. \AJ{The closure-phase parameter} is equal to $0^\circ$ or $180^\circ$ for a central-symmetric surface brightness distribution. In $\omicron$1~Ori, 
there is a small deviation from centrosymmetry  (CSP~$= 8.1^\circ\pm0.8^\circ$ instead of $0^\circ$ in the central-symmetric case). However, this level of  asymmetry could also be caused by the convective  
pattern at the surface of this giant star \citep[see][for a  discussion of convective vs. tidal asymmetries in giant stars]{2014MNRAS.443.3550C,2014SPIE.9146E..33P}.  The CSP value for $\omicron$1~Ori indeed lies at the expected position along  the sequence of increasing  convective asymmetries with increasing pressure scale-heights along the giant branch \citep[see Figs.~4 and 6 of][]{2015MNRAS.446.3277C}.  Therefore, it is likely that $\omicron$1~Ori shows no sign of tidal deformation, meaning that its Roche-filling factor must be closer to 72\% than to 92\%; hence, the most likely masses are those corresponding to the filling factor of 72\%, namely $M_{\rm S} = 2.5$~M$_\odot$ and $M_{\rm WD} = 0.65$~M$_\odot$.

\subsection{HD 98839 = 56 UMa}

For HD~98839 (=56~UMa), we improve upon the orbit published by \citet{2008Obs...128..176G}. Thanks to 25 new HERMES measurements spanning the years 2009 -- 2016 as listed in Table~\ref{Tab:VrBa}, a full orbital cycle has now been covered for this barium star with an orbital period of $45.1\pm0.3$~yr, one of the longest among barium and extrinsic S stars. 

An offset of +0.6~\kms\  was applied to these HERMES measurements to put them in agreement with those used by
Griffin for his orbital solution. Nevertheless, the systemic velocity listed  in Table~\ref{Tab:orbits} has been converted back into the HERMES/IAU system to ensure consistency with the other orbital solutions. In our orbital solution (shown in Fig.~\ref{Fig:Orbit_98839}), we did not include measurements older than JD~$2\;440\;000$, because they  degrade the orbit quality.

\subsection{HD~134698}
\label{Sect:134698}

HD~134698 has a very large eccentricity ($e \sim 0.95$), and it was not possible to converge to a solution taking into account  all the old CORAVEL data points, as shown in Fig.~\ref{Fig:Orbit_134698}, because the data sampling does not cover the periastron passage
sufficiently well. The solution obtained is very sensitive to the choice of the CORAVEL points used to compute the orbit (choices different from the one displayed in Fig.~\ref{Fig:Orbit_134698} generally lead to eccentricities even closer to one), which is a sign that the solution is not robust.

\subsection{HD 196673}
HD 196673 is a visual double star (WDS 20377+3322) with a separation varying between 2.5\arcsec\ in 1828 and 3.2\arcsec\ in 2014. 
According to the Gaia Data Release 2 \citep{2016A&A...595A...1G,2018arXiv180409365G} the B component is about one magnitude fainter than the barium star, and their $Bp-Rp$ are similar (1.275 and 1.187 for A and B, respectively), as are their parallaxes ($\varpi = 1.62\pm0.03$~mas). This indicates that AB is a pair of red giants, separated by $\sim 1850$~au. Assuming a mass of 1.5~M$_\odot$ for both stars leads to an orbital period of $5\times 10^4$~ yr.
One radial-velocity measurement of  HD~196673B has been obtained (-25.5~\kms; Table~\ref{Tab:VrBa}), close to the systemic velocity of the Aa spectroscopic pair (-24.2~\kms; Table~\ref{Tab:orbits}), further confirming that the visual pair is physical.

\subsection{T Sgr = HD 180196}

\begin{figure}
\includegraphics[width=9cm]{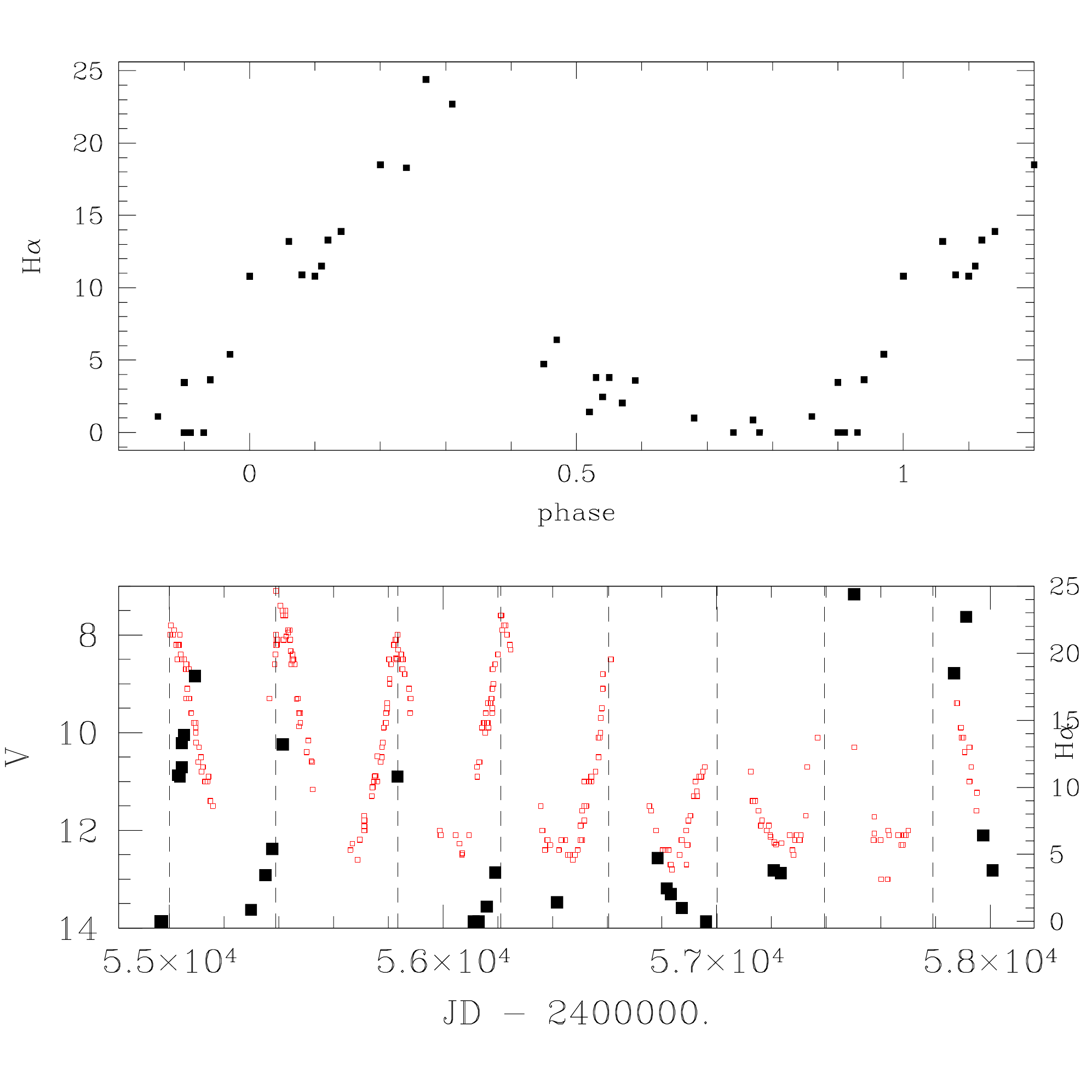}
\caption[]{Bottom panel: Light curve for the S star T~Sgr (small red open squares on bottom panel), from the {\it Association of French Variable Star Observers} (AFOEV). The vertical dashed lines mark the light maxima, either directly identified on the light curve (the first four) or inferred from the GCVS period 394.7~d. The strength of the Balmer H$\alpha$ emission line peak value normalised with respect to the continuum) is represented as solid open squares to be read off the right scale. Top panel: Balmer H$\alpha$ line strength as a function of the photometric phase.  
\label{Fig:TSgr_Ha}
}
\end{figure}

\begin{figure}
\includegraphics[width=9cm,height =17cm]{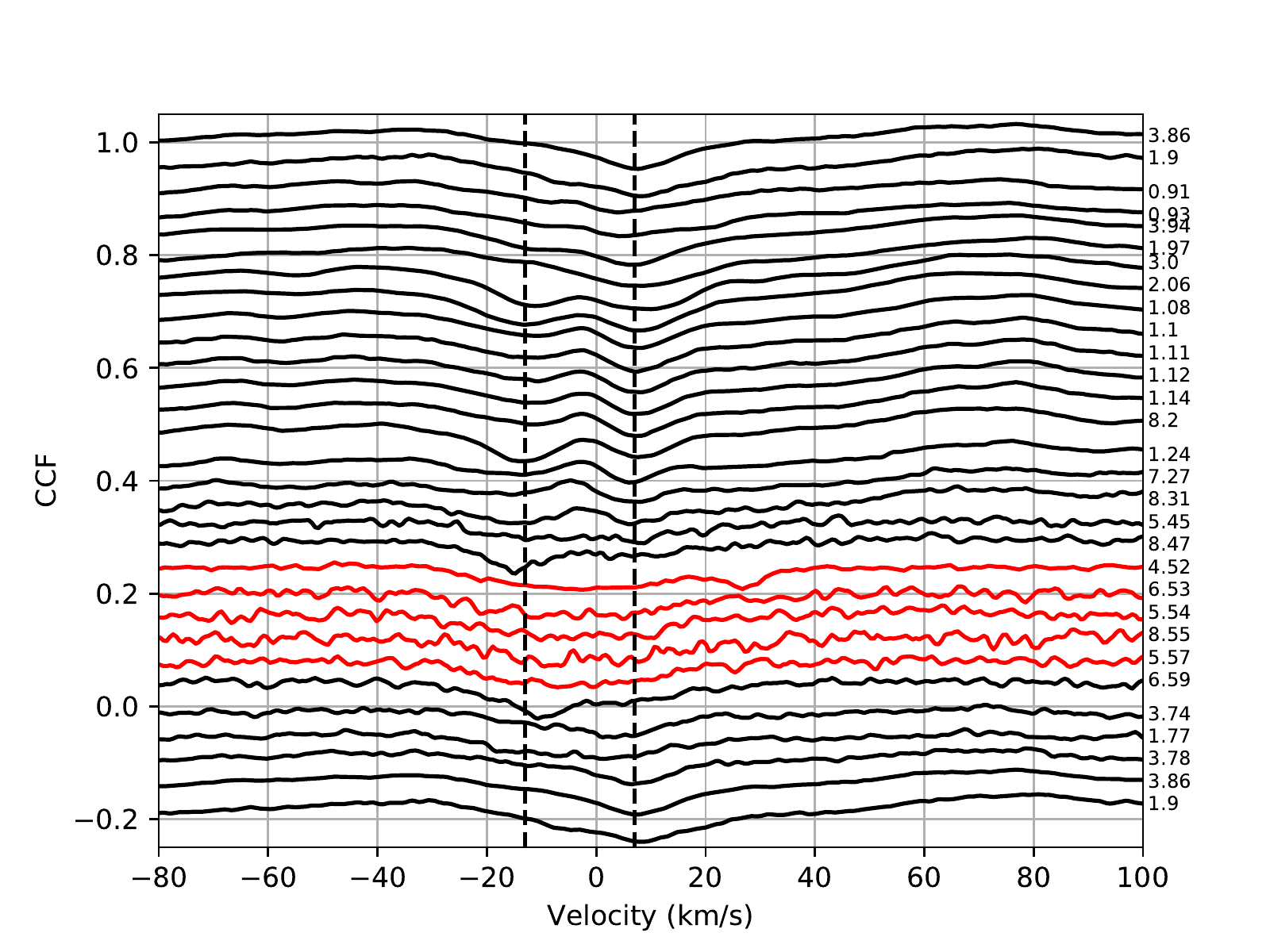}
\caption[]{Cross-correlation functions of the S star T~Sgr (using mask F0) ordered according to  (fractional) photometric phase, counted in cycles (as labelled in the right margin) since the light maximum at JD 2454611 (see Fig.~\protect\ref{Fig:TSgr_Ha}). The vertical dashed lines mark the two peaks appearing in the Mira spectrum between phases -0.1 and 0.3, at velocities -13 and +7~\kms. These peaks exhibit no noticeable orbital drift over the 9 years covered by the radial-velocity monitoring. The noisy CCFs between phases 0.5 and 0.6 (represented in red, and corresponding to the Mira minimum light) reveal a broad single peak, most likely belonging to the F companion, centered around -2~\kms, and with a rotational broadening of about 25~\kms.
\label{Fig:TSgr_CCF_F0}
}
\end{figure}

The Tc-rich S star T Sgr has been included in the monitoring because the star is known to have a composite spectrum, with a F4~IV companion becoming visible near minimum light \citep{1965Herbig,1975ApJ...195L..37C}, and we hoped to detect the velocity drift associated with the orbital motion. However, as we explain below, despite 9 years of monitoring no such drift has been clearly detected, suggesting that the pair must be relatively far apart.

T Sgr is a Mira variable with a period of 394.7~d according to the {\it General Catalogue of Variable Stars} \citep[v. 5.1;][]{2017ARep...61...80S}. This period is confirmed by the AFOEV  which detected light variations with a period varying between 377~d (cycle 3) and 446~d (cycle 2) during the time-span of the HERMES monitoring campaign (bottom panel of Fig.~\ref{Fig:TSgr_Ha}). The photometric phase was computed with an origin at  JD~2454611, and using either the contemporaneous period (when it could be measured from the photometric data, for cycles 1--4) or the GCVS period of 395~d (cycles 5--8). The photometric phase is listed again in the right margin of Fig.~\ref{Fig:TSgr_CCF_F0}, which  shows the series of CCF obtained with a F0 template,  ordered according to the photometric phase.  Along the eight photometric cycles covered, the CCFs have stayed remarkably similar at any given phase, thus showing no sign of orbital drift. 

The Mira has a shock wave traveling through its photosphere around maximum light. This shock wave manifests  as line doubling (between fractional phases -0.1 to 0.3; Fig.~\ref{Fig:TSgr_CCF_F0}), a well-studied behavior known as the Schwarzschild scenario (see \citealt{2000A&A...362..655A} and \citealt{2016ASSL..439..137J} for an illustration of that scenario at work in Mira variables). The red component, corresponding to infalling matter, is the only one present during fractional phases 0.7 -- 0.9, and gives way to an increasingly strong blue component, corresponding to rising matter. In T~Sgr, these two peaks have velocities of about \mbox{-13} and +7~\kms\  (Table~\ref{Tab:F_Vr}).
At the same time, 
 H$\alpha$  in emission gets stronger and stronger (top panel of Fig.~\ref{Fig:TSgr_Ha}).  In that figure, the number characterizing H$\alpha$ emission is simply $[I_{\rm max}(H\alpha)-I(\rm{continuum})]/I(\rm{continuum})$.
 
 \begin{table}[]
     \centering
          \caption{The radial velocities of the F and Mira components of the T~Sgr system as a function of the photometric phase $\phi$, along with the $\sigma$ of the CCF. The last columns list the velocities of the two peaks associated with the shock wave traveling in the Mira photosphere.}
     \label{Tab:F_Vr}
     \begin{tabular}{llllll}
     \hline
     JD & $\phi$ & $Vr$ (F) & $\sigma$(CCF) & $Vr1$ (Mira) & $Vr2$ (Mira)\\
        &       & (\kms) & (\kms) &(\kms) &(\kms)\\
     \hline\\
     2455297.71 & 1.77 & $-0.8\pm0.3$ & 16.3 & - & - \\
     2455413.57 & 2.06 & - & - & $-13.9\pm0.1$ & $8.3\pm0.1$ \\
     2456159.46 & 3.86 & - & - & - & $6.9\pm0.2$ \\
     2456190.45 & 3.94 & - & - & $-12.9\pm0.3$ & $6.4\pm0.2$ \\
     2456416.70 & 4.52 & $-1.8\pm0.4$ & 14.6 & - & - \\
     2456784.70 & 5.45 & $-2.7\pm0.6$ & 20.5 & - & - \\
     2456817.65 & 5.54 & $-3.2\pm0.4$ & 19.0 & - & - \\ 
     2456832.61 & 5.57 & $-4.2\pm0.4$ & 15.0 & - & - \\ 
     2457208.54 & 6.53 & $-3.8\pm0.5$ & 15.0 & - & - \\ 
     2457867.75 & 8.20 &  - & - & $-13.7\pm0.1$ & $8.7\pm0.1$ \\
     2458008.39 & 8.55 & $-2.7\pm0.5$ & 15.8 & - & - \\ 
          \hline\\
     \end{tabular}
 \end{table}
 
 Around phase 0.5 (minimum light), these double peaks give way to a broad single peak, and this feature is especially visible when performing the correlation of the observed spectrum with a F0 mask. The corresponding velocities are listed in Table~\ref{Tab:F_Vr}, which 
 reveals a drift, but its significance is weakened by the broadness of the CCF (on the order of $\sigma \sim 15$~\kms, associated with  a rotational velocity $V_{\rm rot} \sin i \sim 25$~\kms).
 The Mira velocity peaks do not confirm this drift, although a supplementary complication here comes from the fact that the shock-wave velocity may vary from cycle to cycle.
 
 We note as well that the F-star velocity falls almost exactly at mid-range between the two Mira peaks, which is surprising; either the two stars are now going through a conjunction on a very long orbit, or the velocity amplitude of their orbit is small (a few \kms), or indeed the broad CCF seen at minimum light is not at all related to the F star.

\subsection{HD 218634 = 57 Peg}

The preliminary orbit of the S star 57 Peg (HD 218634) stands out, with its orbital period of the order of 500~yr (Table~\ref{Tab:orbits} and Fig.~\ref{Fig:Orbit_57Peg}), the longest period known so far for a chemically peculiar red giant, and probably even among spectroscopic binaries as a whole \citep{2008Obs...128..176G}. To better constrain it, it was necessary to add the old measurements from \citet{1974Obs....94..188G}.
The period is not well constrained, and we do not exclude however that the orbital period could turn out to be shorter (100~yrs?; see the dashed line in Fig.~\ref{Fig:Orbit_57Peg}) when evaluated with measurements spanning a longer time interval.

The Tc-poor star 57~Peg (see bottom panel of Fig.~\ref{Fig:57Peg_Tc}) is special in many respects. First, it has a rather high luminosity ($M_{\rm bol} = -4.3$, with a small uncertainty on its Hipparcos parallax $\sigma_{\varpi}/\varpi = 0.21$; Hipparcos and Gaia DR2 parallaxes for 57~Peg are consistent with each other, with the Gaia parallax being only twice more precise: $\sigma_{\varpi}/\varpi = 0.10$), and it falls along the $Z = 0.2$ evolutionary track of a 3~M$_\odot$ star \citep[Fig.~6 of][]{1998A&A...329..971V}. Second, according to the detailed analysis of its UV colours presented in the Appendix of \citet{1998A&A...329..971V}, it has a composite spectrum with an A6V companion instead of the WD companion expected for extrinsic S stars in the framework of the binary paradigm. Adopting a mass of 1.9~M$_\odot$ for such an A6V companion yields a  $Q$ value  of 0.286~M$_\odot$ (with $Q = M_{\rm A}^3/(M_{\rm S}+M_{\rm A})^2$), assuming a mass of 3~M$_\odot$ for the S star. Incidentally, the S star primary must have evolved faster and should therefore be more massive than 1.9~M$_\odot$, which is consistent with its position along the 3~M$_\odot$ track in the HR diagram. 
Despite the fact that the orbit is not yet fully constrained, it yielded a mass function of $f(M_{\rm S},M_{\rm A}) = 0.38\pm0.22$~M$_\odot$, compatible (within the error bars) with the above-predicted value for a 1.6~M$_\odot$ companion (since $\sin^3i = f(M_{\rm S},M_{\rm A}) / Q$, and therefore $f(M_{\rm S},M_{\rm A})$ should be smaller than $Q$).  
However, a WD companion with a mass of 0.7~M$_\odot$, which would yield $Q = 0.025$~M$_\odot$, seems incompatible with the observed mass function and the condition $f(M_{\rm S},M_{\rm A}) \le Q$.
All evidence therefore suggests that the companion is on the main sequence. The star 57~Peg thus adds to the small set of Tc-poor S stars (HD~191589, HDE~332077) with a main sequence companion \citep{1992A&A...260..115J,1992ASPC...26..579A,1994ASPC...64..678A,1998A&A...332..877J}.

Faced with such strong evidence, possibilities to resolve this puzzle within the framework of the
binary paradigm include scenarios where (i) 57 Peg is a triple system (S+A6V+WD), (ii) the companion is an accreting WD mimicking a main
sequence spectrum, (iii) 57~Peg is an intrinsic (Tc-rich) rather than a Tc-poor S star, and finally (iv) 57 Peg is not an S star at all. 

Possibility (i) is incompatible with the long-period orbit of the system, since a triple system needs to be hierarchical to be stable, with a period ratio on the order of ten. Since the 500 yr orbit is that of the main-sequence companion (as derived from the mass function), a 5000 yr orbit  is implied for the WD companion (a 50 yr orbit is not possible, since it would be detected first, having a larger velocity amplitude). However, a 5000 yr orbit (corresponding to $1.8\times10^6$~d) would never yield large enough pollution levels to transform the accretor into an extrinsic S star, since the longest orbital periods among our representative samples of extrinsic stars do not exceed $4\times10^4$~d (see also Fig.~\ref{Fig:Period_abundance} showing how [s/Fe] decreases with increasing orbital periods).

 Possibility (ii) is neither supported by the mass function, which calls for a genuine A6V star rather than a rejuvenated WD, nor the IUE SWP
spectrum, which carries no sign of binary interaction (no \ion{C}{IV}~$\lambda$ 155.0~nm emission for instance).

Possibility (iii) is refuted by Fig.~\ref{Fig:57Peg_Tc}, which clearly demonstrates the Tc-poor nature of 57~Peg, based on a HERMES spectrum around the \ion{Tc}{I} $\lambda$~423.82~nm, 426.23, and 429.71 lines.

\begin{figure*}
\vspace*{-2cm}
    \includegraphics[width=15cm,height=20cm,angle=-90]{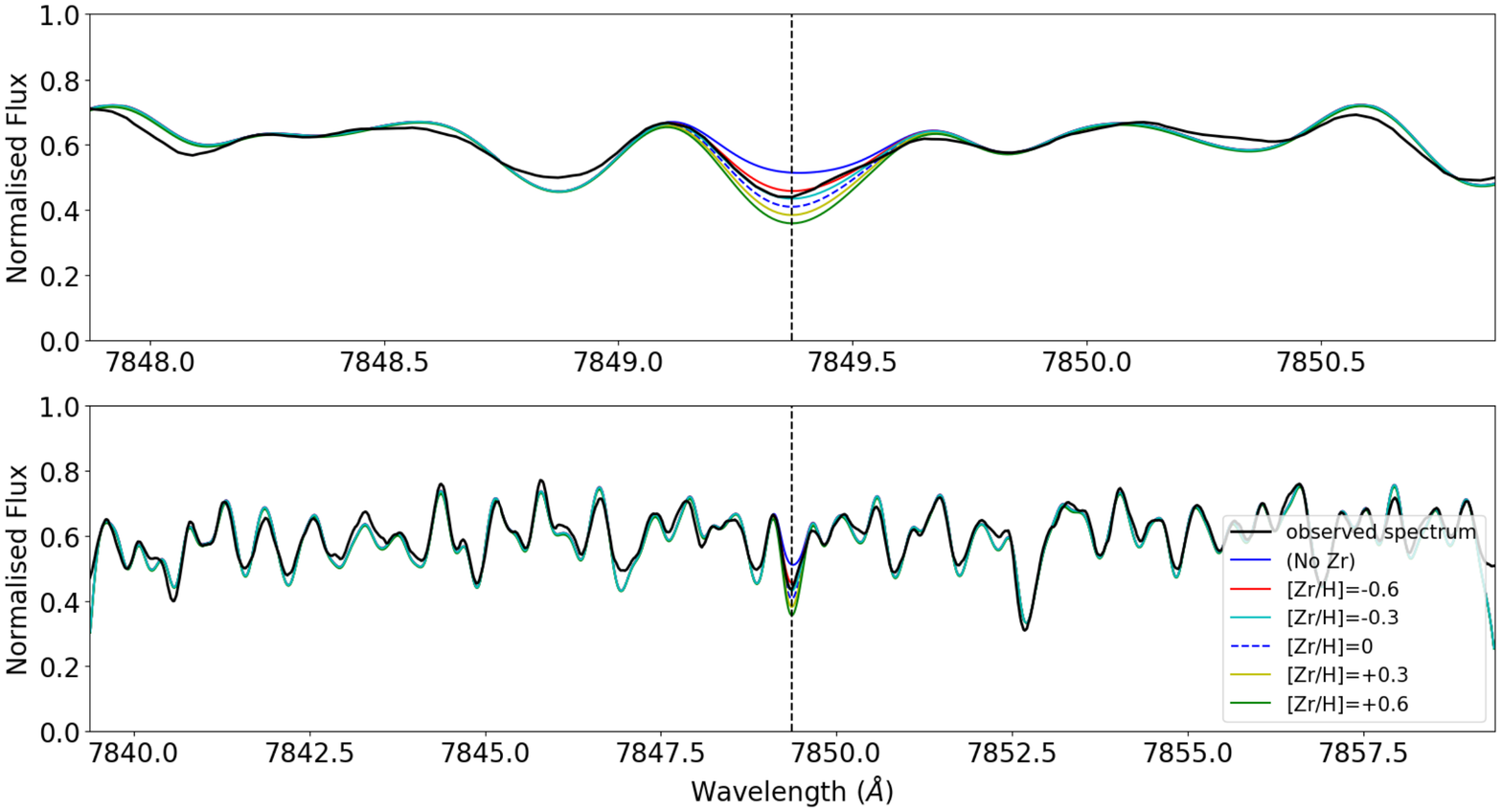}
    \vspace*{-2cm}
    \caption{As in Fig.~\protect\ref{omiOri_Zr1} but for 57~Peg.}
    \label{57Peg_Zr1}
\end{figure*}

Possibility (iv) -- that 57~Peg is not an S star -- was already suggested by \citet{1988ApJ...333..219S}. This hypothesis can be tested from an abundance analysis of the s-process elements in 57~Peg. For this purpose, we use an HERMES spectrum of 57~Peg obtained on September 6, 2009 (JD~2455080.640), with a signal-to-noise ratio of 150 in the $V$ band. The atmospheric parameters of 57~Peg were derived following the method described by \citet{2017A&A...601A..10V} and 
\citet{Shetye2018}. The adopted model parameters are listed in the bottom panel of Table~\ref{Tab:abundances_omiOri}. 
The abundance analysis has been performed using the same iron and s-process lines as in \citet{Shetye2018}. We note especially that the lines used were located far enough in the red not to be contaminated by light from the A-type companion. Figure~\ref{57Peg_Zr1} 
presents the good match between the synthetic and observed spectra around a \ion{Zr}{I} 
line. Table~\ref{Tab:abundances_omiOri} reveals that all the heavy elements studied have an abundance compatible with the solar-scaled value. The error bars quoted in column [X/Fe] of that table include, on top of the line-to-line scatter,  the uncertainty propagating from the model-atmosphere uncertainties. The latter was estimated by  Shetye et al. (2018; their Table~8) for the S star V915~Aql and applied here to 57~Peg, since both stars have similar atmospheric parameters (same $T_{\rm eff}$ but $\log g$ differing by 1~dex).

There is however a surprising N overabundance \citep[larger than expected after the first dredge-up, if taken at face value; on that topic, see also][]{2018A&A...MMMA.NNNK}. Therefore, our conclusion regarding 57~Peg is  that this star has been misclassified as an S star, in line with the suggestion by  
\citet{1988ApJ...333..219S}.

\section{The eccentricity--period diagram}
\label{Sect:P-e}

\begin{figure*}
\vspace{-2.5cm}
\includegraphics[width=15cm]{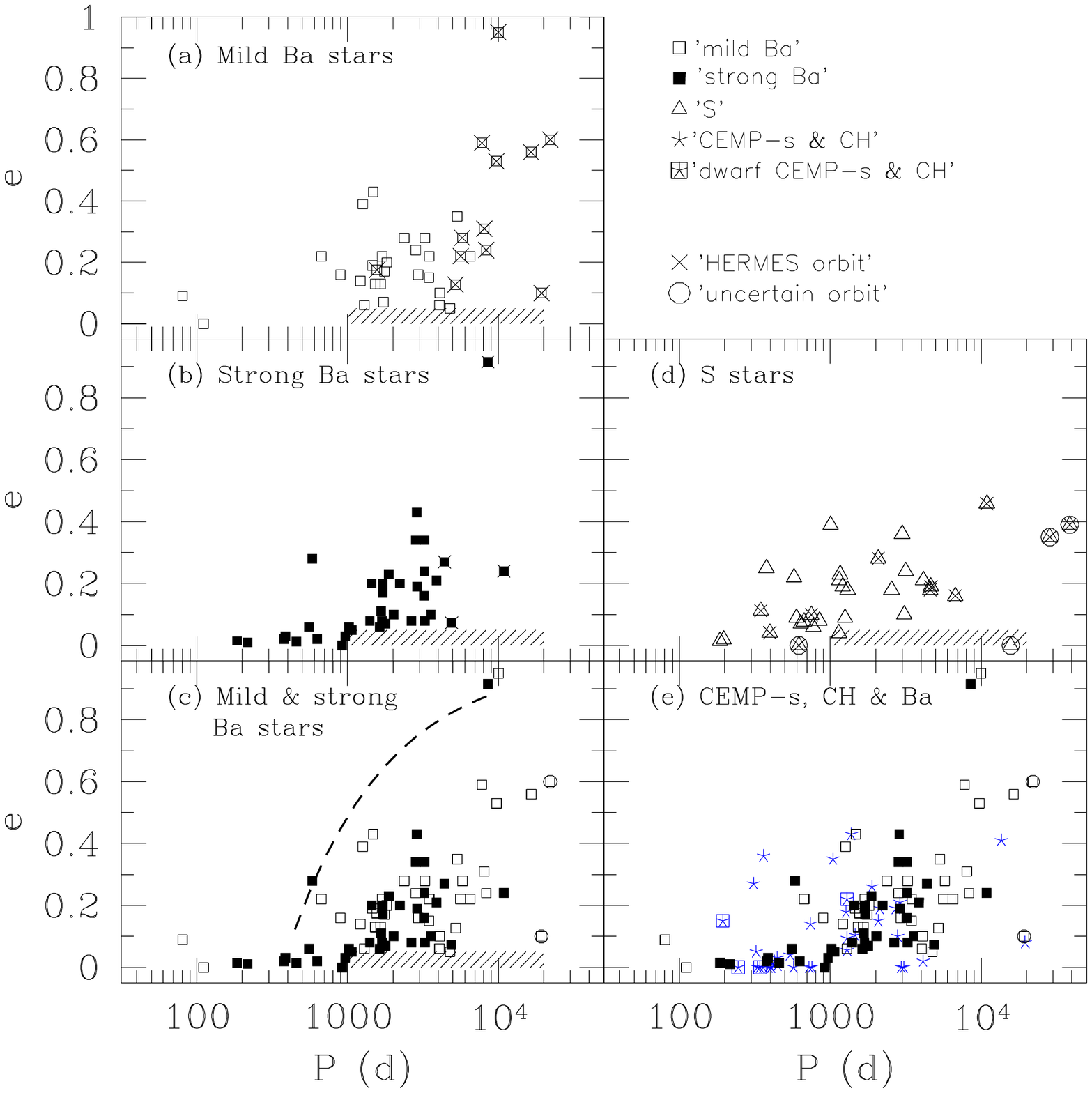}
\vspace{-2.5cm}
\caption{\label{Fig:ePBa} 
Eccentricity--period diagrams for various samples. The meaning of the various symbols is given in the upper right panel.  (a) Mild barium stars;  (b) Strong barium stars; (c) Mild (open squares) and strong (filled squares) barium stars plotted together.  The dashed line corresponds to the upper envelope of the data points, well represented by the condition 143~R$_\odot = R_{\rm Roche} = A \;(1-e)\; [0.38 + 0.2 \log (M_{\rm Ba} / M_{\rm WD})]$, corresponding to RLOF occurring at periastron for a star of radius 143~R$_\odot$. $A$ is the semi-major axis of the orbit, linked to the orbital period $P$ through the third Kepler law, adopting component masses of  $M_{\rm Ba} = 2$~M$_\odot$ and $M_{\rm WD} = 0.65$~M$_\odot$;
 (d) S stars (triangles). The S star HD~184185, with $P \sim  15723$~d and $e \sim 0$, falls in the low $e$ -- long $P$ gap (represented by the hatched area), probably as a consequence of its still-uncertain orbital parameters;
(e) As in (c),  adding CEMP-s and CH stars \citep[blue 5-branch crosses;
squared crosses correspond to carbon dwarfs from ][]{2016A&A...586A.158J}. The dwarf CEMP star  HE~0024-2523 with an orbital period of  3.4~d  falls outside the graph boundaries.
}
\end{figure*}

With the addition of the  new HERMES orbits to the existing sample \citep{1998A&A...332..877J,2017A&A...597A..68V},  the number of giant barium and S stars with orbital elements available now amounts to 105 systems (36 strong Ba, 37 mild Ba, and 32 S stars).  In the remainder of this paper, we analyze this rich data set, starting with the $e - P$ diagram (Fig.~\ref{Fig:ePBa}). This diagram reveals distinctive features that may be used as benchmarks
for binary-evolution models:
\begin{itemize}
\item the upper left threshold (represented by the dashed line in panel (c) of Fig.~\ref{Fig:ePBa}), due to tidal evolution or periastron mass transfer;
\item the lower right gap (represented by the hatched area in Fig.~\ref{Fig:ePBa}), survival from initial conditions observed in young binaries like pre-main sequence stars;
\item the existence of two populations: a population with (nearly) circular, short-period ($P < 10^3$~d) orbits (almost exclusively found among strong barium stars), and a population of eccentric systems with intermediate ($10^3 \le P({\rm d}) \le 10^4$) and   very long periods ($P > 10^4$~d).
\end{itemize}

With all orbits now available, including the very long-period ones, we can state that the longest orbital period where s-process pollution through mass transfer may produce an extrinsic star is about $4\times10^4$~d ($\sim 110$~yr), since no system with a period longer than this value is found in our samples. This period however is the post-mass-transfer value, which certainly differs from the initial value. This maximum  period provides an interesting constraint for binary mass-transfer models,  which often predict the possibility of forming extrinsic systems with even longer periods \citep[as long as a few $10^5$~d; see][and references therein]{2015A&A...581A..62A,2018A&A...620A..63A}.

Extrinsic S stars (Panel d of Fig.~\ref{Fig:ePBa}) do not add new features or structure in the $e - P$ diagram; they confirm the division of Ba stars in two populations in the $e - P$ diagram.
The maximum eccentricity at a given period is similar for barium and S stars (if one excepts the presence of two barium stars at $P \sim 10^4$~d with large eccentricities -- $e > 0.9$, with no equivalent among S stars, but this may result from small-number statistics). Extrinsic S stars
are thus fully identical to barium stars in terms of their orbits. 

The population of (almost) circular barium stars with $P < 1000$~d is likely to contain objects that were circularized by tidal effects while the current barium star was ascending the first giant branch \citep[supposing that most barium stars are currently located in the red-giant clump, as suggested by the analysis of their Hertzsprung-Russell diagram;][]{2017A&A...608A.100E}. 
This circularisation process is posterior to and independent from the mass-transfer process.  We justify the above statement by the fact that 
S stars, which are still on the RGB, are not yet fully circularized in the same period range, as indicated by the large clump of S stars observed around 
$P \sim 700$~d and $e \sim 0.08$. A similar argument holds for the $e - P$ diagram of post-AGB and dwarf barium stars, which also include short-period noncircular systems \citep{2019Oomen,2019Escorza}. One may actually wonder why these \AJ{post-AGB, dwarf-barium, and S systems with short orbital periods} are not circularized as they have hosted a large AGB star in the past. Several authors argued that an as yet not fully identified process (e.g. periastron mass transfer, tidal interaction with a circumbinary disk, or a momentum kick associated with the white-dwarf formation) must have been at work during the mass-transfer process to counteract the circularisation process and to pump the eccentricity up \citep{2000A&A...357..557S,2010A&A...523A..10I,2013A&A...551A..50D}.

We stress moreover that the population of (almost) circular barium stars with $P < 1000$~d is almost absent among mild barium stars (with the exception of HD~77247 and HD~218356). Therefore, there must be a link between the mass-transfer properties and the resulting s-process overabundances to account for the near absence of short-period systems among systems with mild s-process overabundances. This is further discussed in Sect.~\ref{Sect:abundances}. As a corollary, we note that HD~199939, with $P = 585$~d and $e = 0.28$, is an outlier among strong barium stars, having a large eccentricity for its period (panel b of  Fig.~\ref{Fig:ePBa}). Its orbital elements were obtained by \citet{1990ApJ...352..709M}, and a closer look at their orbital solution does not reveal any anomaly (like the presence of a third companion) that could account for its outlying nature. Finally, we stress that the segregation between mild and strong barium stars, used to draw panels (a) and (b) of Fig.~\ref{Fig:ePBa}, although initially based on the Warner visual index of the strength of the \ion{Ba}{II} line \citep{1965MNRAS.129..263W}, is generally confirmed by a detailed abundance analysis (as further discussed in Sect.~\ref{Sect:abundances}). Adopting [La/Fe] and [Ce/Fe] values of 1~dex as threshold between mild and strong barium stars, we find only two stars that need to be reclassified: 
HD~183915 (from mild to strong), and NGC~2420-173 (from strong to mild; see Table~\ref{Tab:abundances}).

The CH and CEMP stars \citep[panel (e) of  Fig.~\ref{Fig:ePBa} from][]{2016A&A...586A.158J}, which are the low-metallicity counterparts of barium stars, behave exactly as barium stars, in particular regarding the circular nature of most of the short-period orbits. The only notable difference is the presence of a very short-period ($P < 10$~d) orbit, but this one is associated with a dwarf carbon star. The few CH/CEMP systems falling in the low-eccentricity gap probably have inaccurate values for the eccentricity, as discussed by \citet{2016A&A...586A.158J}.

The correlation between abundances and location in the $e - P$ diagram is discussed in Sect.~\ref{Sect:abundances}.

\section{Mass distribution of barium stars and their white dwarf companions}
\label{Sect:mass}

\subsection{Methods}
\label{Sect:methods7}

In previous papers, we obtained the mass \citep{2017A&A...608A.100E} and mass-ratio \citep{2017A&A...597A..68V} distributions of barium stars. The formerly  obtained mass distribution  however had only a statistical value, since it was based on the average metallicity\footnote{For the sake of simplicity, in the remainder of this paper, we do not differentiate between metallicity (usually denoted $Z$ in the context of stellar evolution) and [Fe/H], thus neglecting any possible decorrelation between these two quantities due to possible enrichments of N and C in barium stars. Their due consideration would require the use of grids of models accounting for specific C and N abundances, which is beyond the scope of this paper. The current STAREVOL models use the solar photosphere abundance table of \citet{Asplund2009} with [Fe/H] = 0 corresponding to $Z = 0.0134$.} of barium stars ([Fe/H] = -0.25), rather than on their individual values. 
Masses derived from the comparison between evolutionary tracks and location in the Hertzsprung-Russell diagram (HRD), as done by \citet{2017A&A...608A.100E}, are sensitive to the metallicity and therefore do require prior derivation of individual metallicities to reach the ultimate accuracy. Metallicities of barium stars have now been collected from the literature, and when not available were derived from high-resolution HERMES spectra \citep{2011A&A...526A..69R}. The derivation of the atmospheric parameters was performed as described in \citet{2018A&A...MMMA.NNNK}, and the luminosities as described in \citet{2017A&A...608A.100E}.  The Fe line list used is given in Table~\ref{Tab:Felinelist}, with metallicities in Table~\ref{Tab:abundances}. Masses are then derived by matching the position of the barium star in the HRD with STAREVOL evolutionary tracks of the same metallicity \citep{2017A&A...597A..68V,2017A&A...608A.100E}. In case of ambiguities, when tracks of different masses pass close to the location of the star in the HRD, 
we use a statistical criterion that compares the speed of evolution along the different  tracks at a given location in the HRD and select the slowest one \citep[see the discussion around Eqs.~2 and 3 in][]{2017A&A...608A.100E}. The resulting metallicities and masses  
are listed in Table~\ref{Tab:abundances}, along with  heavy-element abundances derived as outlined by \citet{2018A&A...MMMA.NNNK}.

We used Gaia DR2 parallaxes \citep{2018arXiv180409365G}  to derive the distances and  luminosities following the method outlined in \citet{2017A&A...608A.100E}. Gaia DR2 parallaxes result exclusively from single-star solutions. As shown by \citet{2000A&AS..145..161P}, the absence of binary processing by the astrometric pipeline could lead to incorrect parallaxes only when the orbital motion with a period close to 1 yr confuses the parallactic motion. In our sample, only two stars match this criterion (DM~$-64^\circ$4333, HD~24035; see Table~\ref{Tab:abundances}) and therefore their masses listed in Table 8 are subject to caution. Nevertheless, the corresponding WD masses for these two stars do not look peculiar.

The distribution of mass ratios\footnote{In this paper, we use the notation $M_{\rm Ba}$ or $M_{\rm S}$ to designate the barium-star or S-star mass (i.e., the primary component of the current system), and $M_{\rm WD}$ to designate the companion mass (i.e., the secondary component of the current system), even though the demonstration that the companion is indeed a WD only comes in the present section.} ($q = M_{\rm WD} / M_{\rm Ba}$) is obtained from the distributions of primary masses $M_{\rm Ba}$ and mass functions $f(M_{\rm Ba},M_{\rm WD})$ under the assumption that  the orbital inclination is randomly distributed according to $g(i) = \sin i$, since 
\begin{equation}
f(M_{\rm Ba},M_{\rm WD}) = \frac{M_{\rm WD}^3}{(M_{\rm Ba}+M_{\rm WD})^2}\;\sin^3 i = M_{\rm Ba} \frac{q^3}{(1+q)^2} \;\sin^3 i.
\end{equation}
To derive the distribution of $q$, we use the method designed by \citet{1992btsf.work...26B}, which relies on a Richardson-Lucy deconvolution and has proven to be very robust and reliable \citep[see][]{1993A&A...271..125B,1994InvPr..10..533C,2004RMxAC..21..265P,2010A&A...524A..14B,2012ocpd.conf...41B,2017A&A...597A..68V}.

In principle, the distribution of mass ratios $q$ has only a statistical meaning and it is difficult to attribute a given mass ratio to a specific system.
Nevertheless, this may be attempted under two different hypotheses: (i)  finding the {\it most peaked} distribution of the companion masses, as expected if they are WD companions, or (ii) finding the distribution corresponding to a constant $Q = M^3_{\rm WD}/(M_{\rm Ba} + M_{\rm WD})^2$ value, separately for mild and strong barium stars. The two methods are discussed in turn in what follows.

Since $M_{\rm WD} = q \times M_{\rm Ba}$, the most peaked $M_{\rm WD}$ distribution may be obtained by performing the $q \times M_{\rm Ba}$ product with $q$ and  $M_{\rm Ba}$ sorted in opposite order (i.e., largest $q$ combined with smallest $M_{\rm Ba}$, and so on). The number of occurrence of each $q$ in this list of products is fixed by its frequency distribution $f(q)$ (Fig.~\ref{Fig:q_distribution}) multiplied by the sample size of $M_{\rm Ba}$. To limit the round-off errors due to the small sample size (only $N = 24$ strong barium stars), the sample size has been multiplied by ten, meaning that each individual $M_{\rm Ba}$ value appears ten times in the list, while each $q$ bin value appears $f(q) \times N \times 10$ times. 

\begin{figure}
\includegraphics[width=9cm]{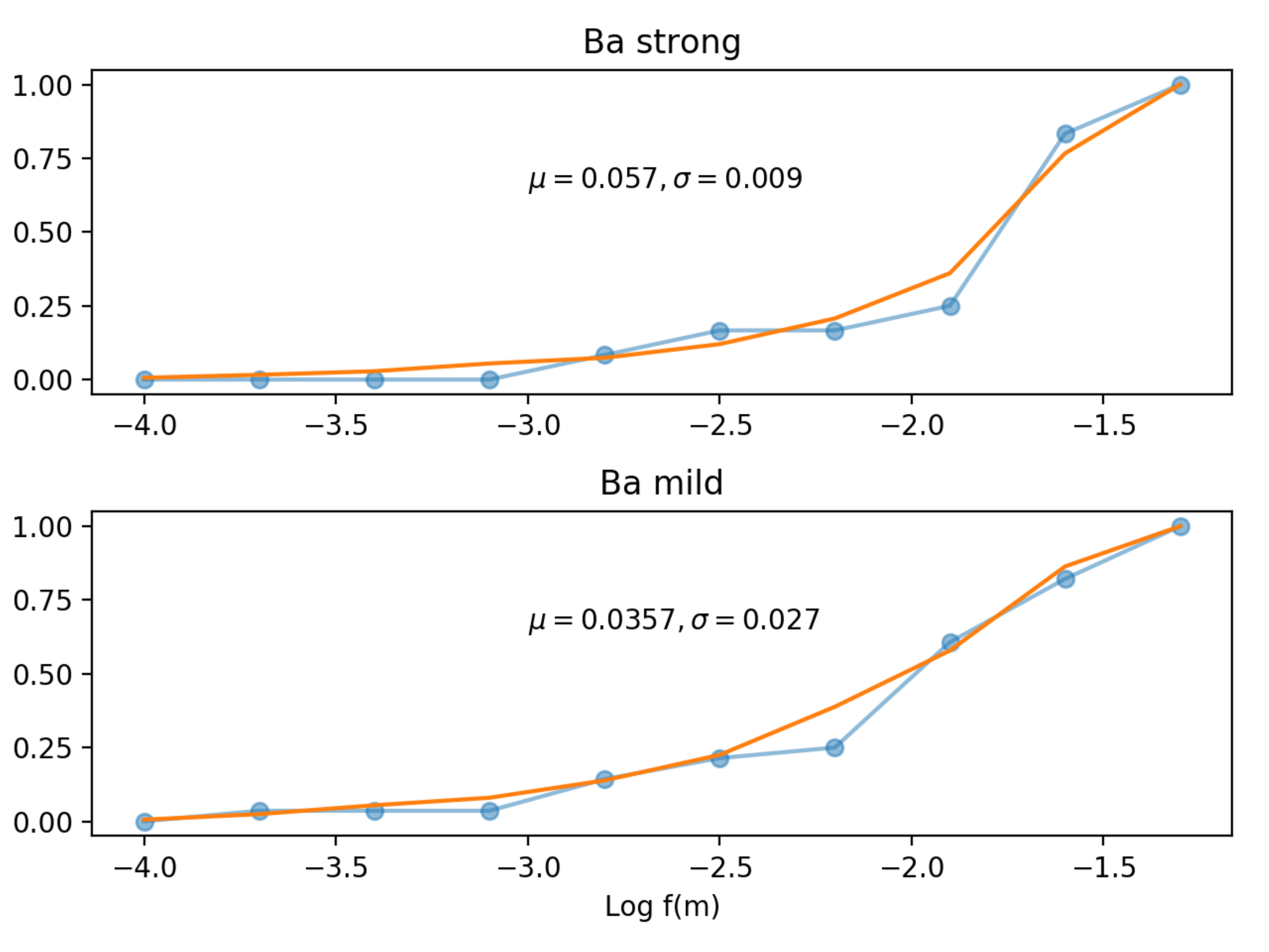}
\caption{\label{Fig:Q} 
Cumulative mass-function distributions for mild and strong barium stars, as compared to those inferred from a Gaussian distribution of $Q = M_{\rm Ba} \; q^3/(1+q)^2$ (with the Gaussian parameters as mentioned in the figure: $\mu$ is the Gaussian average and $\sigma$ its standard deviation, both in M$_\odot$), convolved with random orbital inclinations.
}
\end{figure}

An alternative method to derive the mass-distribution of WDs is based on the assumption of constant $Q = M^3_{\rm WD}/(M_{\rm Ba} + M_{\rm WD})^2$.
\citet{1988covp.conf..403W} showed that the distribution of $f(M_{\rm Ba},M_{\rm WD}) \equiv Q \sin^3 i$ known at the time for barium stars was compatible with a single value of $Q = M_{\rm Ba} \; q^3/(1+q)^2$.
This correlation between the masses of the barium star and its WD companion is understandable since the more massive the barium star is, the more massive its companion had to be (and hence its WD progeny).
Looking at the current sample, it appears that for the strong barium stars $Q$ is indeed sharply peaked at $0.057 \pm 0.009$~M$_\odot$, whereas for mild barium stars the distribution of $Q$ may be approximated by a somewhat wider Gaussian ($0.036 \pm 0.027$~M$_\odot$; Fig.~\ref{Fig:Q}). The good agreement between observed and modeled distributions is confirmed by a Kolmogorov-Smirnov test. A similar analysis on an earlier sample of barium-star orbits \citep{1998A&A...332..877J} yielded similar values ($0.049$~M$_\odot$ and $0.035$~M$_\odot$ for strong and mild barium stars, respectively).
Considering the result that $Q$ is basically fixed (a very good approximation at least for strong barium stars), it will be possible to extract $M_{\rm WD}$ from $Q$ and $M_{\rm Ba}$ for each system. 

\subsection{Results}
\label{Sect:results7}

\begin{figure}
\vspace{-2cm}
\includegraphics[width=9cm]{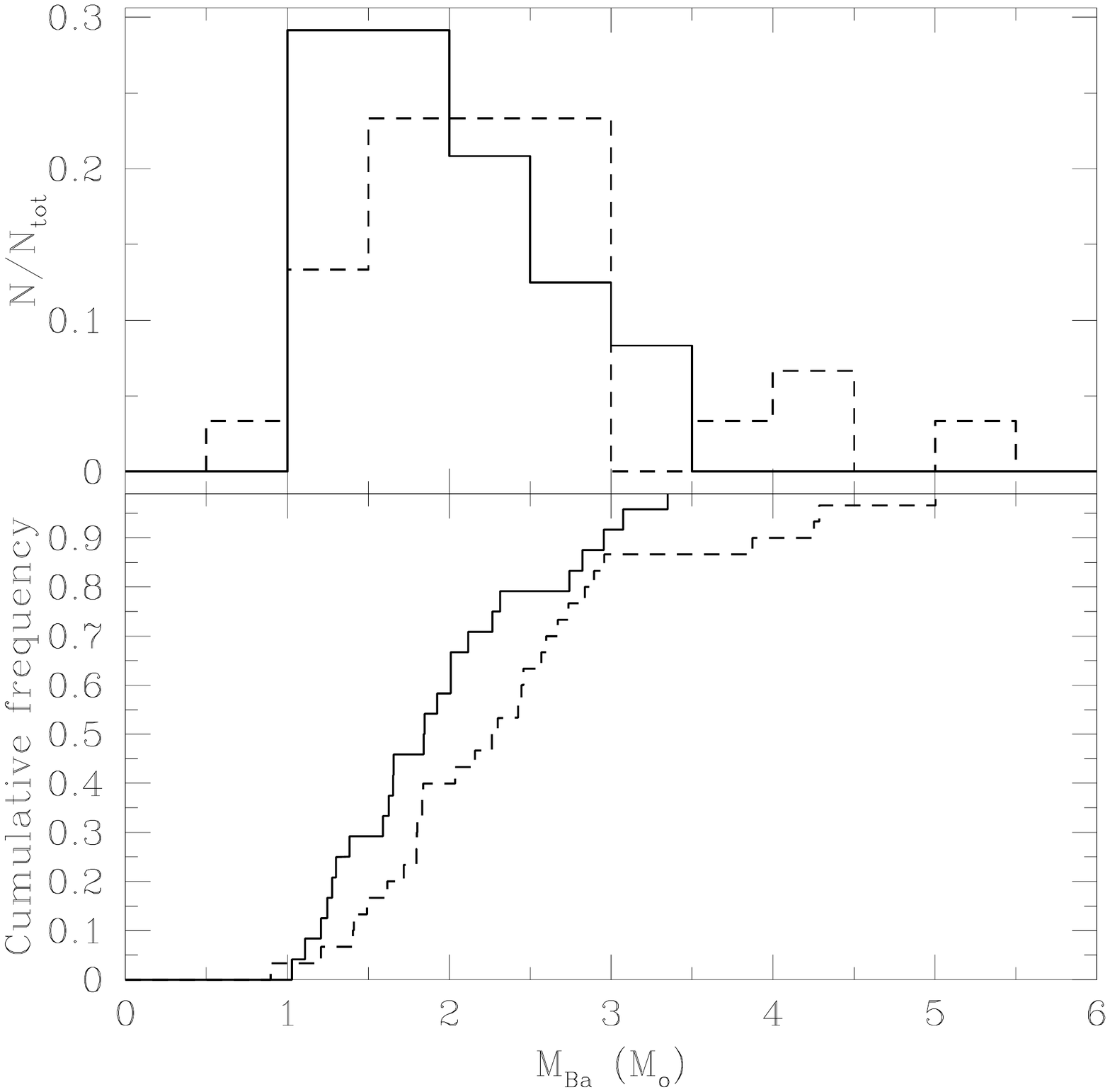}
\vspace{-2cm}
\caption{\label{Fig:mass_distribution} 
Top panel: Mass distributions for mild (dashed lines) and strong (solid lines) barium stars. Bottom panel: Cumulative  mass distributions. The maximum vertical difference between the two distributions amounts to 0.29.
}
\end{figure}

\begin{figure}
\vspace{-2cm}
\includegraphics[width=9cm]{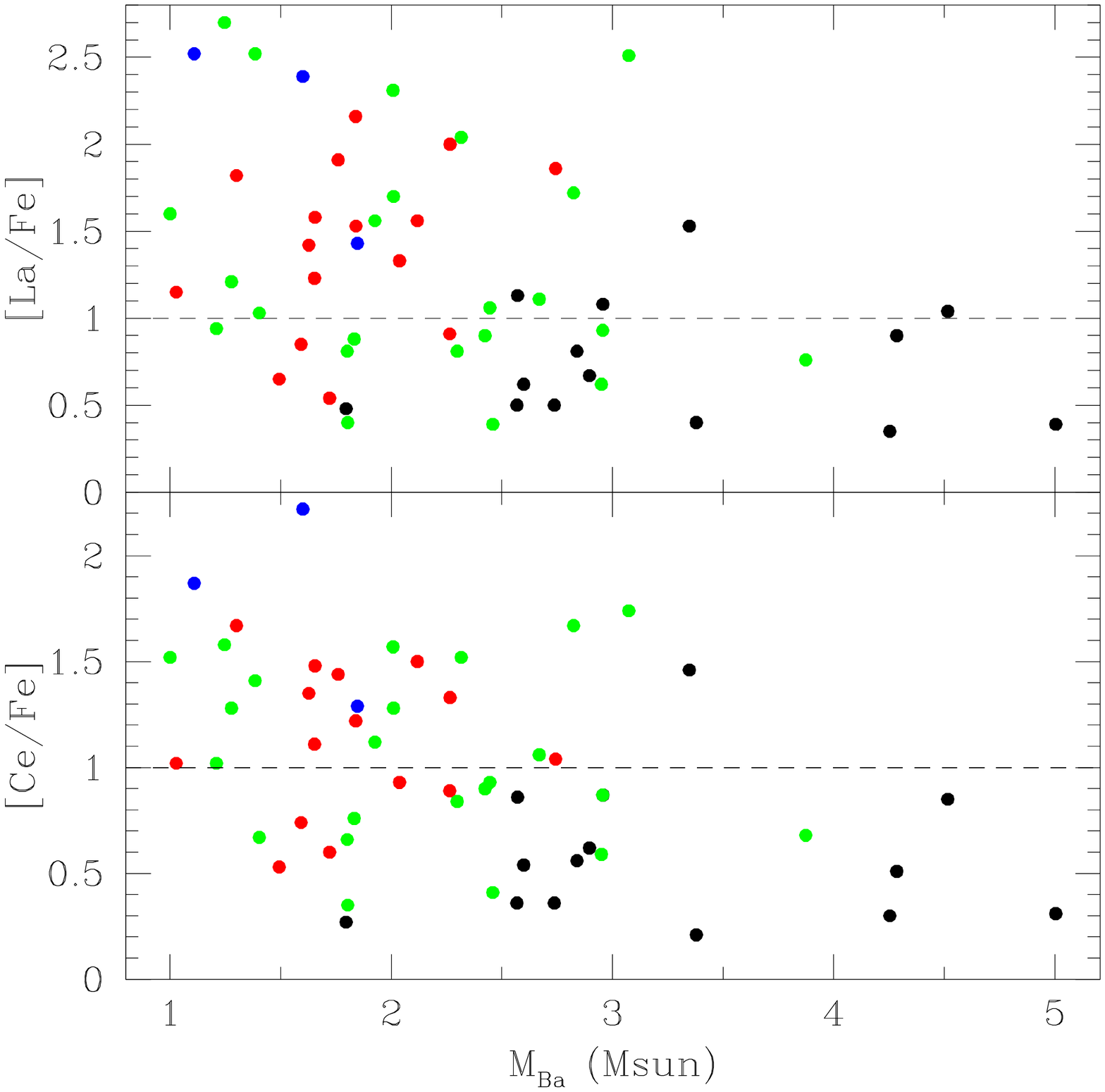}
\vspace{-2cm}
\caption{\label{Fig:M1_LaCe} 
Relationship between barium-star masses and [La/Fe] and [Ce/Fe] abundances, color-coded according to metallicity:
Blue ([Fe/H]~$\le -0.6$), red ($-0.6 <$~[Fe/H]~$\le -0.3$), 
green ($-0.3 <$~[Fe/H]~$\le -0.1$), and
black ($-0.1 <$~[Fe/H]). 
}
\end{figure}

The mass distribution is shown in Fig.~\ref{Fig:mass_distribution}, separately for mild and strong barium stars. Mild barium stars exhibit a clear tail towards masses up to 5~M$_\odot$, whereas strong barium stars are restricted to about 3.5~M$_\odot$. 
Figure~\ref{Fig:M1_LaCe} confirms that if the threshold between mild and strong barium stars is set at 1~dex for both [La/Fe] and [Ce/Fe] (a reasonable value as revealed by Table~\ref{Tab:abundances}), mild barium stars indeed include a tail of high-mass ($M > 3$~M$_\odot$), high-metallicity ([Fe/H]~$ > -0.1$) stars. If this high-mass tail is removed, any correlation between mass and abundances disappears.

The statistical significance of the apparent difference between the mass distributions for mild and strong barium stars may be evaluated from a Kolmogorov-Smirnov test \citep[e.g.,][]{NumRecipes}. The maximum difference between the two cumulative frequency distributions amounts to 0.29 (bottom panel of Fig.~\ref{Fig:mass_distribution}). Considering that the samples comprise $m = 30$ mild barium stars and $n = 24$ strong barium stars, resulting in an effective sample size of $m \times n / (m + n) = 13$, the observed difference  translates into a significance level of 79\%. This implies that the first-kind error  of erroneously rejecting the null hypothesis that the two distributions are similar is 21\%. The difference between the mass distributions of mild and strong barium stars -- albeit not very significant -- clearly originates from the presence of a high-mass tail among mild barium stars. The fact that in our previous paper \citep[Fig.~14 of][]{2017A&A...608A.100E}, this high-mass tail among mild barium stars was not as clear as here, rather than contradicting our present results, reveals the limitations of the analysis of that former paper, where the same metallicity for all barium stars was adopted.  
Similar claims that mild barium stars extend towards larger masses than strong barium stars were formerly put forward  by 
\citet{1977MNRAS.181..391C}, \citet{1997A&A...326..722M}, and \citet{1998A&A...332..877J}.

\begin{figure}
\vspace{-2cm}
\includegraphics[width=9cm]{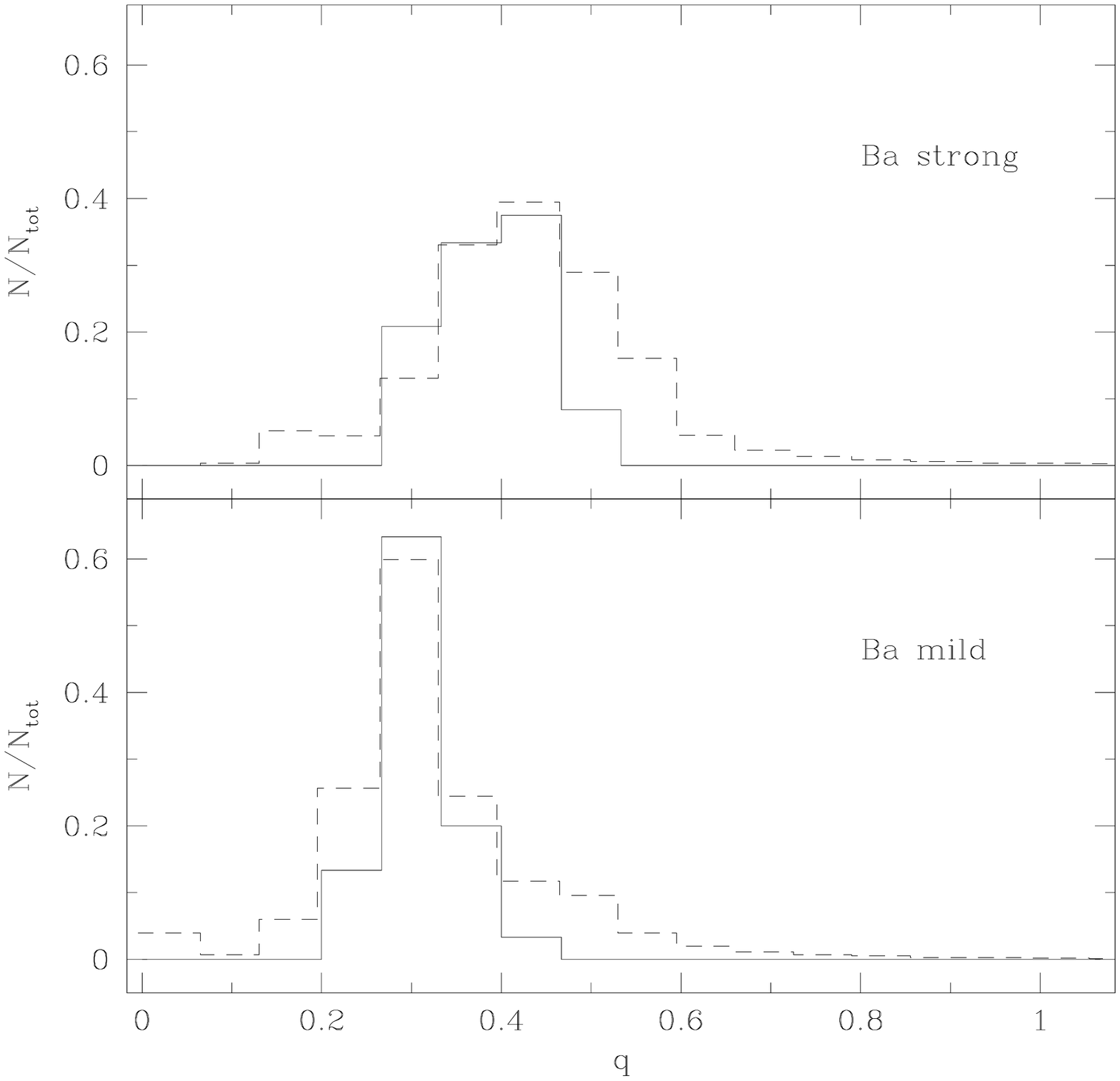}
\vspace{-2.5cm}\\
\caption{\label{Fig:q_distribution} 
Mass-ratio distributions  for mild  (bottom panel) and strong (top panel) barium stars. The dashed lines correspond to the $q$ distribution obtained  from the Richardson-Lucy inversion (see text), whereas the solid line corresponds to the $q$ distribution rederived from the individual $M_{\rm Ba}$ and $M_{\rm WD}$ estimates  (Table~\ref{Tab:abundances}) under the assumption of a constant $Q$ (different for mild and strong barium stars).
}
\end{figure}

\begin{figure}
\vspace{-2cm}
\includegraphics[width=9cm]{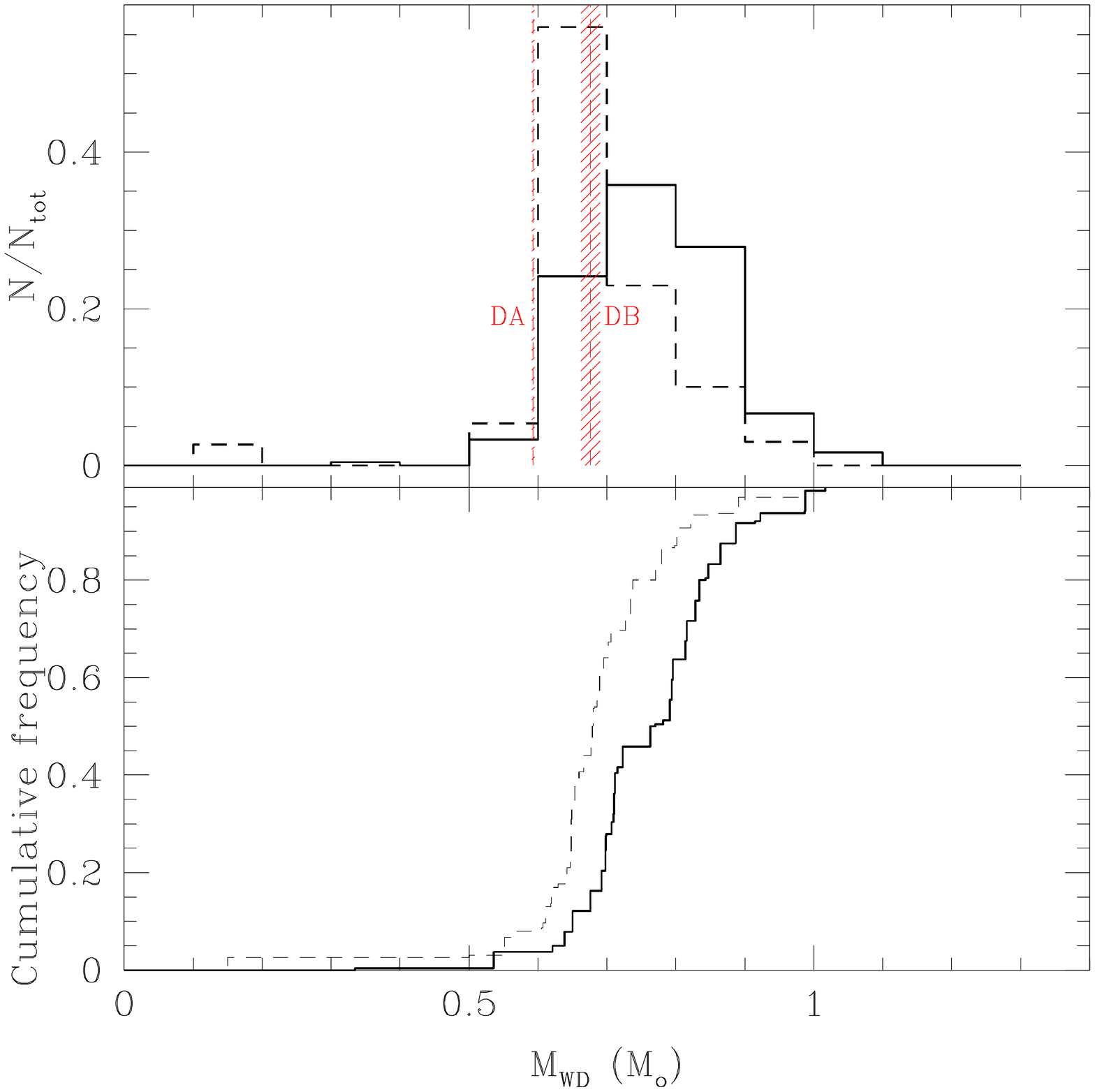}
\vspace{-2cm}
\caption{\label{Fig:M2} 
Top panel: Mass distributions of WDs orbiting mild and strong barium stars (dashed and solid lines, respectively) under the hypothesis of maximum concentration. The red shaded areas labeled DA and DB correspond to the average masses ($\pm 1\sigma$) for field WDs. Bottom panel: Cumulative mass distributions corresponding to the two samples from the top panel.
The maximum vertical distance between the two curves is 0.45.
}
\end{figure}

\begin{figure}
\vspace{-2cm}
\includegraphics[width=9cm]{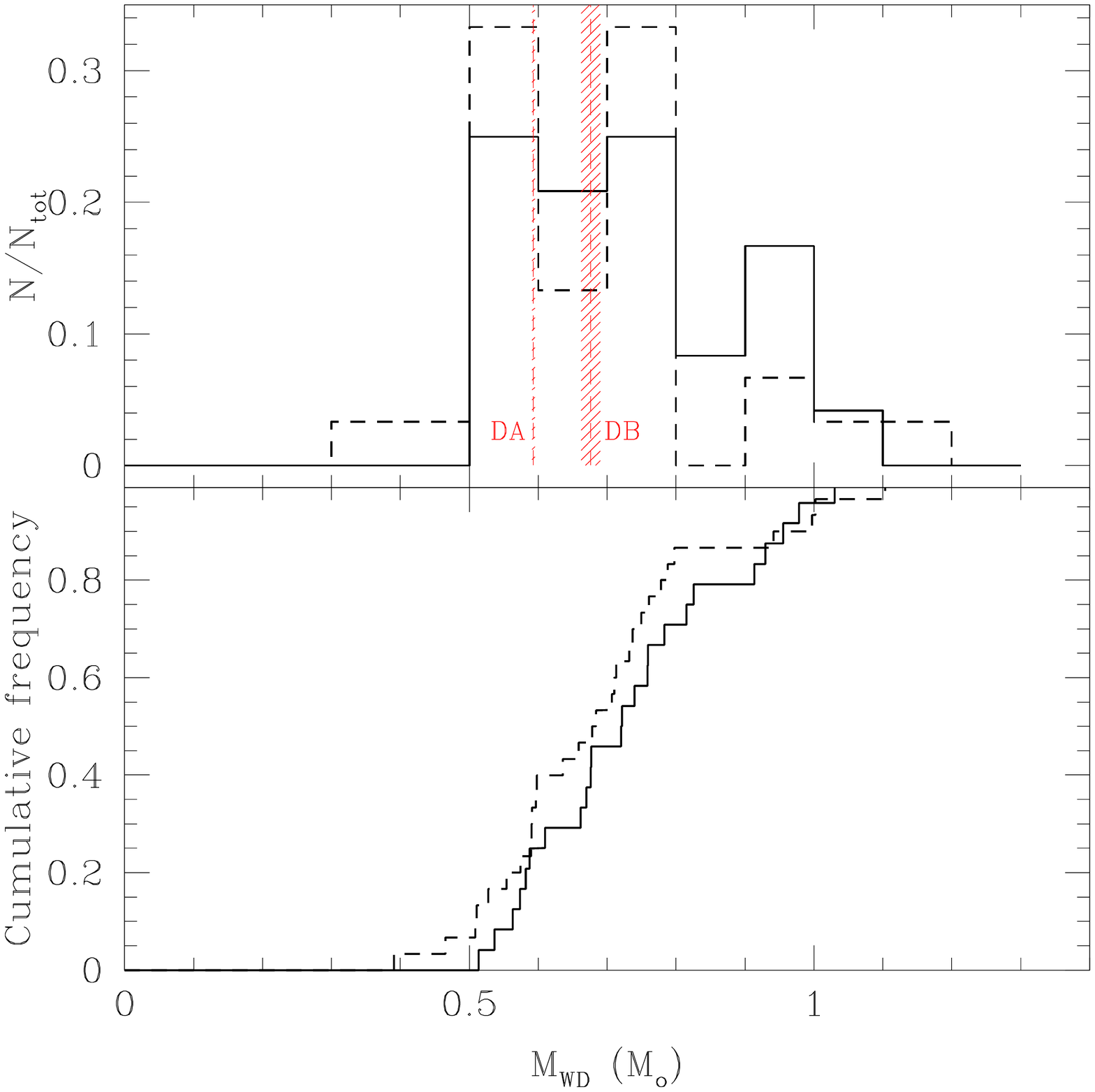}
\vspace{-2cm}
\caption{\label{Fig:M2Q} 
As in Fig.~\ref{Fig:M2} but derived under the assumption of constant $Q$. In the bottom panel, the maximum difference between the two curves is 0.18.
}
\end{figure}

The $q$~$(= M_{\rm WD} / M_{\rm Ba})$ distributions for mild and strong barium stars are shown in Fig.~\ref{Fig:q_distribution}. 
The resulting $M_{\rm WD}$ distribution is shown in Fig.~\ref{Fig:M2} under the most-peaked assumption, as explained in Sect.~\ref{Sect:methods7}, and in Fig.~\ref{Fig:M2Q} for the constant-$Q$ assumption.

The distributions of Fig.~\ref{Fig:M2} are by construction strongly peaked (with the exception of a few non-physical "WDs" around 0.2 and 0.45~M$_\odot$) at 0.6 -- 0.7~M$_\odot$ 
for WDs around mild barium stars, and at 0.6 -- 0.9~M$_\odot$ for WDs around strong barium stars. As expected, the distributions shown in Fig.~\ref{Fig:M2Q} 
for the constant-$Q$ assumption are somewhat broader than the limiting cases displayed in Fig.~\ref{Fig:M2}, and the small-mass outliers have disappeared. 
The consistency of this WD mass distribution obtained under the assumption of a constant $Q$ is further checked by comparing in Fig.~\ref{Fig:q_distribution}  the mass ratios $q$ obtained from these WD masses and the paired barium masses  (as listed in Table~\ref{Tab:abundances}), with the $q$ distribution obtained directly from the Lucy-Richardson inversion. 
Both $q$ distributions are in good agreement, as they differ only by the presence of sparsely populated bins in the Richardson-Lucy results (dashed lines in Fig.~\ref{Fig:q_distribution}).

Another independent check of the WD masses obtained above may be performed for the few barium stars which were found to be astrometric binaries based on Hipparcos data \citep{2000A&AS..145..161P}. 
A subsequent study \citep{2005A&A...442..365J} assessed the quality of the astrometric orbital elements derived by \citet{2000A&AS..145..161P} and concluded that only HD~46407 (HIP~31205) and HD~101013 (HIP~56731) marginally satisfy the orbital quality checks (see their Table~5). Relevant data for these two systems are collected in Table~\ref{Tab:astrometry}, which reveals an agreement between the two mass values within $2\sigma$. 

Moreover, in the case of HD~204075 ($\zeta$~Cap), the WD companion has been detected directly from its UV radiation, using the IUE satellite \citep{1980ApJ...239L..79B}, and the mass estimated from the observed spectrum is of the order of 1~M$_\odot$, in perfect agreement with the "constant-$Q$" value WD mass.

For the sake of  completeness, Table~\ref{Tab:abundances} also lists HD~121447, although that star was not included in the luminosity determination using Gaia DR2 parallaxes.
This star is suspected to be an ellipsoidal variable \citep{1995A&A...301..707J}. The photometric analysis of the system has yielded  masses of $1.6\pm0.1$~M$_\odot$ and $0.6\pm0.1$~M$_\odot$ for the barium star and its WD companion, respectively.

\begin{table}[]
\renewcommand{\arraystretch}{1.5}
    \centering
        \caption{WD masses derived from the astrometrically based $\sin i$  \citep{2000A&AS..145..161P} and from the assumption of constant $Q$. The error bars for the WD mass derived from astrometry correspond to the error propagation from the inclination. The column labeled `Ba' lists whether the star is a mild (`m') or strong (`s') barium star.}
    \label{Tab:astrometry}
    \begin{tabular}{rrrrrrrrrr}
    \hline
    HD & HIP & Ba & $M_{\rm Ba}$ & $f(M)$ & \multicolumn{1}{c}{$i$} & \multicolumn{2}{c}{$M_{\rm WD}$}\\
    \cline{7-8}
        &&&             & & & \multicolumn{1}{c}{($i$)} & \multicolumn{1}{c}{($Q$)}\\
        &&& (M$_\odot$) & \multicolumn{1}{c}{(M$_\odot$)}  & \multicolumn{1}{c}{($^\circ$)} & \multicolumn{1}{c}{(M$_\odot$)} & (M$_\odot$)\\
        \hline\\
         46407 & 31205 &s& 2.12 & 0.035 &$80\pm10$ & 0.71${+0.04\atop -0.02}$ & 0.78\\
         101013& 56731 &s& 1.65 & 0.037 &$78\pm26$ & 0.59${+0.18\atop-0.02}$ & 0.68\\
         \hline\\
    \end{tabular}

\end{table}

It is now possible to compare the masses of the WD companions of barium stars with field WDs. Current estimates for the average mass of the latter (represented by the red vertical dashed lines on Figs.~\ref{Fig:M2} and \ref{Fig:M2Q}) is $0.593 \pm  0.002$~M$_\odot$ for DA~WDs and $0.676 \pm 0.014$~M$_\odot$ for DB~WDs \citep{2013ApJS..204....5K}. The mass distribution of WD companions of barium stars appears to have a tail extending toward  masses larger than those of field WDs. 
Moreover, there is a hint that WDs around strong barium stars may be more massive on average than WDs around mild barium stars. This trend is relatively significant for the "peaked" distributions, where a Kolmogorov-Smirnov test  (bottom panel of Fig.~\ref{Fig:M2}) 
yields a first-risk error of rejecting the null hypothesis of equality between the two mass distributions of only  0.44\%, considering the number of stars in the sample (30 mild barium stars and 24 strong barium stars, resulting in an effective sample size of 13, as computed above) and a maximum vertical distance of 0.45 between the two cumulative frequency distributions. 
The difference between the mass distributions of WDs orbiting around mild and strong barium stars is also visible in the case of WD masses computed under the "constant-$Q$" assumption (Fig.~\ref{Fig:M2Q}). In this case however the difference is much less significant. The maximum vertical difference between these two cumulative distributions (bottom panel of Fig.~\ref{Fig:M2Q}) is only 0.18. Consequently, the first-kind risk of erroneously rejecting the null hypothesis of equality between the two distributions is now as large as  22\%. 

A correlation between  the final s-process overabundance in barium stars (i.e., mild vs. strong barium stars) and the WD-companion mass would have its origin in the relation between the maximum luminosity reached by the AGB progenitor (through the core mass--luminosity relation) and the efficiency of s-process nucleosynthesis up to that luminosity. As we show in Sect.~\ref{Sect:abundances}, there are however many other parameters  controlling the final level of s-process abundances in the barium stars that could therefore blur the above correlation and account for its observed weakness.

The parameters controlling the final s-process abundances in a barium star include the dilution factor of the accreted matter in the barium-star envelope (which depends on both its mass and the amount of accreted matter, which in turn depends on the orbital separation). These parameters also include the level of s-process overabundance in the accreted matter, which reflects the ability of the AGB companion to efficiently synthesize the s-process. This in turn depends on its metallicity, mass, and at a given mass, on the number of  thermal pulses and third dredge-ups experienced by the AGB star (considering that the AGB evolution may have been truncated prematurely due to Roche-lobe overflow). The covariance analysis presented in Sect.~\ref{Sect:abundances} is a first attempt at disentangling the impact of these intricated effects on the final overabundance level.

\begin{figure}
\vspace{-2cm}
\includegraphics[width=9cm]{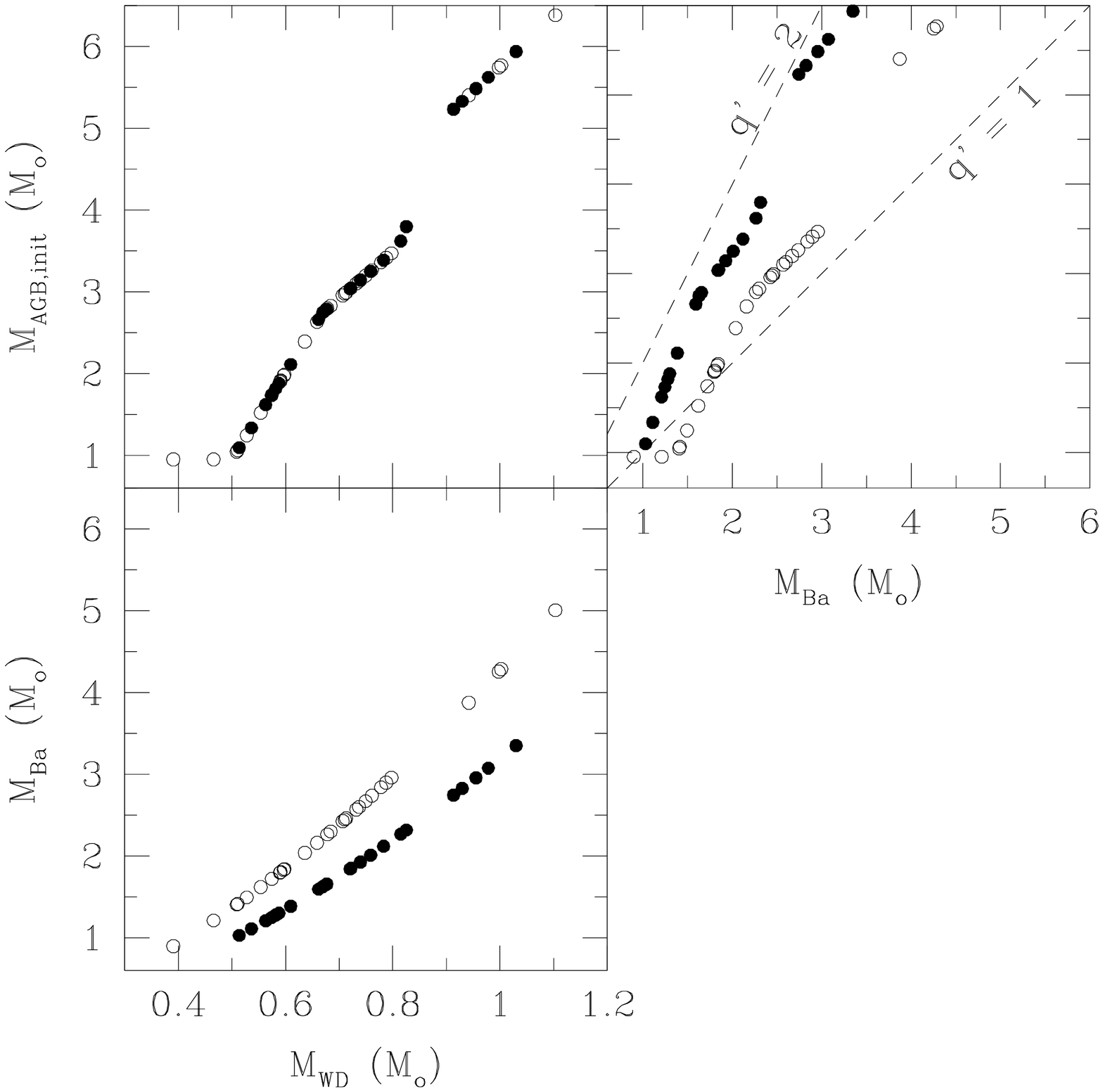}
\vspace{-4cm}\\
\includegraphics[width=9cm]{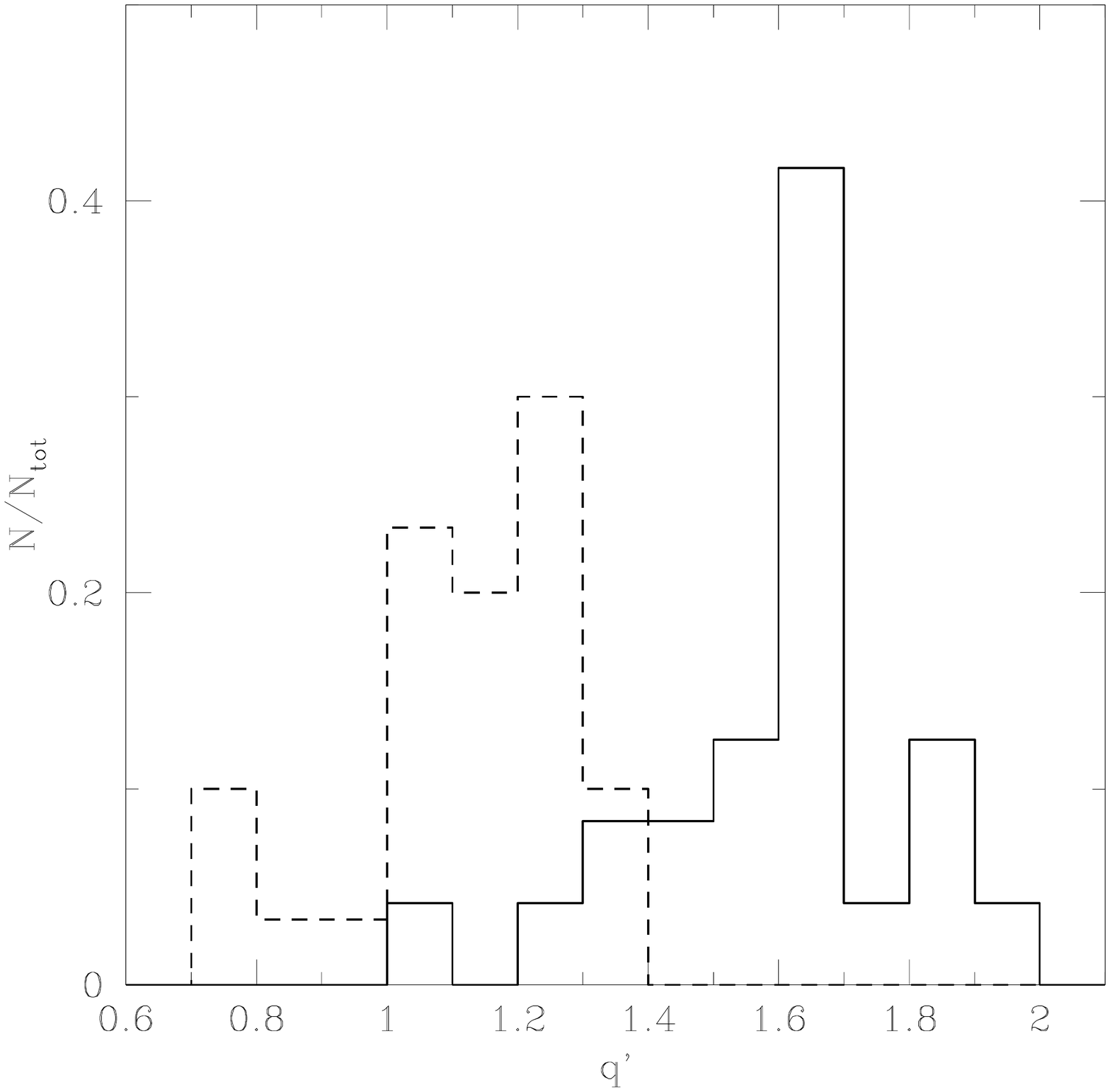}
\vspace{-2cm}
\caption{\label{Fig:q_init} 
Top panel: Relationships between the current barium-star mass ($M_{\rm Ba}$), the WD mass ($M_{\rm WD}$), and the WD-progenitor mass ($M_{\rm AGB,init}$). Mild and strong barium stars are represented by open and solid dots, respectively. Their differing sequences are caused by the adoption of two different values for $Q = M^3_{\rm WD}/(M_{\rm WD} + M_{\rm Ba})^2$.
Bottom panel: Distribution of the "initial" mass ratio $q' = M_{\rm AGB,init}/M_{\rm Ba}$, which should in principle be larger than unity. Mild and strong barium stars are represented by dashed and solid lines, respectively.
}
\end{figure}

\renewcommand{\arraystretch}{1.5}
\setlength{\tabcolsep}{2pt}
\begin{table*}
\caption[]{\label{Tab:abundances}Abundances for s-process elements in barium stars, from various sources, as listed in the column Ref. The second column, labeled Ba, lists whether the star is considered as a mild (m) or strong (s) barium star. Assignment shifts (based on the abundances; see text) are indicated by an arrow. For completeness, the other columns list the component masses and orbital elements. Possibly inaccurate masses for the two stars with orbital periods close to 1~yr are listed in italics (see text). }

\begin{tabular}{ll|llrrr|ll|lll|rllllrr}
\hline
HD/DM  &  Ba & \multicolumn{1}{c}{$T_{\rm eff}$} & $\log g$ & $L_{\rm min}$ & \multicolumn{1}{c}{$L$} & $L_{\rm max}$ & $M_{\rm {Ba}}$ &  $M_{\rm WD}$ &   $P$        & $e$   & $f(M)$ &    [Fe/H] &   [Y/Fe]&   [Zr/Fe] & [La/Fe]&   [Ce/Fe]   &  [hs/ls]  & Ref\\
\cline{5-7}
             &          & \multicolumn{1}{c}{(K)} & &
             \multicolumn{3}{c}{(L$_{\odot}$)} & (M$_{\odot}$) & (M$_{\odot}$) & (d) & & (M$_{\odot}$)\\
\hline
$-64^\circ$4333&s    & $4900\pm100$  &  2.6   & 34  & 37  & 40  & {\sl 1.4$+0.1\atop-0.1$} &  {\sl 0.61}&  386    & 0.03  &  0.068 &  -0.10  &         1.13  &    1.12  &    2.52   &    1.41    &  0.84     &       2\\
$-42^\circ$2048&s    & $4400\pm100$ &   1.6     & 170 & 234 & 303 &1.9$+0.7\atop-0.5$ &  0.74 & 3260    & 0.08  &  0.065 &  -0.23  &          0.95  &    0.96  &    1.56   &    1.12   &  0.38     &        2\\
$-14^\circ$2678&m    & $5200\pm100$ &$3.1\pm0.2$& 57  & 73  & 92  & 3.0$+0.2\atop-0.2$ &  0.80 & 3470    & 0.22  &  0.023 &   0.01  &          1.02  &    0.85  &    1.08   &    0.87   &  0.04     &        2   \\
$-01^\circ$3022&m    & $4832\pm25$ &$2.7\pm0.4$& 51 & 56 & 61 & 1.6$+0.1\atop-0.1$ &  0.55  & 3253    & 0.28  &  0.016 &  -0.14  &         0.58 &    0.71  &    0.44   &    0.33     &  -0.26    &        4\\
 5424&s       &$4728\pm80$ &  $2.5\pm0.0$ &33&60        &90&1.3$+0.4\atop-0.3$ &  0.59& 1881    & 0.23  &  0.005 &  -0.43  &         1.30  &    1.05  &    1.82   &    1.67    &  0.57     &       1\\
 16458&s      &$4550\pm25$  & $1.8\pm0.2$ &205  &217    &229 &1.9$+0.1\atop-0.1$ &  0.72 & 2018        & 0.1   &  0.041 &  -0.64  &     1.06  &    1.29  &    1.43   &    1.29    &  0.19     &       1\\
 18182&m      &$4858\pm31$  & $2.5\pm0.4$ &60   &65     &71 &1.8$+0.2\atop-0.1$ &  0.59 & 8059        & 0.31  &  0.0002    &  -0.17  &     0.50  &    0.35  &    0.40   &    0.35    &  -0.05    &       2\\
 20394&s      &$4926\pm17$  & $2.5\pm0.0$ &578  &69     &82 &2.0$+0.2\atop-0.2$ &  0.76 & 2226        & 0.2   &  0.002 &  -0.27  &     1.00  &    1.14  &    1.70   &    1.28    &  0.42     &       2\\
 24035&s      &$4700\pm100$ &$2.5\pm0.2$&13&26  &39 &{\sl 1.3}$+0.3\atop-0.2$ &  {\sl 0.57} & 377.8   & 0.3       &  0.047 &  -0.23  &     1.35  &    1.20  &    2.70   &    1.58    &  0.87     &       2\\
 27271&m      &$5022\pm40$ &  $2.9\pm0.5$ &68   &82     &98 &2.9$+0.2\atop-0.2$ &  0.79 & 1693        & 0.22  &  0.024 &  -0.07  &     0.77  &    0.79  &    0.67   &    0.62    &  -0.13    &       1\\
 31487&s      &$4960\pm50$  & $3.1\pm0.2$ &124  &141    &160 &3.4$+0.2\atop-0.3$    &  1.03& 1066        & 0.05  &  0.038 &  -0.04  &     1.23  &    1.11  &    1.53   &    1.46    &  0.32     &       1\\
 40430&m      &$4930\pm29$  & $2.4\pm0.2$ &74   &84     &95 &2.3$+0.2\atop-0.2$    &  0.68& 5609        & 0.22  &  0.0025&  -0.34  &     0.76  &    0.58  &    0.91   &    0.89    &  0.23     &       2\\
 43389&s      &$4000\pm50$  & $2.0\pm0.5$ &196  &260    &330 &1.8$+0.4\atop-0.3$    &  0.72& 1689        & 0.08  &  0.043 &  -0.35  &     0.91  &    0.32  &    1.53   &    1.22    &  0.76     &       1\\
 44896&s      &$4300\pm100$&  0.7        &526   &676    &841 &3.0$+1.2\atop-1.0$    &  0.96& 629     & 0.02      &  0.048 &  -0.25  &     1.16  &    0.81  &    0.93   &    0.87    &  -0.09    &       11\\
 46407&s      &$4854\pm100$ & $2.2\pm0.4$ &35   &83     &135 &2.1$+0.6\atop-0.7$    &  0.78& 457     & 0.013     &  0.035 &  -0.35  &     1.15  &    1.28  &    1.56   &    1.50    &  0.31     &       1\\
 49641&s      &$4400\pm100$           &  $1.5\pm0.2$           &345     &457         &579 &2.7$+1.2\atop-0.8$    &  0.91& 1785    & 0.07  &  0.003 &  -0.3   &         0.89  &    0.53  &    1.86   &    1.04    &  0.74     &       2\\
 49841&m      &5200           &   3.2          &49      &61     &74 &2.8$+0.2\atop-0.2$    &  0.78& 897         & 0.16      &  0.032 &   0.2   & 0.85  &    0.65  &    0.81   &    0.56    &  -0.06    &       9\\
 50082&s      &$4789\pm100$ & $2.4\pm0.5$ &60   &63     &66 2&1.6$+0.3\atop-0.2$    &  0.67& 2896        & 0.19  &  0.027 &  -0.32  &     0.86  &    1.04  &    1.42   &    1.35    &  0.44     &       1\\
 51959&m      &$4814\pm34$  & $3.2\pm0.2$ &11   &13     &16 &1.2$+0.1\atop-0.1$    &  0.47& 9718        & 0.53  &  0.0005&  -0.21  &     0.98  &    1.25  &    0.94   &    1.02    &  -0.13    &       4\\
 53199&m      &$5119\pm28$  & $2.9\pm0.2$ &47   &55     &64 1&2.5$+0.1\atop-0.1$    &  0.71& 8314        & 0.24  &  0.028 &  -0.20  &     0.68  &    0.70  &    1.06   &    0.93    &  0.31     &       2\\
 58121&m      &$4600\pm100$& $1.8\pm0.2$&121    &142    &163    &2.6$+0.5\atop-0.4$    &  0.73 & 1214    & 0.14  &  0.015 &  -0.01  &     0.41  &    0.26  &    0.50   &    0.36    &  0.10     &       2\\
 58368&m      &$5000\pm100$& $2.6\pm0.2$ &37    &43     &49     &2.6$+0.1\atop-0.2$    & 0.73     & 672     & 0.22      &  0.021 &   0.04  &     0.85  &    0.60  &    1.13   &    0.86    &  0.27     &       2\\
 59852&m      &$5000\pm100$& $2.2\pm0.2$ &81    &90     &100 &2.5$+0.2\atop-0.3$    &  0.71& 3464        & 0.15  &  0.0022&  -0.22  &     0.40  &    0.27  &    0.39   &    0.41    &  0.06     &       2\\
 77247&m      &$5050\pm100$ & $2.5\pm0.5$ &306  &346    &388    &3.9$+0.1\atop-0.2$    &  0.94  & 80      & 0.09      &  0.005 &  -0.13  &     0.73  &    0.75  &    0.76   &    0.68    &  -0.02    &       4\\
 84678&s      &$4600\pm100$& $1.7\pm0.2$ &154   &189    &227 &2.3$+0.6\atop-0.5$    &  0.83  & 1630        & 0.06  &  0.062 &  -0.13  &     1.09  &    1.21  &    2.04   &    1.52    &  0.63     &       2\\
 88562&s      &$4000\pm50$  & $2.0\pm0.5$ &240  &277    &321 &1.0$+0.1\atop-0.1$    &  0.51  & 1445        & 0.2   &  0.048 &  -0.53  &     0.93  &    0.43  &    1.15   &    1.02    &  0.41     &       1\\
 91208&m      &$5093\pm67$  & $2.9\pm0.3$ &35 9 &39     &43     &2.3$+0.1\atop-0.2$    &  0.68  & 1754        & 0.17  &  0.022 &  -0.16  &     0.94  &    0.61  &    0.81   &    0.84    &  0.05     &       2\\
 92626&s      &$4800\pm100$&   2.3          &171        &214    &259 &3.1$+0.4\atop-0.6$    &  0.98  & 918     & 0.        &  0.042 &  -0.15  &     0.99  &    1.21  &    2.51   &    1.74    &  1.02     &       2\\
 95193&m      &$5008\pm22$  & $2.8\pm0.1$ &59   &69     &79 &2.7$+0.1\atop-0.1$    &  0.76  & 1653        & 0.13  &  0.026 &  -0.04  &     0.75  &    0.26  &    0.50   &    0.36    &  -0.07    &       2\\
 98839&m      &$4917\pm34$  & $2.3\pm0.6$ &276  &332    &395 &4.3$+0.2\atop-0.2$    &  1.00  & 16471   & 0.56      &  0.056 &  -0.05  &     0.10  &    0.17  &    0.35   &    0.30    &  0.19     &       4\\
 101013&s     &$4722\pm32$  & $2.3\pm0.2$ &77   &108    &141    &1.7$+0.3\atop-0.3$    &  0.68 & 1711        & 0.2   &  0.037 &  -0.40  &     1.17  &    0.97  &    1.23   &    1.11    &  0.10     &       4\\
 104979&m     &$5100\pm100$& $2.7\pm0.2$&80     &95     &111 &2.7$+0.1\atop-0.2$    &  0.75 & 19295   & 0.08      &  -     &  -0.26  &     0.71  &    0.85  &    1.11   &    1.06    &  0.31     &      6 \\
 107541&s     &$5000\pm100$&  $3.2\pm0.2$ &8.9  &11     &14 &1.1$+0.2\atop-0.1$    &  0.54 & 3570        & 0.1   &  0.029 &  -0.63  &     1.53  &    1.35  &    2.52   &    1.87    &  0.75     &      2\\
 119185&m     &$4919\pm18$ &  $2.5\pm0.0$ &65   &77     &90. &1.7$+0.2\atop-0.2$    &  0.57& 22065   & 0.6       &  -     &  -0.42  &     0.30  &    0.21  &    0.54   &    0.60    &  0.32     &       2\\
 121447&s     &$4000\pm50$ &  $1.0\pm0.5$ &                  - &     -             &        -        &1.6$+0.1\atop-0.1$                      &  0.6    & 185.7   & 0.015 &  0.025 &  -0.90  &     1.35  &    1.57  &    2.39   &    2.22    &  0.84     &       1\\
 \hline
 \end{tabular}
 \end{table*}
 
 \addtocounter{table}{-1}
 
\begin{table*}
\caption[]{Continued. }

\begin{tabular}{ll|llrrr|ll|lll|rllllrr}
\hline
HD/DM  &  Ba & \multicolumn{1}{c}{$T_{\rm eff}$} & $\log g$ & $L_{\rm min}$ & $L$ & $L_{\rm max}$ & $M_{\rm {Ba}}$ &  $M_{\rm WD}$ &   $P$        & $e$   & $f(M)$ &    [Fe/H] &   [Y/Fe]&   [Zr/Fe] & [La/Fe]&   [Ce/Fe]   &  [hs/ls]  & Ref\\
\cline{5-7}
             &          & \multicolumn{1}{c}{(K)} & &
             \multicolumn{3}{c}{(L$_{\odot}$)} & (M$_{\odot}$) & (M$_{\odot}$) & (d) & & (M$_{\odot}$)\\
\hline
 123949&s    &$4378\pm80$ & $1.8\pm0.5$ &59     &92     &128  &1.3$+0.3\atop-0.1$    &  0.58& 8523        & 0.92  &  0.046 &  -0.23  &     0.91  &    0.88  &    1.21   &    1.28    &  0.35     &       1\\
 134698&m    &$4438\pm30$ & $1.7\pm0.3$ &163    &192&225  &1.5$+0.2\atop-0.2$    &  0.53& 10005   & 0.95      &  0.054 &  -0.57  &     0.56  &    0.59  &    0.65   &    0.53    &  0.02     &        2\\
 139195&m    &$5029\pm29$ & $3.1\pm0.2$  &38    &44     &514  &2.6$+0.1\atop-0.1$    &  0.74& 5324        & 0.35  &  0.026 &  -0.07  &     0.72  &    0.79  &    0.62   &    0.54    &  -0.17    &       4\\
 143899&m    &$5144\pm26$ &  $2.9\pm0.3$ &43    &50     &67  &2.4$+0.1\atop-0.1$    &  0.71& 1461        & 0.19  &  0.017 &  -0.29  &     0.86  &    0.57  &    0.90   &    0.90    &  0.19     &        2\\
 154430&s    &$4200\pm100$ &  $1.2\pm0.2$&382   &685    &1046     &2.3$+1.4\atop-0.7$    &  0.81& 1668        & 0.11  &  0.034 &  -0.36  &     0.93  &    0.97  &    2.00   &    1.33    &  0.71     &       2\\
 178717&s    &$3800\pm50$  & $1.0\pm0.5$&156    &1617   &3189         &1.6$+0.9\atop-0.7$    &  0.66& 2866        & 0.43  &  0.006 &  -0.52  &     0.79  &    0.44  &    0.85   &    0.74    &  0.18     &       1\\
 180622&m    &$4600\pm100$ & $2.2\pm0.2$ &59    &63     &68       &1.8$+0.3\atop-0.2$    &  0.59& 4049        & 0.06  &  0.07  &   0.03  &     0.61  &    0.41  &    0.48   &    0.27    &  -0.13    &       2\\
 183915&m$\rightarrow$s&$4494\pm130$&$1.6\pm0.4$&153    &266    &386&1.8$+1.0\atop-0.6$    &  0.60 & 4382        & 0.27  &  7E-05 &  -0.59  &     0.88  &    0.68  &    2.16   &    1.22    &  0.91     &        2\\
 196673&m    &$4914\pm9$   & $2.5\pm0.3$ &618   &900    &1206 &5.0$+0.0\atop-0.1$    &  1.10 & 7780    & 0.59      &  0.020 &  0.12  &     0.00  &    0.25  &    0.39   &    0.31    &  0.23     &        4\\
 199939&s    &$4710\pm9$   & $2.4\pm0.4$ &159   &214    &271 &2.8$+0.4\atop-0.4$    &  0.93 & 584.9   & 0.28      &  0.025 &  -0.22  &     1.38  &    1.19  &    1.72   &    1.67    &  0.41     &       1\\
 200063&m    &$4100\pm100$ & $1.1\pm0.2$   &206 &753    &1349      &2.0$+1.3\atop-0.9$    &  0.64 & 1735        & 0.07  &  0.058 &  -0.34  &     0.88  &    0.62  &    1.33   &    0.93    &  0.38     &       2    \\
 201657&s    &$4700\pm100$ & $2.2\pm0.2$ &63    &80     &100 &1.8$+0.5\atop-0.4$    &   0.70    & 1710        & 0.17  &  0.004 &  -0.34  &     0.72  &    0.98  &    1.91   &    1.44    &  0.82     &       2\\
 201824&s    &$4937\pm52$  & $2.6\pm0.2$ &51    &69     &90 &1.7$+0.4\atop-0.2$    &  0.68 & 2837        & 0.34  &  0.04  &  -0.40   &     0.91  &    0.87  &    1.58   &    1.48    &  0.64     &        2\\
 202109&m    &4700  &  2.4            &120      &147        &176 &3.4$+0.2\atop-0.4$    &  0.87     & 6489        & 0.22  &  0.023 &  -0.03  &     0.42  &    0.39  &    0.40   &    0.21    &  -0.10    &       9\\
 204075&m    &$5269\pm53$  & $1.7\pm0.3$ &418   &561    &741 &4.5$+0.3\atop-0.2$    &  1.03     & 2378        & 0.28  &  0.004 &  -0.09  &     1.37  &    1.37  &    1.04   &    0.85    &  -0.43    &        5\\
 205011&m    &$4803\pm21$  & $2.5\pm0.1$ &59    &73     &88 &1.8$+0.3\atop-0.3$    &  0.60 & 2837        & 0.24  &  0.034 &  -0.26  &     0.82  &    0.86  &    0.88   &    0.76    &  -0.02    &        4\\
 210946&m    &$4780\pm76$  & $2.4\pm0.2$ &50    &74     &99 &1.8$+0.5\atop-0.4$    &  0.59 & 1529        & 0.13  &  0.041 &  -0.29  &     0.77  &    0.56  &    0.81   &    0.66    &  0.07     &        2\\
 211594&s    &$4947\pm57$  & $2.6\pm0.1$ &49    &63     &77 &2.0$+0.3\atop-0.2$    &  0.76 & 1019        & 0.06  &  0.014 &  -0.29  &     1.20  &    1.18  &    2.31   &    1.57    &  0.75     &        2\\
 218356&m    &$4500\pm100$ &  1.8     &543      &733    &976 &4.3$+0.2\atop-1.1$    &  1.00 & 111         &   0&  4E-05 &  -0.06 &     0.45  &    0.26  &    0.90   &    0.51    &  0.35     &        7\\
 223617&m    &$4560\pm20$  & $2.2\pm0.1$ &61    &65         &70 &1.4$+0.1\atop-0.1$    &    0.51   & 1294        & 0.06  &  0.0064&  -0.20  &     0.70  &    0.73  &    1.03   &    0.67    &  0.14     &        4\\
 NGC \\
 2420-173&s$\rightarrow$m&$5150\pm100$&2.2&95   &128    &173&3.0$+0.3\atop-0.4$& 0.80& 1479 & 0.43 & 0.008 & -0.26 & 1.00 & 0.72 & 0.62 & 0.59 & -0.26 & 10 \\
\hline\\

\end{tabular}

References: (1) Karinkuzhi et al. (2018); (2) \citet{2016MNRAS.459.4299D}; (3) \citet{2006A&A...454..895A}; (4) This work; (5) \citet{2016A&A...586A.151M}; (6) \citet{2015MNRAS.446.2348K};
(7) \citet{2014AJ....147..137L}
(8) \citet{1984A&A...132..326S}; (9) \citet{2011A&A...533A..51P}; (10) \citet{2017A&A...597A..68V}; (11) \citet{1984A&A...132..326S}

\end{table*}
\renewcommand{\arraystretch}{1}
\setlength{\tabcolsep}{3pt}

\section{Mass distribution of the WD progenitor and initial mass ratio of the system}
\label{Sect:initial}

In this section, we derive the mass distribution of the WD progenitor (denoted $M_{\rm AGB,ini}$) and compare it with the mass of the barium star ($M_{\rm Ba}$), expecting that $q' \equiv M_{\rm AGB,ini} / M_{\rm Ba} > 1$, unless mass accretion by the barium star has substantially increased its current mass over its initial value. For the purpose of deriving $M_{\rm AGB,ini}$, we use the most recent full-range initial--final mass relationship (IFMR) as derived by \citet{2018ApJ...860L..17E} from the Gaia DR2. The IFMR has been applied to the WD masses listed in Table~\ref{Tab:abundances} to get $M_{\rm AGB,ini}$. The resulting mass ratio $q'$ is shown in the bottom panel of Fig.~\ref{Fig:q_init}. 
This "initial" $q'$ distribution appears to be very different among mild and strong barium stars. As shown in the top panel of Fig.~\ref{Fig:q_init}, this difference may ultimately be traced back to the difference between the $Q$ values characterizing mild and strong barium stars (Fig.~\ref{Fig:Q}). Nevertheless, this constitutes a clear difference among mild and strong barium stars, and one may wonder whether it could be the cause of their different levels of chemical pollution. This question is addressed in Sect.~\ref{Sect:abundances}.

The procedure used here to derive $M_{\rm AGB,ini}$ assumes that the binary evolution did not affect the IFMR, but there is no guarantee that this statement is true, quite the contrary. Still, most of the barium systems displayed in the upper right panel of Fig.~\ref{Fig:q_init} have $q' > 1$ as expected, the only exceptions being the mild barium stars with the lowest masses (open circles falling below the $q' = 1$ line). Some of these have WDs with masses lower than 0.5~\Msun, which is unphysical since this corresponds to progenitors which never reached the TP-AGB \citep[see, e.g.,][]{2016A&A...586A.151M}, and therefore could not trigger the s-process whose ashes are responsible for making the barium star. 

Some of the WD companions to barium stars have  large masses (with a few just above 1~M$_\odot$), pointing towards initial AGB masses larger than 5~M$_\odot$  (top panel of Fig.~\ref{Fig:q_init}). It is worth noting that AGB stars of such large masses and with solar (or slightly subsolar) metallicities are not able to produce substantial s-process enrichments  \citep[see, e.g.,][]{2015ApJS..219...40C,2016ApJ...825...26K,2018A&A...620A.146C}. For that reason, these WD masses above 1~M$_\odot$ derived under the assumption of constant $Q$  are likely somewhat overestimated; we note that the most-peaked WD distribution in Fig.~\ref{Fig:M2} is in that respect preferable, as it contains just one WD with a mass just above 1~M$_\odot$. Except for those extreme cases however the WD mass distributions  presented in Figs.~\ref{Fig:M2} and \ref{Fig:M2Q} are compatible with current expectations from AGB s-process nucleosynthesis.

\section{The period--mass--metallicity--abundance connection}
\label{Sect:abundances}

In this section, we investigate the correlation between abundances, orbital periods, metallicities, and masses (barium star and WD companion).  So far, the overabundances of s-process elements in barium stars were tested for possible correlation primarily with orbital periods \citep{1992btsf.work..110J,1994A&A...291..811B,2004MmSAI..75..760B,2015A&A...576A.118A,2018A&A...MMMA.NNNK}, and to a lesser extent with metallicity \citep{2016A&A...586A.158J,2016A&A...586A.151M}. Our current analysis (especially Sect.~\ref{Sect:results7}) advocates the addition of barium-star and WD masses to the analysis \citep[see also][]{2016A&A...586A.151M}. 

\begin{figure}
\vspace{-2cm}
\includegraphics[width=9cm]{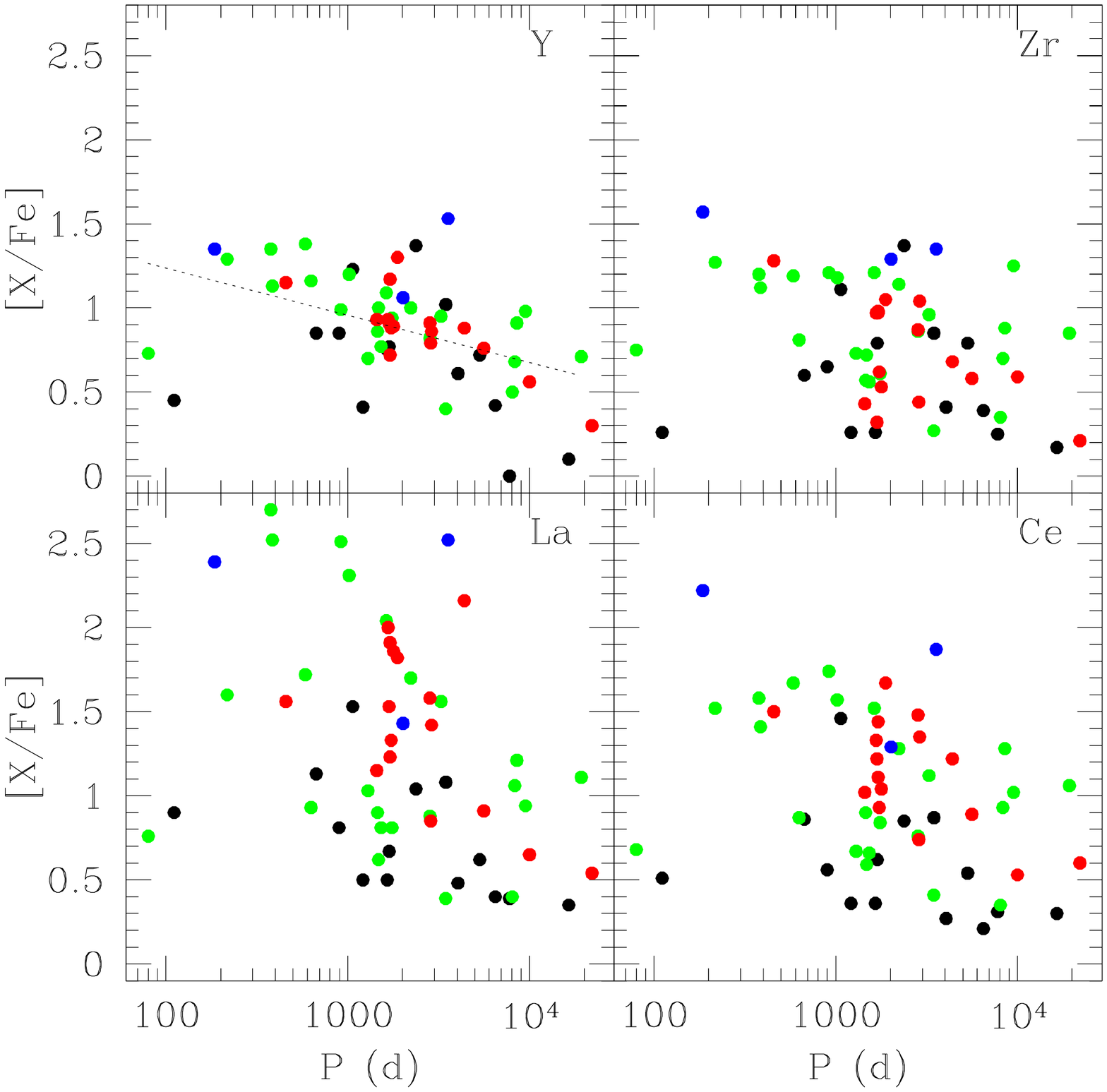}
\vspace{-2cm}\\
\caption{\label{Fig:Period_abundance} 
Period--abundances relationship for s-process elements Y, Zr, La, and Ce. In the panel corresponding to Y, the dotted line is a least-square fit to the data, illustrating the trend existing with orbital period.
Blue ([Fe/H] $< -0.6$), red (from $-0.6$ to $-0.3$), 
green (from $-0.3$ to $-0.1$), and
black ([Fe/H] $\ge -0.1$) symbols denote stars of increasing metallicities. 
}
\end{figure}

 \begin{figure}
\vspace{-2cm}
\includegraphics[width=9cm]{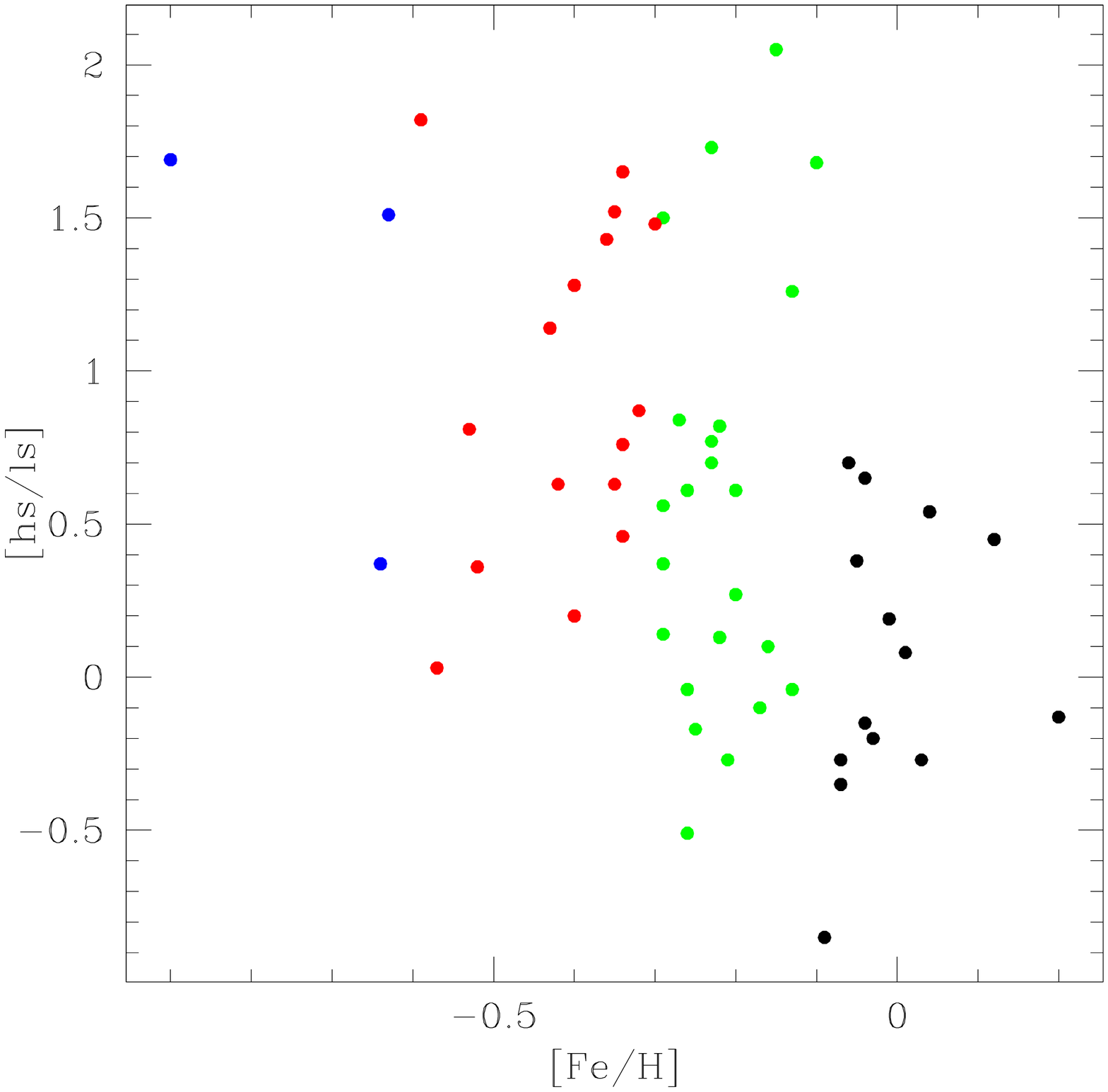}
\vspace{-2cm}
\caption{\label{Fig:hsls} 
Efficiency of the s-process expressed as [hs/ls]~$ \equiv \mathrm{([La/Fe] + [Ce/Fe])} - \mathrm{([Y/Fe] + [Zr/Fe])}$ as a function of metallicity [Fe/H] and color-coded as in Fig.~\ref{Fig:Period_abundance}. 
}
\end{figure}

\begin{figure}
\vspace{-2cm}
\includegraphics[width=9cm]{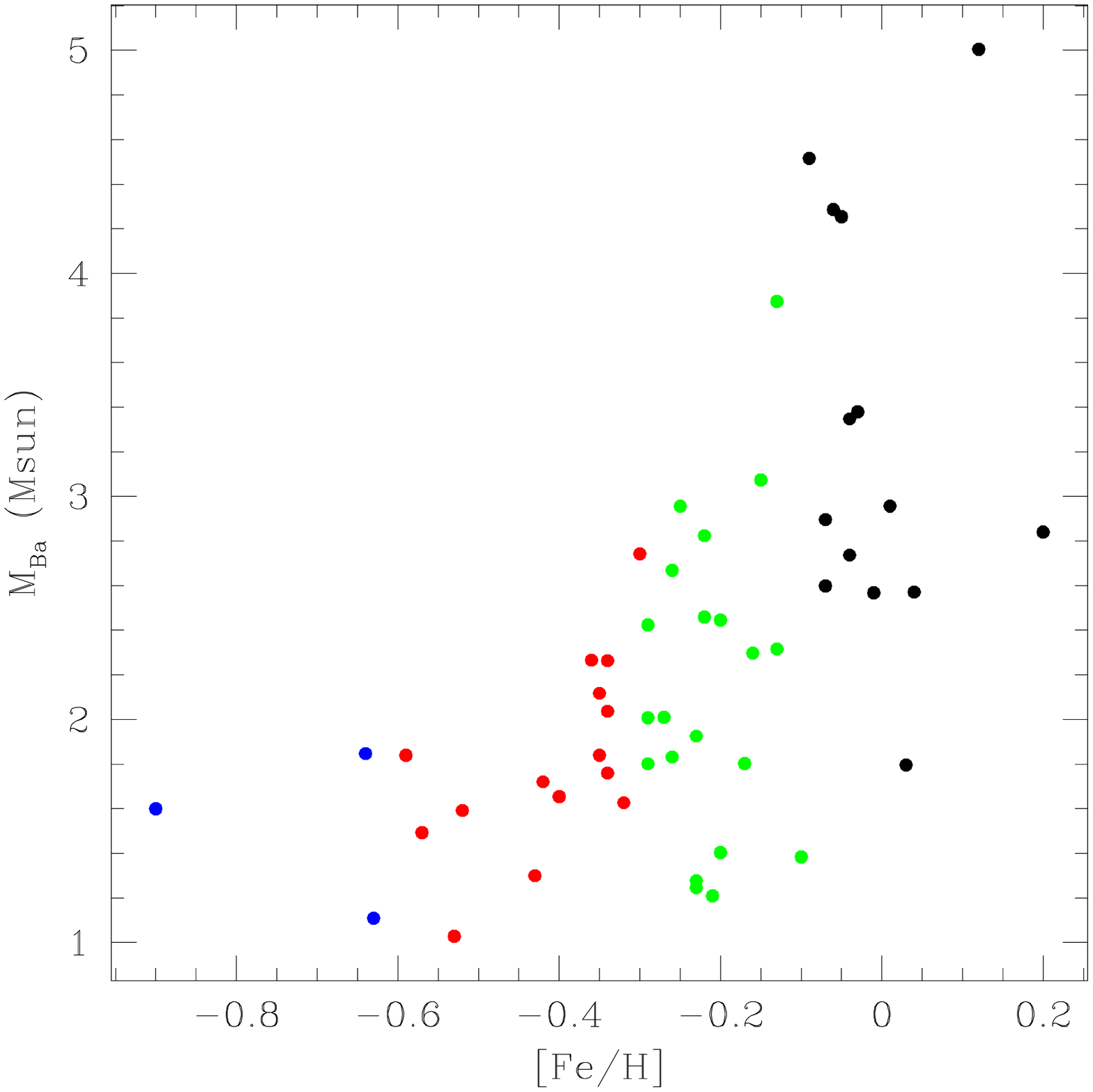}
\vspace{-2cm}
\caption{\label{Fig:Mass_FeH} 
Barium-star mass vs. [Fe/H], color-coded  as in Fig.~\ref{Fig:Period_abundance}. 
}
\end{figure}

Abundances for the barium stars were derived as described in \citet{2018A&A...MMMA.NNNK}, and are listed in Table~\ref{Tab:abundances}; they are also 
displayed in Fig.~\ref{Fig:Period_abundance} as a function of the orbital period. Earlier studies (as listed above) claimed the presence of a general trend of decreasing s-process overabundance with increasing orbital period. In our data sample, this trend is visible only for Y. For Zr, La, and Ce, the trend, if any,  is blurred by a large scatter. As shown by the color sequence in Fig.~\ref{Fig:Period_abundance} (black -- green -- red -- blue, corresponding to stars of decreasing metallicities; see caption of Fig.~\ref{Fig:Period_abundance}), this scatter is partly due to metallicity, since high-metallicity stars (black points) appear mostly at the bottom of the cloud, whereas low-metallicity stars (blue points) appear mostly at its top. The role of metallicity is best revealed by Fig.~\ref{Fig:hsls}, which displays the s-process efficiency expressed as [hs/ls]~$ \equiv \mathrm{([La/Fe] + [Ce/Fe])} - \mathrm{([Y/Fe] + [Zr/Fe])}$ as a function of metallicity. The trend seen on that plot is
not surprising given that the efficiency of the s-process nucleosynthesis, when controlled by the $^{13}$C($\alpha$,n)$^{16}$O neutron source, has been shown to increase with decreasing metallicity 
\citep[e.g.,][and references therein]{1988MNRAS.234....1C,2018A&A...620A.146C}. 

The barium-star mass will play a role as well, since (i) barium-star mass and metallicity vary together (lower masses corresponding to lower metallicities, as expected; see Fig.~\ref{Fig:Mass_FeH}), and (ii) larger barium-star masses imply larger envelope masses, and therefore higher dilution of the accreted matter in the envelope \citep[at least as long as the envelope is convective, notwithstanding any influence of a possible thermohaline mixing erasing the influence of the envelope mass; e.g.,][]{2009PASA...26..176H}.

To summarize the findings of this and previous sections,  Table~\ref{Tab:matrix} shows the Pearson's correlation-coefficient\footnote{Strictly speaking, the Pearson's correlation coefficient (which tests the linearity of the correlation) requires
 that each dataset be normally distributed. The nonparametric Spearman's rank correlation coefficients have therefore been computed as well, but do not differ meaningfully from the  Pearson's correlation coefficients listed in Table~\ref{Tab:matrix}.} matrix of the variables discussed so far, and they reveal in a quantitative way most of the results discussed so far:
\begin{itemize}
\item{(i)} $M_{\rm Ba}$ and $M_{\rm WD}$ are the most strongly correlated variables by construction, since $M_{\rm WD}$ has been derived from $M_{\rm Ba}$ under the assumption of constant $Q \equiv M_{\rm WD}^3/(M_{\rm Ba}+M_{\rm WD})^2$ separately for mild and strong barium stars.
\item{(ii)} The strong correlation between $e$ and $P$ is the manifestation of  the so-called `$e - P$ diagram'.
\item{(iii)} $M_{\rm Ba}$ and [Fe/H] are well correlated (Fig.~\ref{Fig:Mass_FeH}).
\item{(iv)} S-process abundances are well correlated with each other, and moderately anti-correlated with $P$ (the anti-correlation with $P$ is the largest for [Y/Fe] and [La/Fe]; Fig.~\ref{Fig:Period_abundance}). However, the strongest correlation between dynamical and chemical parameters is between $q'$ and [La/Fe], [Ce/Fe], as anticipated in Sect.~\ref{Sect:initial}. 
\end{itemize}

The discussion of the implications of these results on the formation scenario of barium stars (and in particular the origin of the mild/strong nature of the barium star) is deferred to a forthcoming paper.

\begin{table*}
\caption[]{\label{Tab:matrix}
Lower left half of the (symmetric) Pearson's correlation-coefficient matrix of the variables $M_{\rm Ba}$, $M_{\rm WD}$, $q'$, $P$, $e$, $f(M_{\rm Ba},M_{\rm WD})$, [Fe/H], [Y/Fe], [Zr/Fe], [La/Fe], [Ce/Fe].  (Nondiagonal) Correlation coefficients larger than 0.4 (in absolute value) are in boldface. This value of the correlation coefficient  corresponds to a two-tailed $p$-value of 0.17\%, meaning that an uncorrelated system
will produce datasets that have a Pearson's correlation coefficient at least as extreme
 as $\pm0.4$ for 0.17\% of the draws. For a correlation coefficient of 0.3, the two-tailed $p$-value rises to 2\%. 
 }
\begin{tabular}{r|rrrrrrrrrrr}
\hline\\
                      & $M_{\rm Ba}$&  $M_{\rm WD}$&  $q'$ & $P$&  $e$&  $f(M_{\rm Ba},M_{\rm WD})$& [Fe/H]& [Y/Fe]& [Zr/Fe]& [La/Fe]& [Ce/Fe]\\
\hline\\
$M_{\rm Ba}$            & 1. \\
$M_{\rm WD}$ & {\bf 0.913}  & 1.   \\
$q'$                & 0.165  & {\bf 0.532}   & 1. \\
$P$                 & -0.058 & -0.174  & -0.284 & 1.  \\
$e$                 & -0.025  & -0.162  & -0.304 & {\bf 0.593}  &1.   \\
$f(M_{\rm Ba},M_{\rm WD})$& -0.207 &-0.086 & 0.154 & -0.126 &-0.079&  1. \\
{[}Fe/H]            &   {\bf 0.583} & {\bf 0.471} &-0.078 & 0.001 &-0.031& 0.032  &1. \\  
{[}Y/Fe]            &  -0.364 &-0.132 &0.351 & {\bf -0.462} &-0.360&  0.202 &-0.373& 1.\\ 
{[}Zr/Fe]           &  -0.285 &-0.078 &0.309 & -0.330 &-0.251 & 0.164 &-0.331 & {\bf 0.827}  &1.   \\
{[}La/Fe]           & -0.363  & -0.089 &{\bf 0.489} & -0.390 &-0.370 & 0.212 &{\bf -0.432} &{\bf 0.718}  &{\bf 0.715}  &1.\\
{[}Ce/Fe]          & {\bf -0.406}  & -0.121  &{\bf 0.498} & -0.357 &-0.297 & 0.186&{\bf -0.539} &{\bf 0.794}  &{\bf 0.811}  &{\bf 0.913}  &1.  \\
\hline\\
\end{tabular}
\end{table*}

\section{Conclusion}

This study completes the radial-velocity monitoring of samples of mild and strong barium stars, and extrinsic S stars initiated in 1984 with the CORAVEL spectrograph. 
All stars monitored (37 strong barium stars, 40 mild barium stars, and 34 extrinsic S stars) turn out to be binaries (except for the mild barium star HD~95345), and provide a first-hand collection of 111 post-mass-transfer systems among which 105 with orbital elements will serve in the future as benchmark systems for binary-evolution models.

Our HERMES/Mercator radial-velocity monitoring delivered the long-period orbits not yet available in the mid-course analysis published in 1998 \citep{1998A&A...332..877J}. We found several orbits with periods in the range  $1 - 4\times10^4$~d ($\sim 110$~yr). With the present study, we clearly show that the wind-accretion scenario invoked to account for the s-process pollution in the widest systems  \citep{1988A&A...205..155B,2015A&A...581A..62A,2018A&A...620A..63A} is no longer efficient in systems with periods in excess of 
$4\times10^4$~d due to the low accretion cross-section of the wind in such systems.

The eccentricity--period diagram further reveals that, on average, barium stars with strong s-process overabundances are restricted to the period range 200 -- 5000~d (with two exceptions at $10^4$~d), whereas mild barium stars are found in the range 700 -- 20000~d (with two exceptions at $\sim 100$~d). The avoidance region ($P > 10^3$~d, $e < 0.07$) is confirmed, and is likely a vestige of a similar avoidance region (albeit extending towards larger eccentricities) in pre-main sequence binaries. Almost all barium systems with periods shorter than $10^3$~d are circular, and this property is likely attributable to the circularisation occurring as the giant star ascends the red giant branch (RGB), as demonstrated by the models of \citet{2019Escorza}. Extrinsic S stars, which are still ascending the RGB confirm this statement since S  systems with $P < 10^3$~d are not necessarily circular, and
this is the only property that they do not share with barium systems.

Thanks to Gaia DR2 parallaxes and spectral-energy distribution fits, our barium-star sample could be located in the Hertzsprung-Russell diagram, and the position of individual stars compared with STAREVOL tracks of the corresponding metallicity  \citep{2000A&A...358..593S,2008A&A...489..395S}. Metallicities for the barium stars were either collected from the literature when available or were derived from Mercator/HERMES high-resolution spectra. This comparison then gives access to the barium-star masses ($M_{\rm Ba}$), which in turn yield the companion masses ($M_{\rm WD}$) under the assumption of a constant $Q = M_{\rm WD}^3/(M_{\rm Ba} + M_{\rm WD})^2$ value (different for mild and strong barium stars). This constancy was  envisioned by \citet{1988covp.conf..403W} and \citet{1990ApJ...352..709M} and is clearly confirmed by our present extensive samples. The cause of that property, which has not been clearly identified, merits a specific discussion; this is deferred to a forthcoming paper.

The companion masses appear to be restricted in the range 0.5 -- 1.1~M$_\odot$, as expected for WDs. The peak of the distribution lies around 0.55 -- 0.70~M$_\odot$, exactly as for field DA and DB WDs. The heaviest WDs around barium stars point at AGB-progenitor masses around 5~M$_\odot$, at the edge of the predictions for efficient s-process AGB nucleosynthesis.

In the hope of disentangling the various parameters involved in fixing the s-process overabundance levels in barium stars (orbital separation, dilution factor in the barium-star envelope, final AGB core mass, metallicity etc.), we performed a correlation analysis involving parameters $M_1, M_2, q',  P, e, f(M_1, M_2)$, [Fe/H], [Y/Fe], [Zr/Fe], [La/Ce], and [Ce/Fe]. 
Significant correlations or anti-correlations (with coefficients in excess of 0.4) were found between $P$ and $e$,  and  between s-process abundances, $P$, $q'$, and metallicity (as expected). More unexpected is the strong correlation observed between $M_{\rm Ba}$ and metallicity. Such a correlation must be a consequence of the age--metallicity relationship, which predicts that giants of low metallicities ([Fe/H]~$\le -0.4$) in the solar neighborhood must be older than about 5~Gyr \citep[e.g., Fig.~3 of][]{2018MNRAS.477.2326F}, and must therefore be of low mass ($\la 1.3$~M$_\odot$). A strong correlation is also found between [La/Fe], [Ce/Fe], and the ``initial'' mass ratio $q'$. This is the strongest link found so far  between dynamical and chemical abundances; it dominates over any effect related to the orbital period. This is clearly the root  of the difference between mild and strong barium stars, which is visible as well in their different current $Q$ values,  implying that strong barium stars originate from systems with a mass ratio above $\sim1.5$. This is the combined result of the masses of strong barium stars being on average smaller than those of mild barium stars, and of the tendency for WDs around strong barium stars to be  more massive on average.  The first effect contributes to reducing the dilution factor of the accreted matter in the barium-star envelope.
This finding will certainly turn out to be a key constraint for the evolutionary models of binary stars aiming at reproducing the properties of barium stars. Initial conditions adopted in these models should certainly conform to our key finding that the initial mass-ratio $q'$ is very far from being uniform, and differs for strong and mild barium stars.

\begin{acknowledgement}
Based on observations obtained with the HERMES spectrograph, which is supported by the Research Foundation - Flanders (FWO), Belgium, the Research Council of KU Leuven, Belgium, the Fonds National de la Recherche Scientifique (F.R.S.-FNRS), Belgium, the Royal Observatory of Belgium, the Observatoire de Gen\`eve, Switzerland and the Th\"uringer Landessternwarte Tautenburg, Germany. This work required a considerable observing effort, and we therefore gratefully thank all observers from the HERMES consortium and from the Instituut voor Sterrenkunde (KULeuven) who contributed to this effort. We thank as well R. Griffin who kindly communicated his radial-velocity data of the long-period barium star 56~UMa = HD~98839. This work has made use of data from the European Space Agency (ESA) mission
{\it Gaia} (\url{https://www.cosmos.esa.int/gaia}), processed by the {\it Gaia}
Data Processing and Analysis Consortium (DPAC,
\url{https://www.cosmos.esa.int/web/gaia/dpac/consortium}). Funding for the DPAC
has been provided by national institutions, in particular the institutions
participating in the {\it Gaia} Multilateral Agreement. This research has been funded by the Belgian Science Policy  Office
under  contract  BR/143/A2/STARLAB, and by the F.W.O. DK acknowledges the support from Science and Engineering research Board (SERB), Department of Science and technology (DST), India,  through the file number PDF/2017/002338. SVE thanks
Fondation ULB for its support.
The {\it Association of French Variable Star Observers} (AFOEV) is acknowledged for providing the light curve of T~Sgr.
\end{acknowledgement}

\appendix

\section{Spectroscopic binaries with no orbital solutions yet}
\label{Sect:SB_no_orbit}

\begin{figure}
\includegraphics[width=9cm]{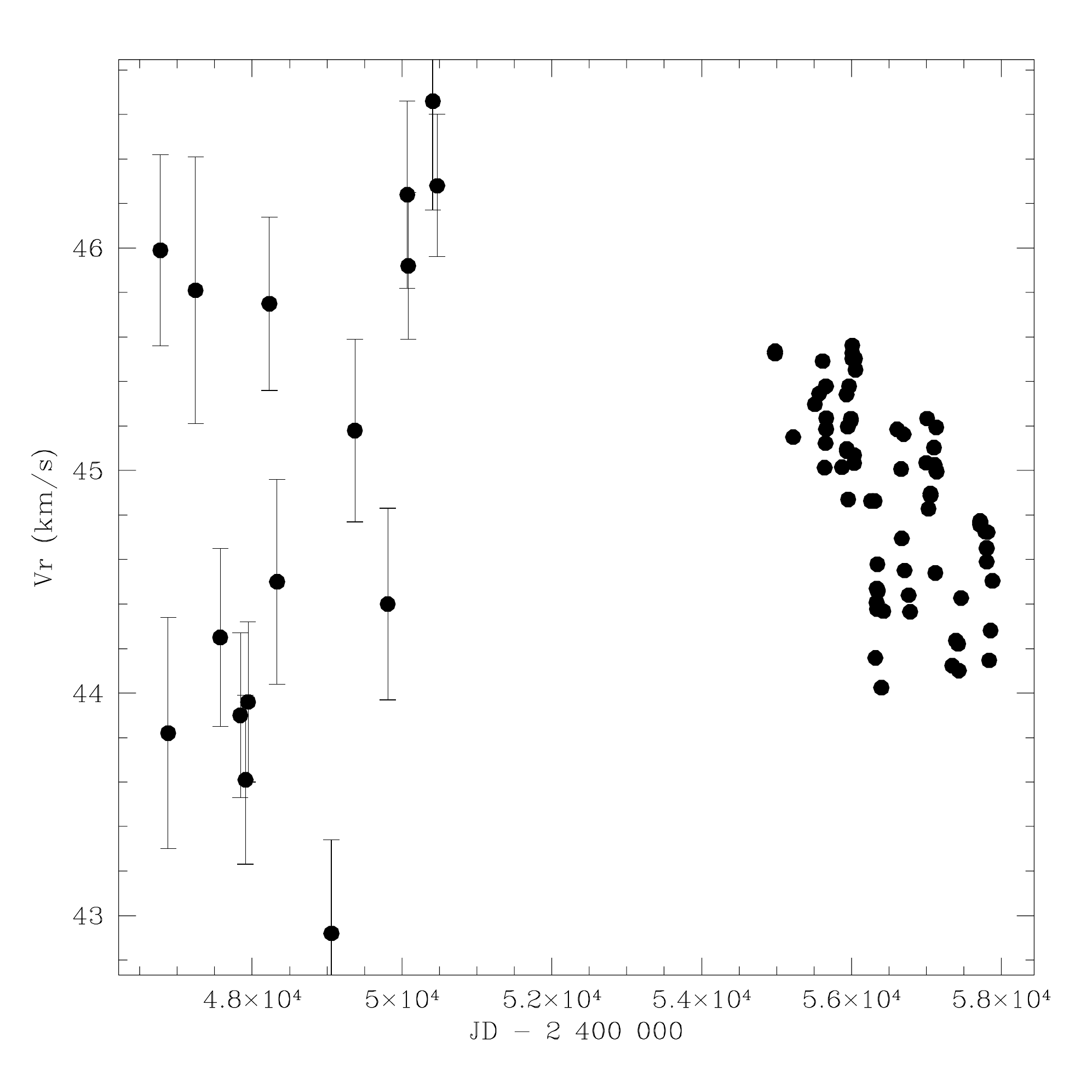}
\includegraphics[width=9cm]{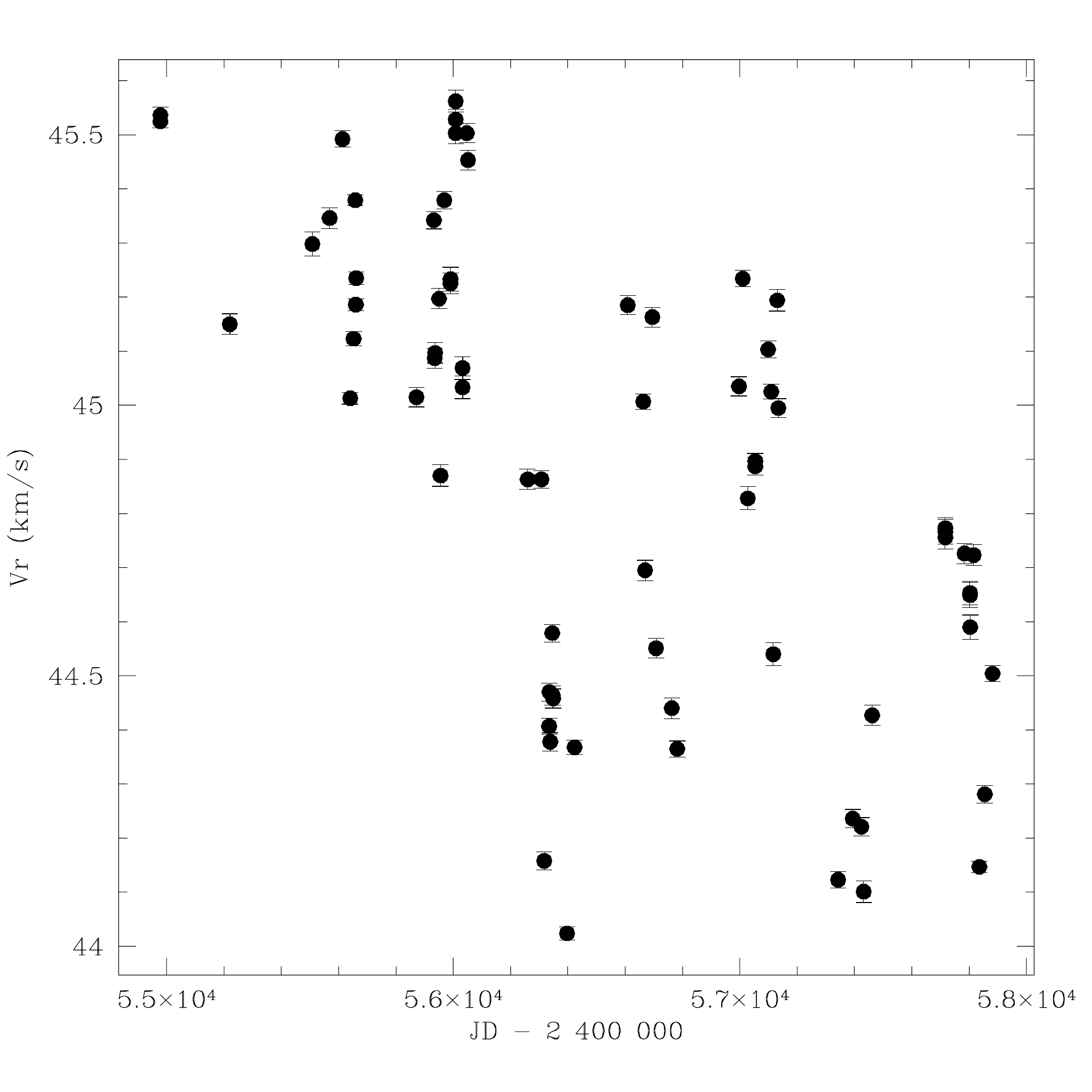}
\caption{\label{Fig:21.2601} 
Top panel: Radial velocities for the S star BD~\mbox{-21$^\circ$2601}. Older data are from CORAVEL, newer from HERMES. No zero-point offset  has been applied to the CORAVEL data.
Bottom panel: Same as top, but for HERMES velocities only.}
\end{figure}

\begin{figure}
\includegraphics[width=9cm]{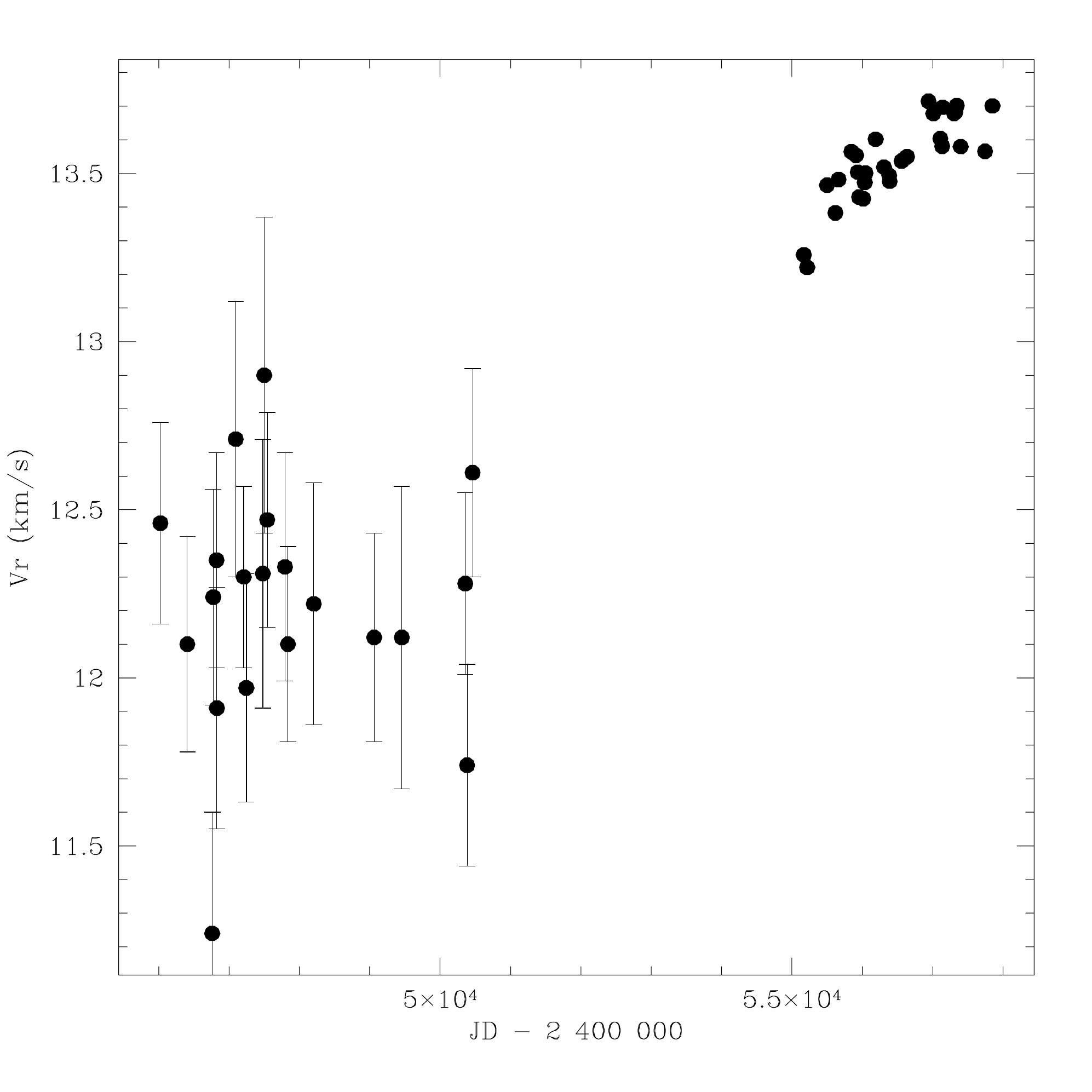}
\caption{\label{Fig:50843} 
Same as Fig.~\ref{Fig:21.2601} but for the mild barium star HD~50843.  No offset  has been applied to the CORAVEL data. 
 }
\end{figure}

\begin{figure}
\includegraphics[width=9cm]{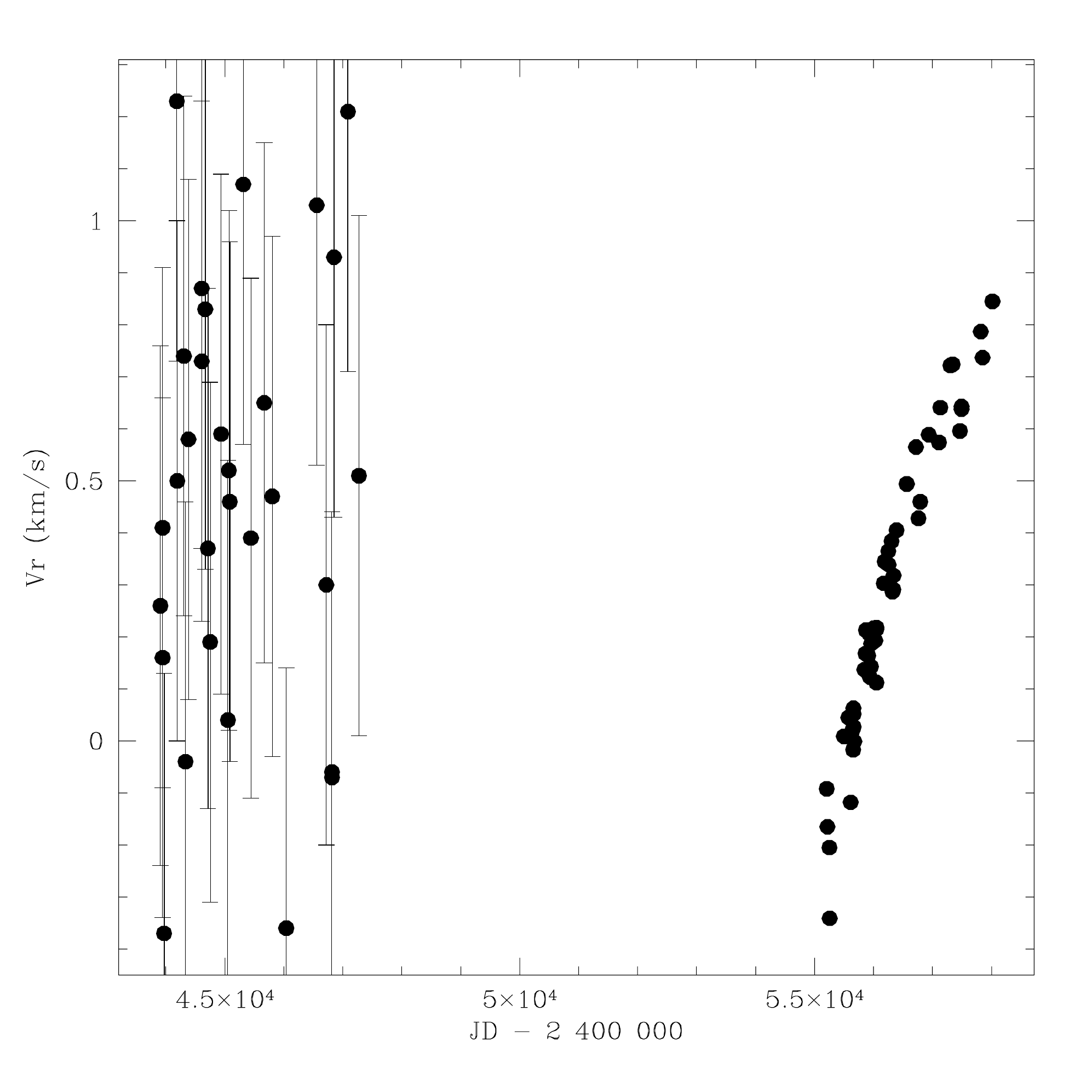}
\caption{\label{Fig:65854} 
Same as Fig.~\ref{Fig:21.2601} but for the mild barium star HD~65854. No zero-point offset  has been applied to the CORAVEL data. }
\end{figure}

\begin{figure}
\includegraphics[width=9cm]{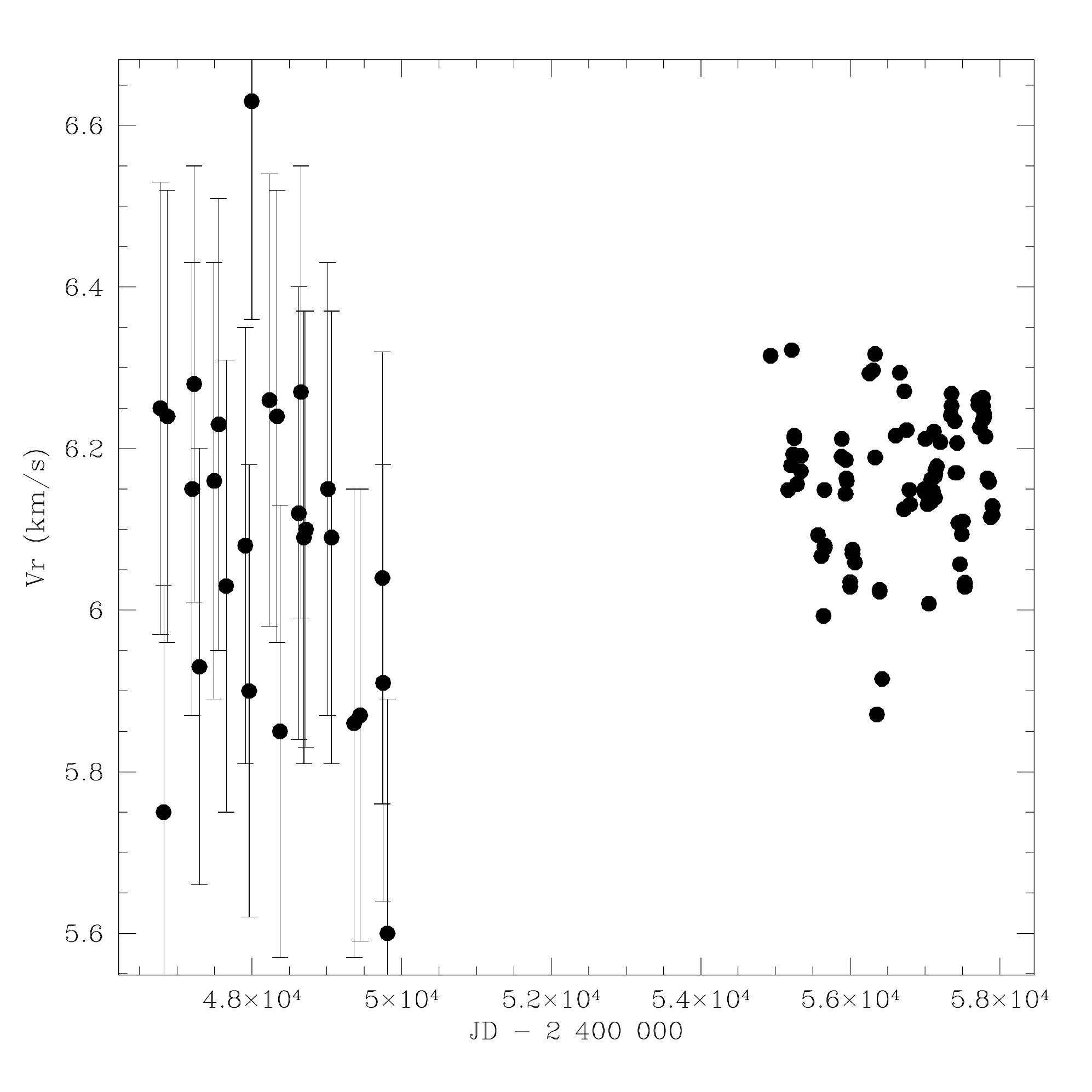}
\includegraphics[width=9cm]{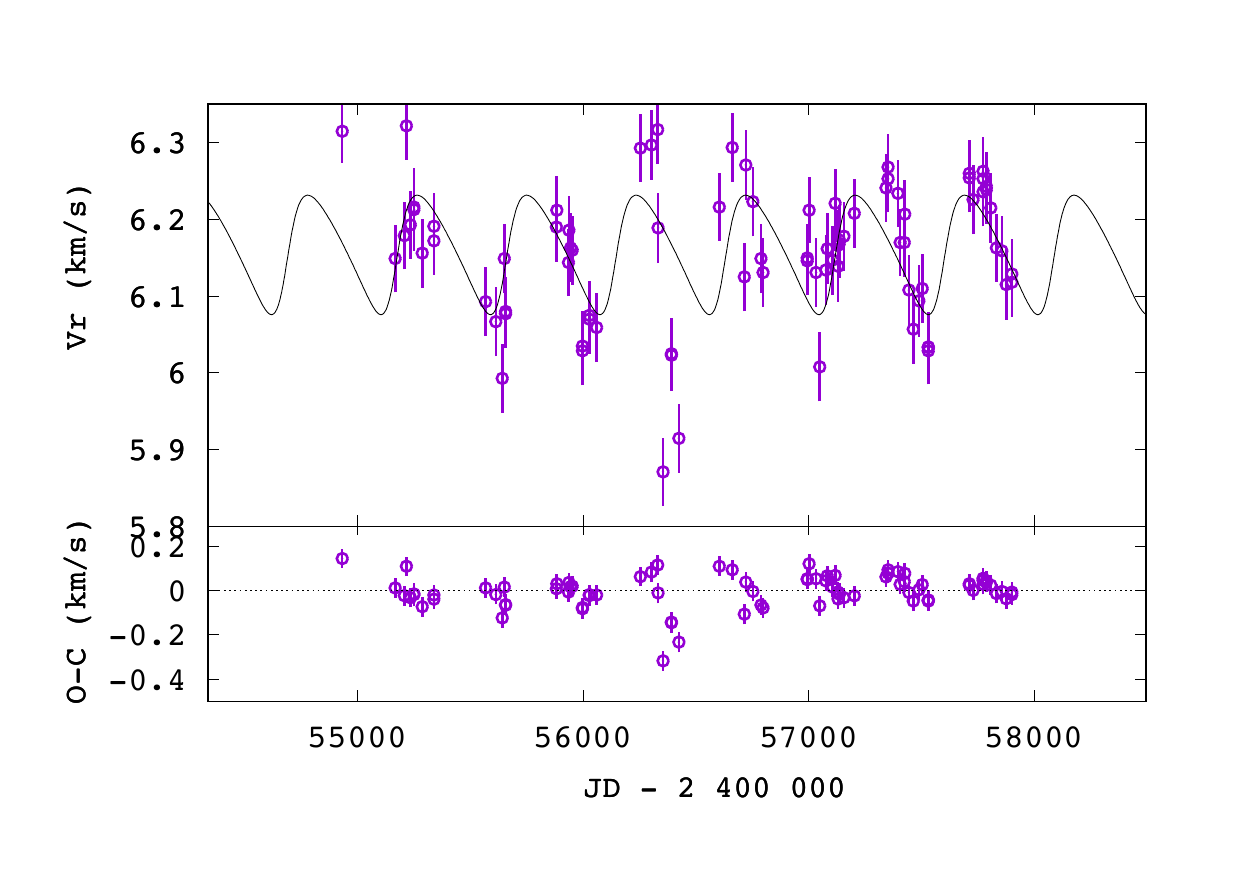}
\caption{\label{Fig:95345} 
Top panel: Same as Fig.~\ref{Fig:21.2601} but for the mild barium star HD~95345.  A zero-point offset of +0.6~km~s$^{-1}$ has been applied to the CORAVEL data. Bottom panel: Tentative orbit based on HERMES data only (see also Table~\ref{Tab:orbits}).
 }
\end{figure}

This section presents velocity curves for the spectroscopic binaries for which orbital solutions are yet to be obtained, namely the extrinsic S star BD~-21$^\circ$2601 (Fig.~\ref{Fig:21.2601}), and the  mild barium stars HD~50843, HD~65854, and HD~95345 (Figs.~\ref{Fig:50843}, \ref{Fig:65854}, and \ref{Fig:95345}). In the figures of this section and the following, all data points posterior to JD~2454900 were obtained with the HERMES spectrograph while the previous ones are all from CORAVEL, the latter being moreover characterized by larger error bars ($\sim 0.3$~\kms).

\section{Orbital solutions}
\label{Sect:Appendix}

This Appendix presents all orbital solutions superimposed  on the velocity data.

\begin{figure}
\includegraphics[width=9cm]{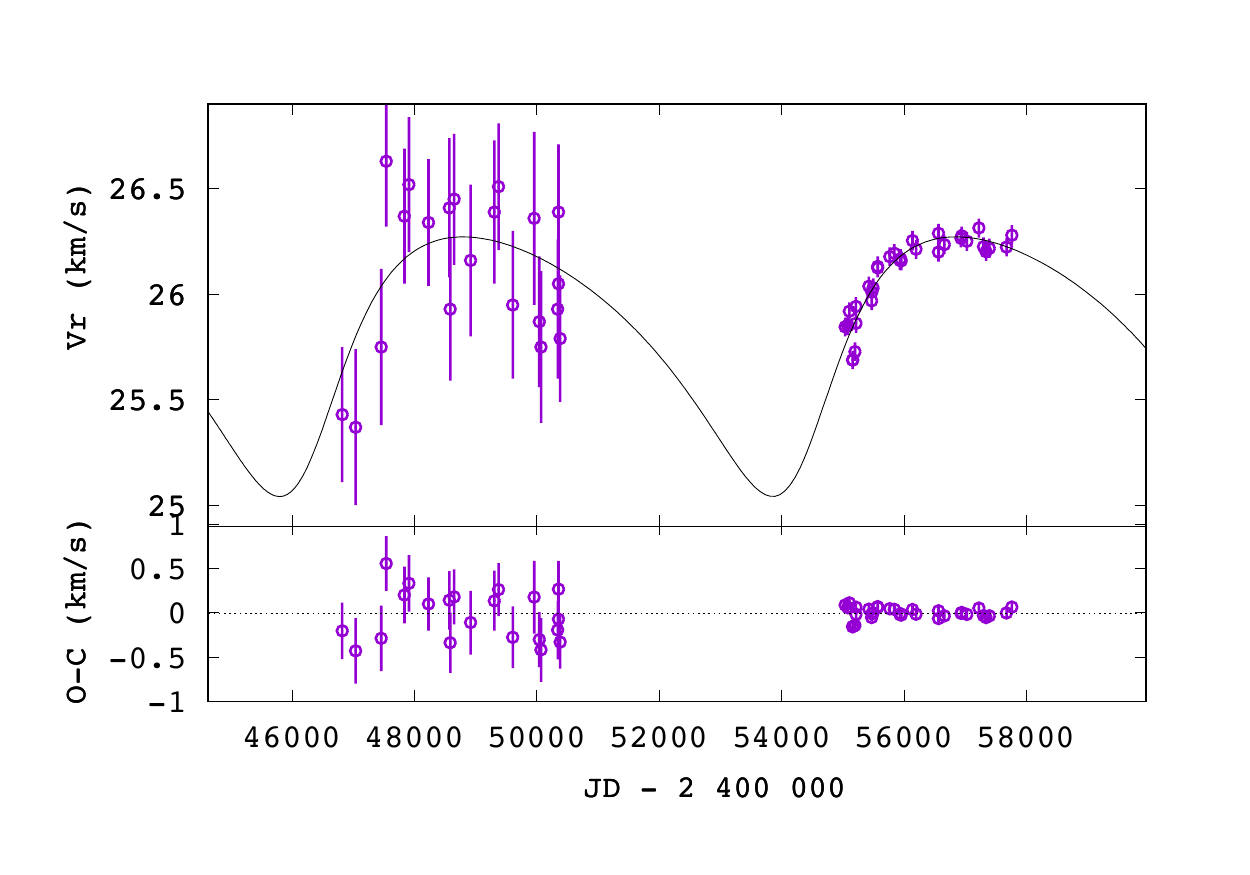}
\caption{\label{Fig:Orbit_18182} 
Upper panel: Radial velocities of the mild barium star HD~18182 and a preliminary orbit with $P = 22$~yr and $e = 0.3$. 
Older data are from CORAVEL, newer from HERMES. A zero point offset of +0.6~\kms\  has been applied to the CORAVEL measurements. Lower panel: O-C residuals.}
\end{figure}

\begin{figure}
\includegraphics[width=9cm]{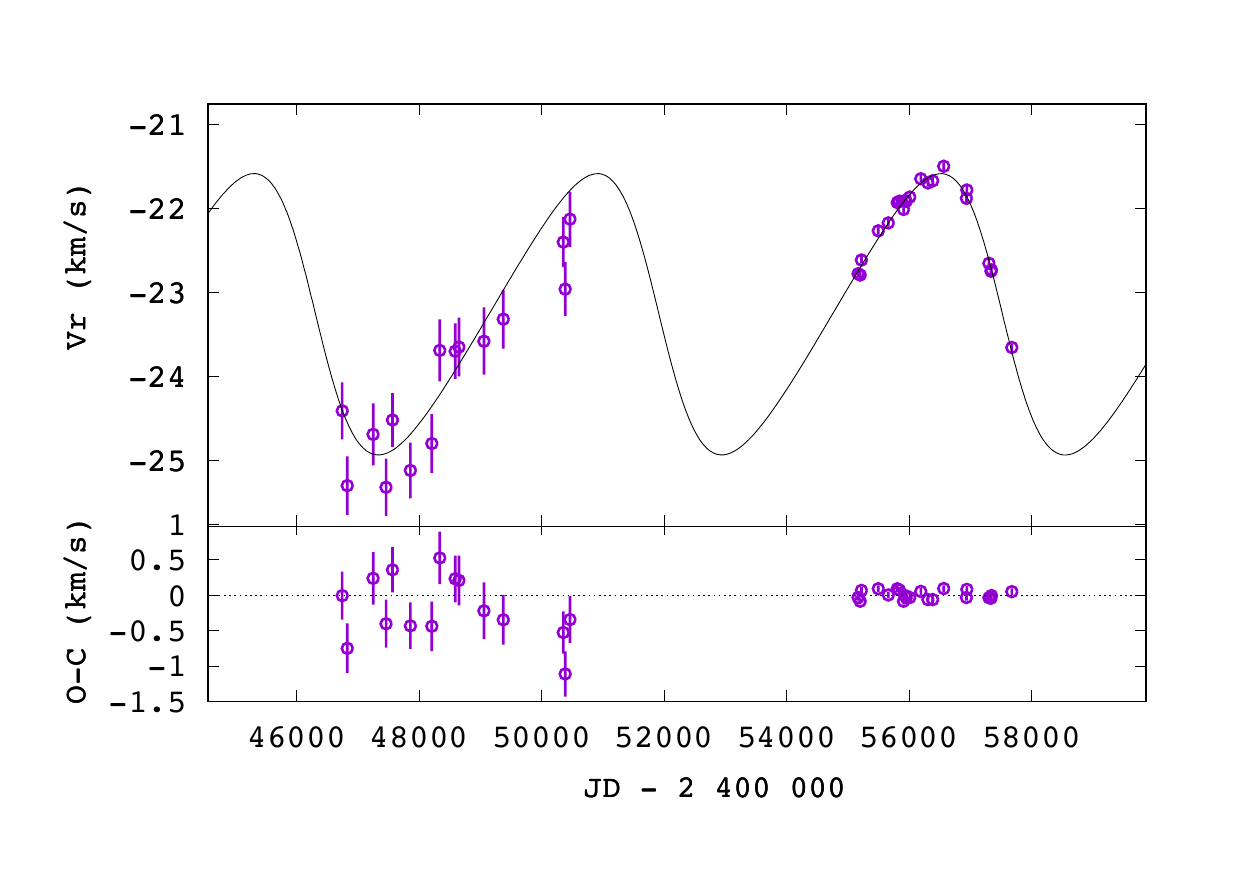}
\caption{\label{Fig:Orbit_40430} 
Upper panel: Radial velocities of the mild barium star HD~40430 and a preliminary orbit with $P = 15$~yr and $e = 0.22$. Older data are from CORAVEL, newer from HERMES. Lower panel: O-C residuals.}
\end{figure}

\begin{figure}
\vspace{-3cm}
\includegraphics[width=9cm]{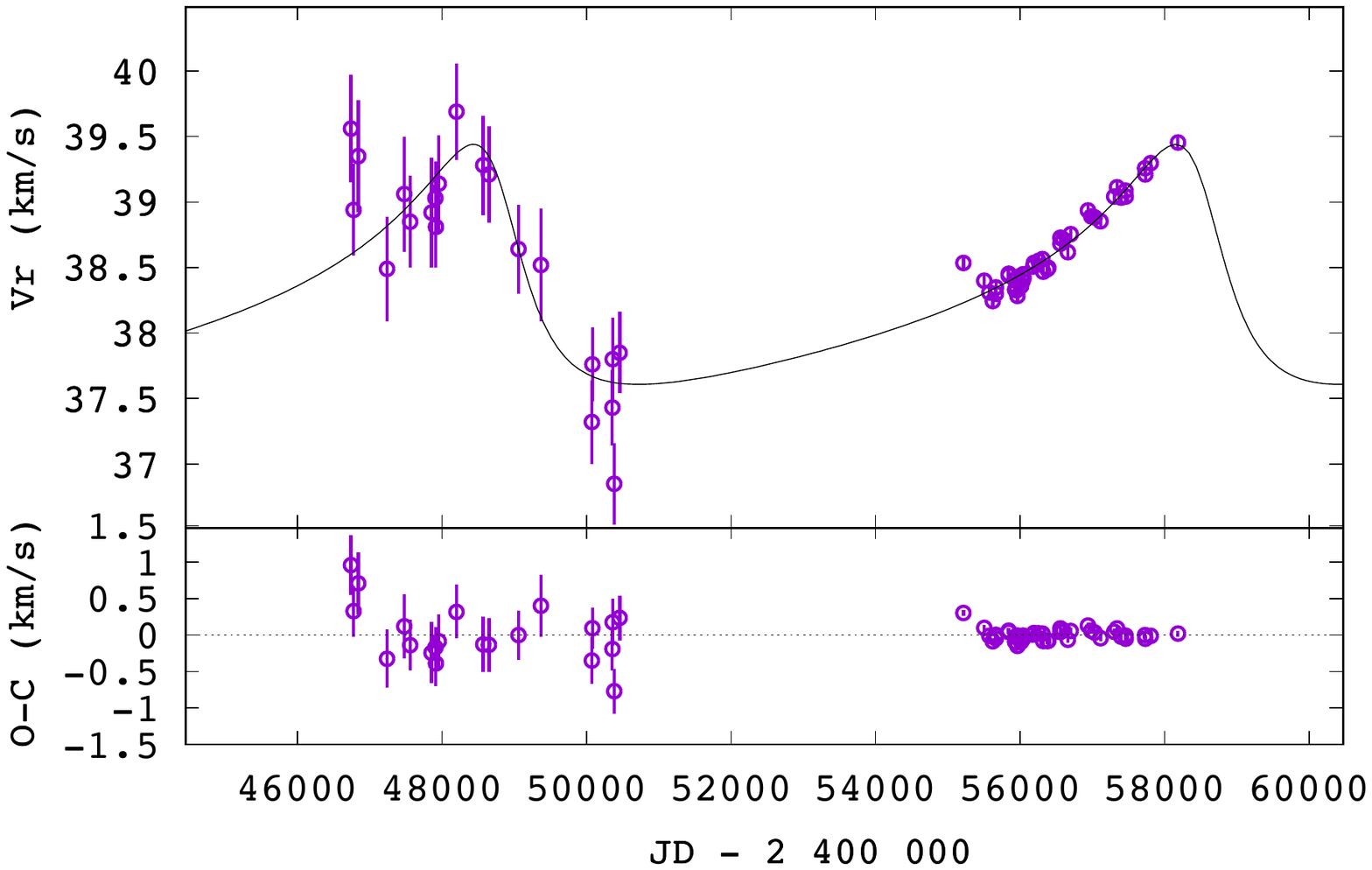}
\vspace{-3cm}
\caption{\label{Fig:Orbit_51959} 
Upper panel: Radial velocities of the mild barium star HD~51959 and a preliminary orbit with $P = 27$~yr and $e = 0.53$. 
Older data are from CORAVEL, newer from HERMES. Lower panel: O-C residuals.}
\end{figure}

\begin{figure}
\includegraphics[width=9cm]{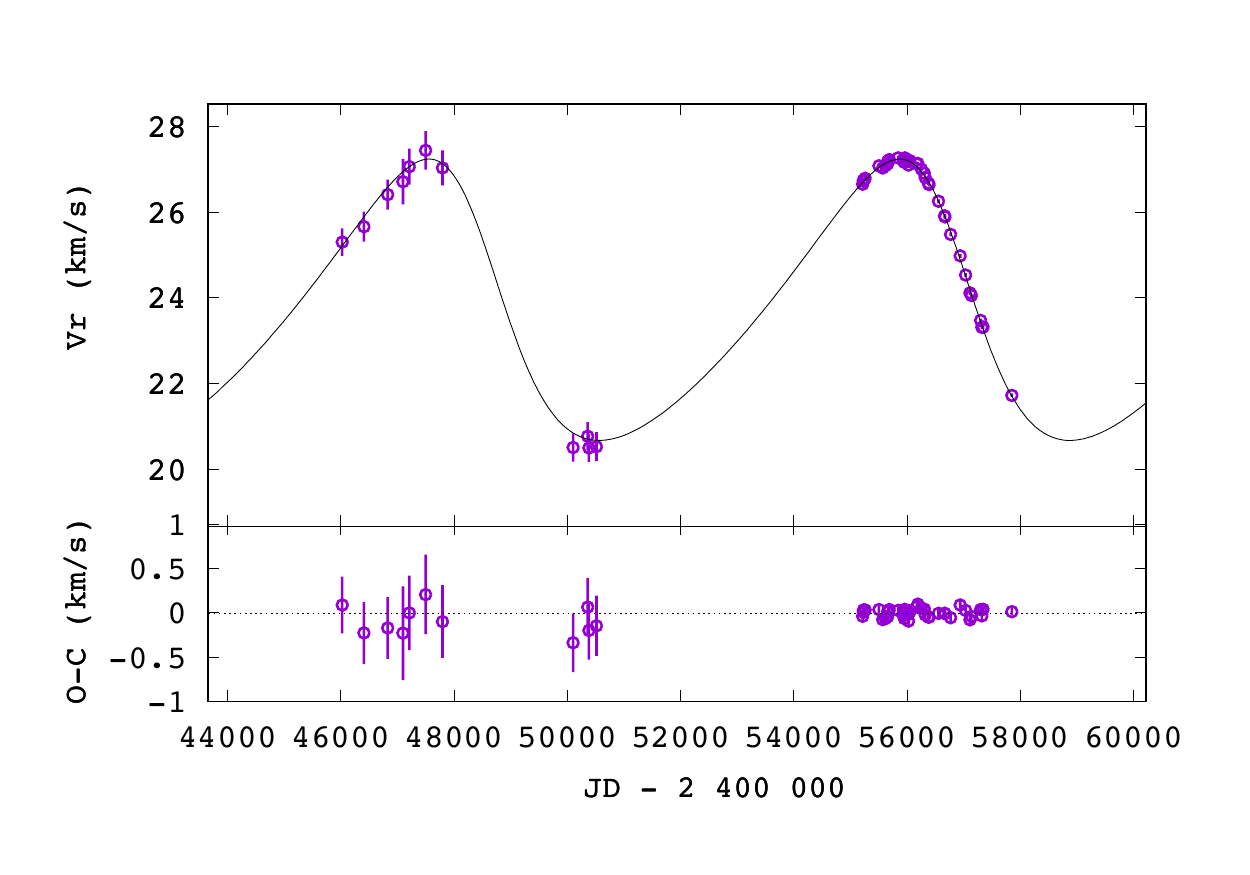}
\caption{\label{Fig:Orbit_53199} 
Upper panel: Radial velocities of the mild barium star HD~53199 and the
associated orbit. Older data are from CORAVEL, newer from HERMES. An offset of +0.4~\kms\  has been applied to the CORAVEL data. Lower panel: O-C residuals.}
\end{figure}

\begin{figure}
\includegraphics[width=9cm]{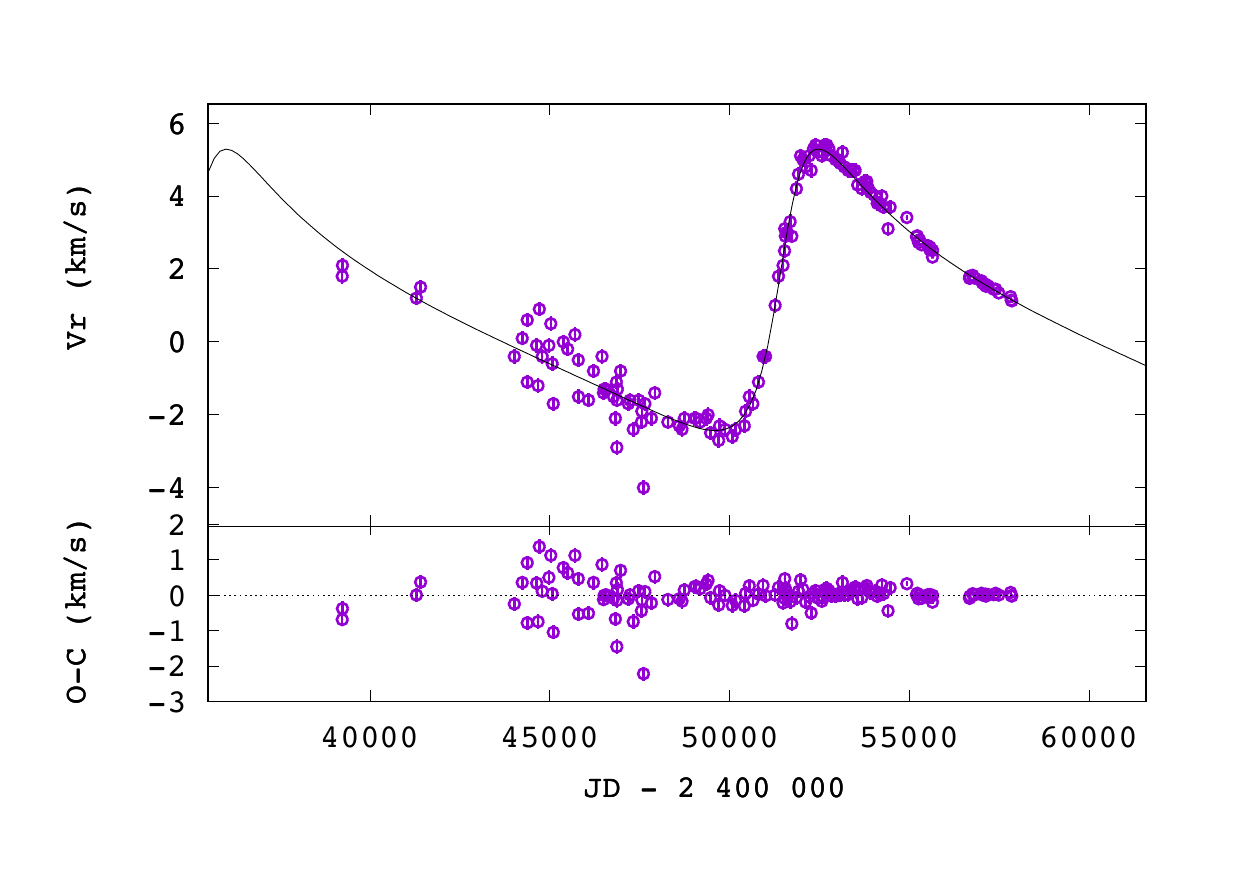}
\caption{\label{Fig:Orbit_98839} 
Upper panel: Radial velocities of the mild barium star HD 98839 (= 56 UMa) and the
associated orbit. Older data are from \citet{2008Obs...128..176G}, newer are from HERMES, according to Table~\ref{Tab:VrS}. In this figure, the
HERMES velocities are offset by +0.6~\kms\  to ensure consistency with Griffin's velocities. Lower panel: O-C residuals.}
\end{figure}

\begin{figure}
\includegraphics[width=9cm]{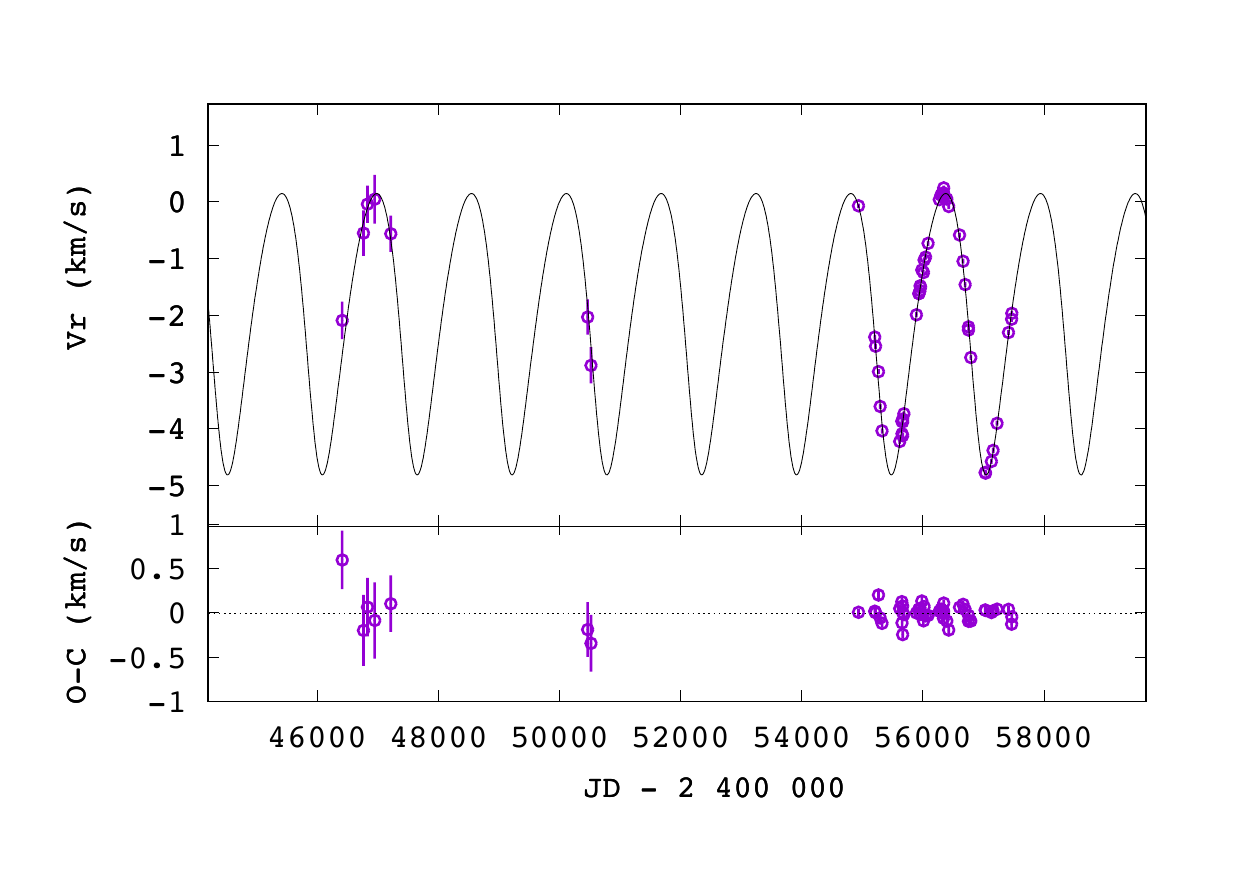}
\caption{\label{Fig:Orbit_101079} 
Upper panel: Radial velocities of the mild barium star HD~101079 and the
associated orbit. Older data are from CORAVEL, newer from HERMES. An offset of +0.5~\kms\  has been applied to the CORAVEL data. Lower panel: O-C residuals.}
\end{figure}

\begin{figure}
\includegraphics[width=9cm]{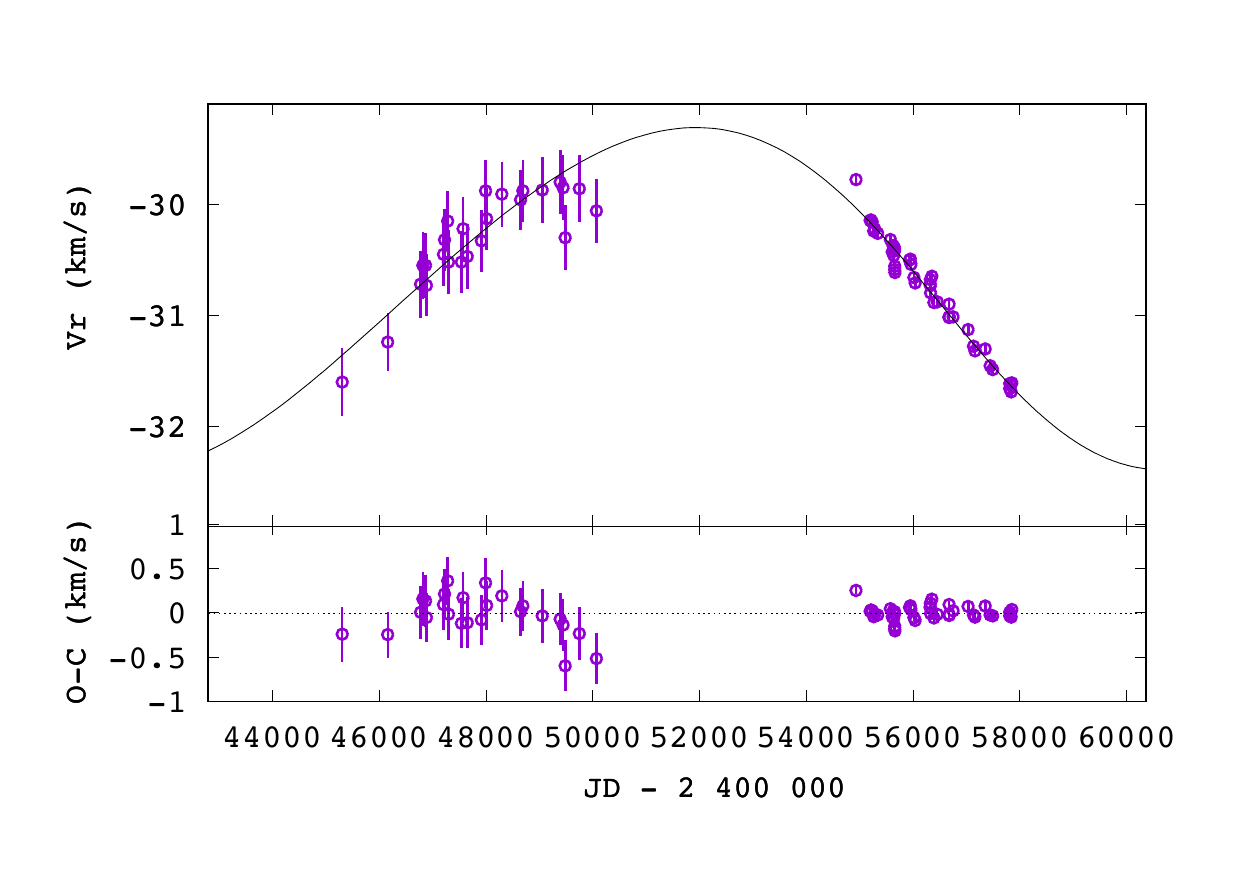}
\caption{\label{Fig:Orbit_104979} 
Same as Fig.~\ref{Fig:Orbit_18182} but for a preliminary orbit of HD~104979 with $P = 53$~yr and $e = 0.1$.
An offset of +0.5~\kms\  has been applied to the CORAVEL data.}
\end{figure}

\begin{figure}
\includegraphics[width=9cm]{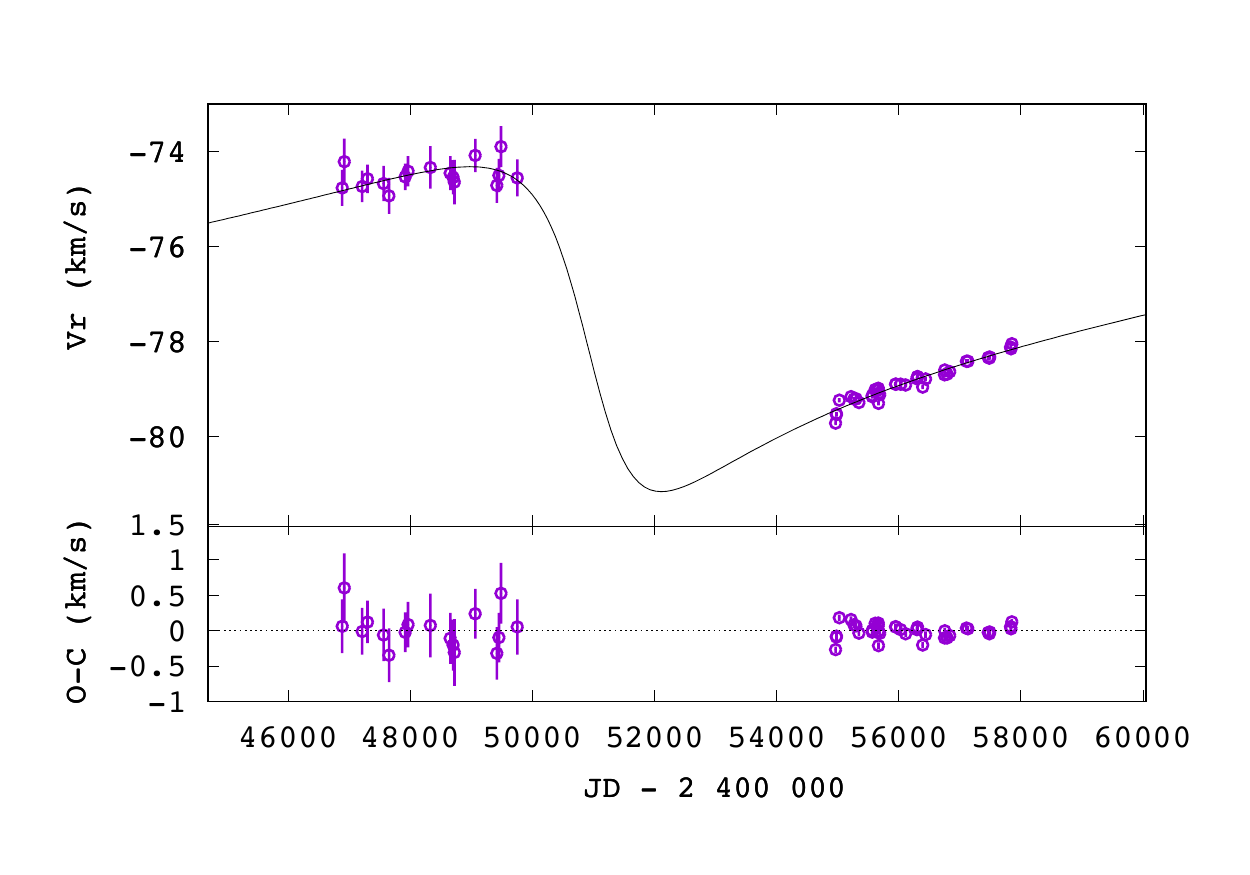}
\caption{\label{Fig:Orbit_119185} 
Upper panel: Radial velocities of the mild barium star HD~119185 and a preliminary orbit with $P = 60$~yr and $e = 0.6$! 
Older data are from CORAVEL, newer from HERMES. Lower panel: O-C residuals.}
\end{figure}

\begin{figure}
\includegraphics[width=9cm]{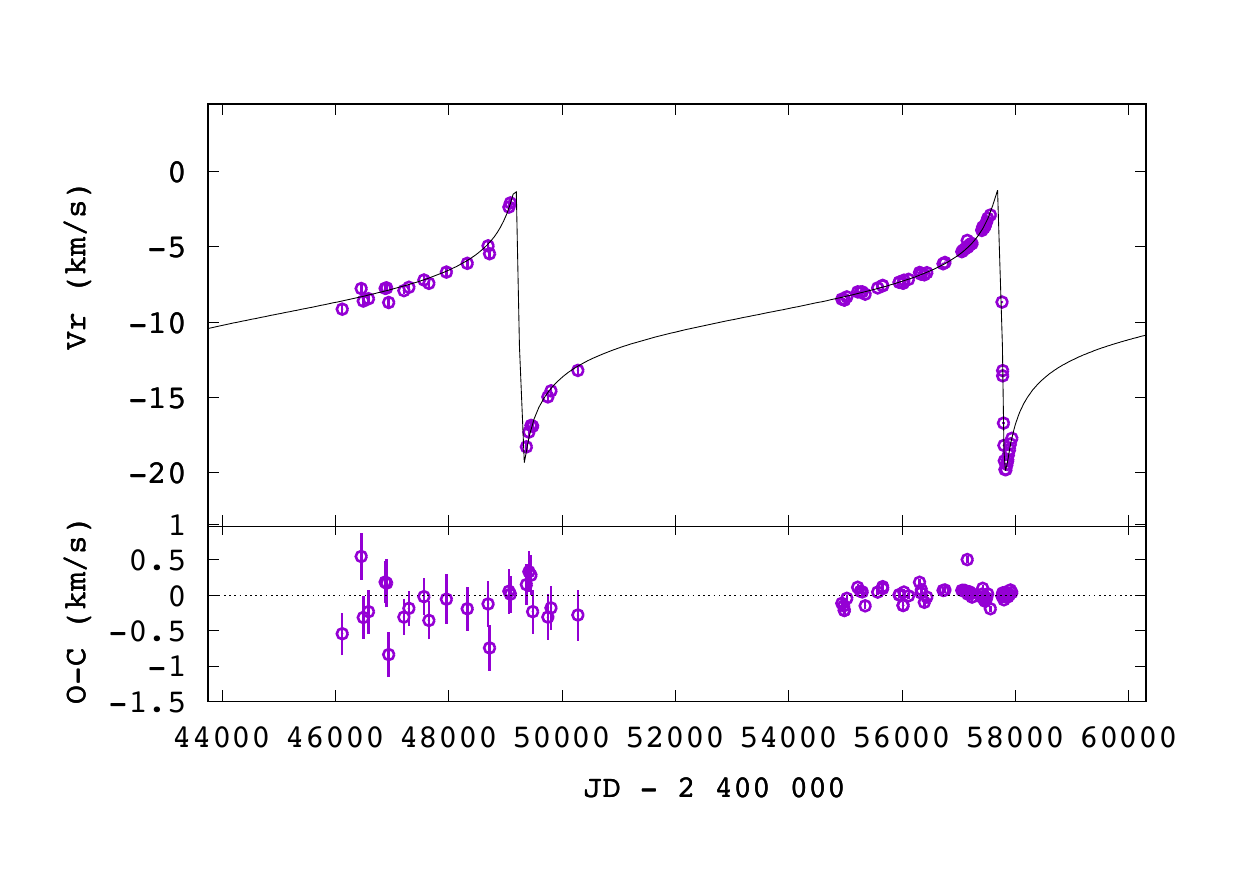}
\caption{\label{Fig:Orbit_123949} 
Upper panel: Radial velocities of the strong barium star HD~123949 and the
associated orbit, having $P = 23.3$~yr and $e = 92$. Older data are from CORAVEL, newer from HERMES. An offset of +0.7~\kms\  has been applied to the CORAVEL data. Lower panel: O-C residuals.}
\end{figure}

\begin{figure}
\includegraphics[width=9cm]{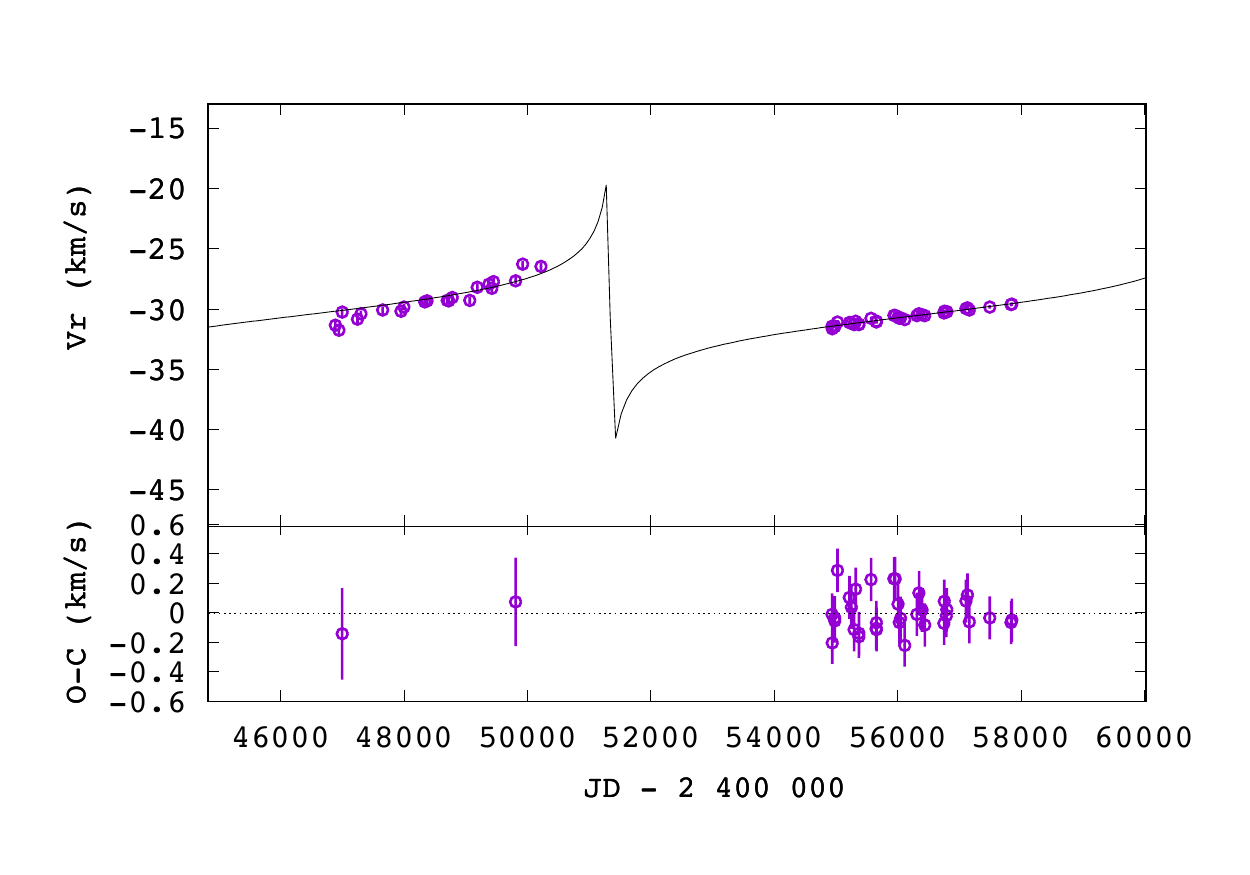}
\caption{\label{Fig:Orbit_134698} 
Upper panel: Radial velocities of the mild barium star HD~134698 and a preliminary orbit with $P = 27$~yr and $e = 0.95$.
Older data are from CORAVEL, newer from HERMES. An offset of +0.5~\kms\  has been applied to the CORAVEL data. Lower panel: O-C residuals, only shown for the data points used in the orbit derivation (see Sect.~\ref{Sect:134698}). The other CORAVEL data are displayed as blue crosses in the upper panel.}
\end{figure}

\clearpage

\begin{figure}
\includegraphics[width=9cm]{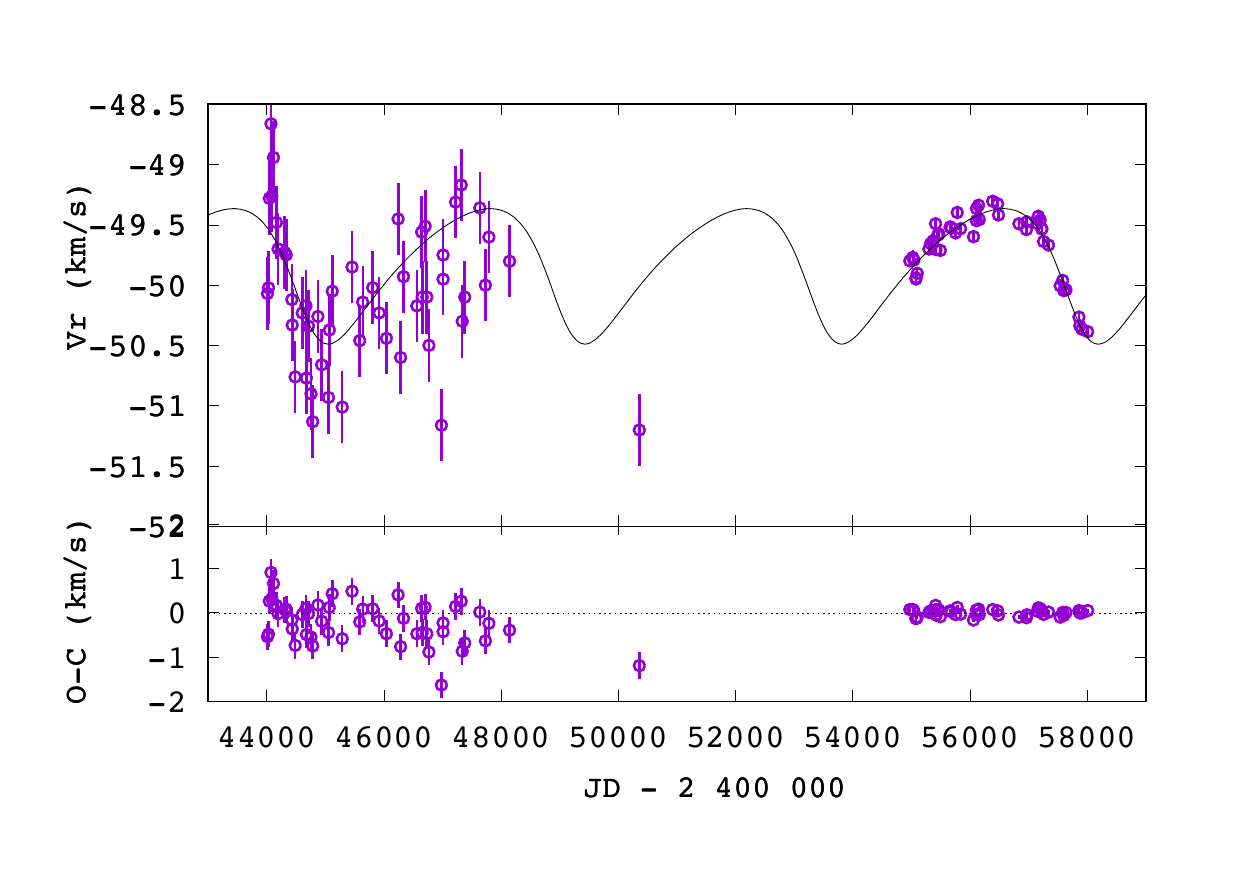}
\caption{\label{Fig:Orbit_183915} 
Upper panel: Radial velocities of the mild barium star HD~183915 and the
associated orbit. Older data are from CORAVEL, newer from HERMES.  Lower panel: O-C residuals.}
\end{figure}

\begin{figure}
\includegraphics[width=9cm]{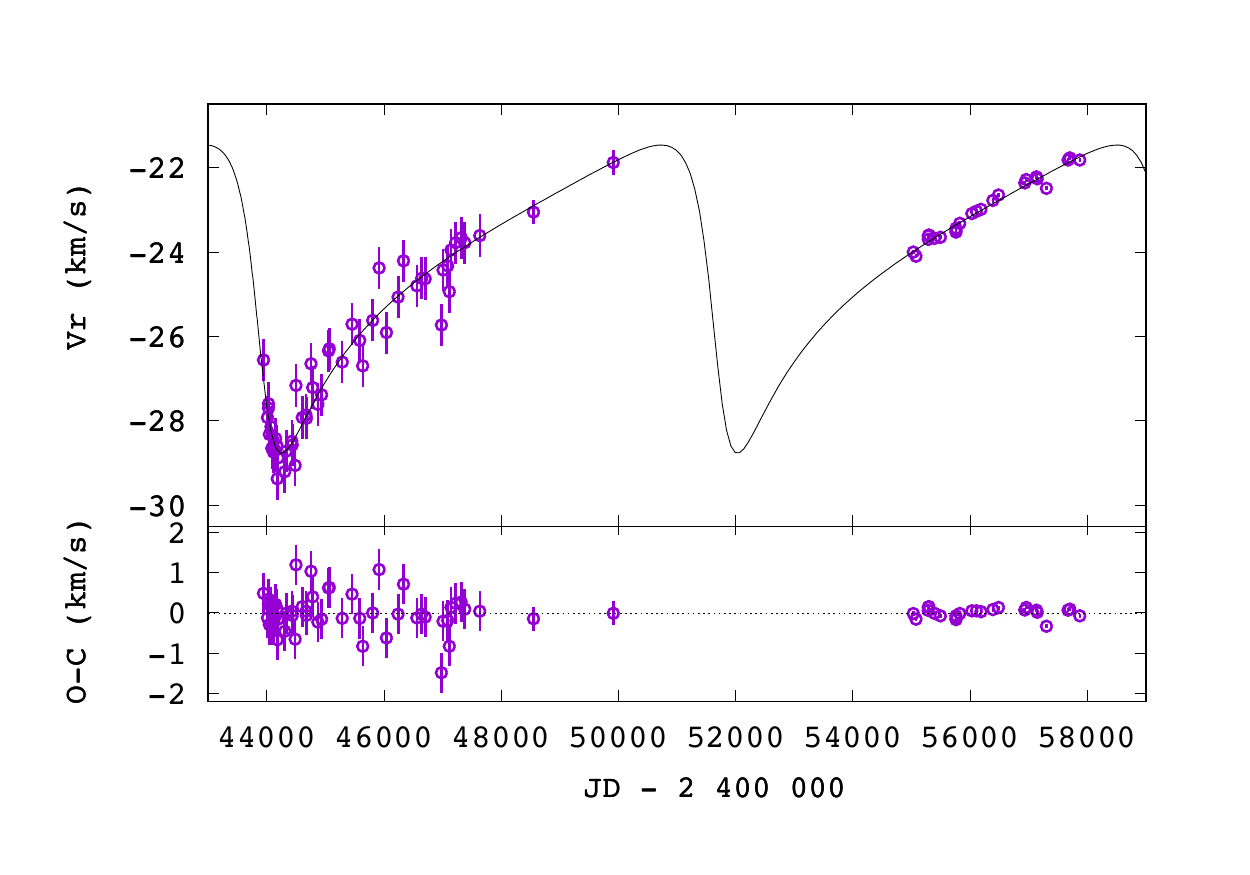}
\caption{\label{Fig:Orbit_196673} 
Upper panel: Radial velocities of the mild barium star HD~196673 and the
associated orbit. Older data are from CORAVEL, newer from HERMES. Lower panel: O-C residuals.}
\end{figure}

\begin{figure}
\vspace{-3cm}
\includegraphics[width=9cm]{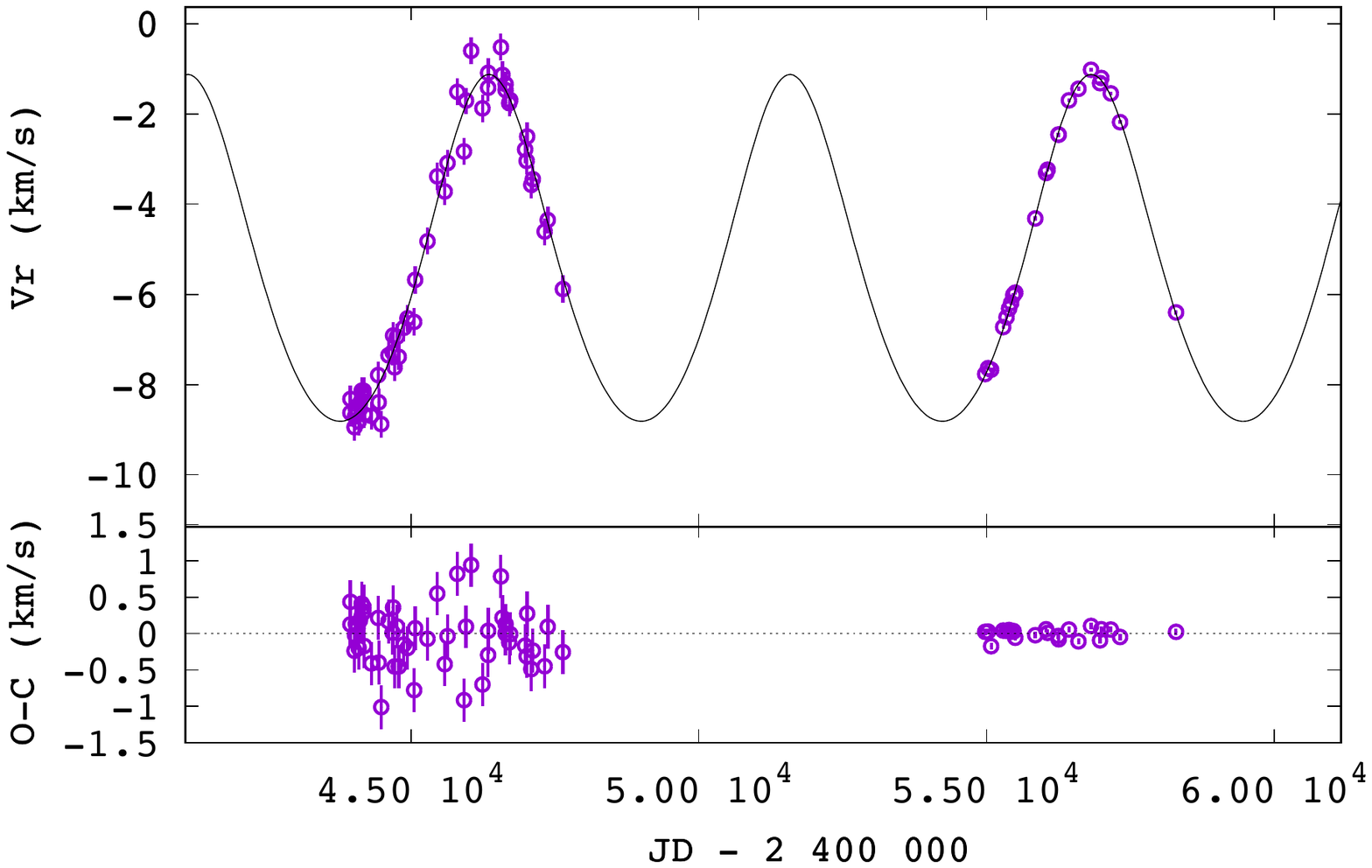}
\vspace{-7cm}\\
\vspace{-3cm}
\includegraphics[width=9cm]{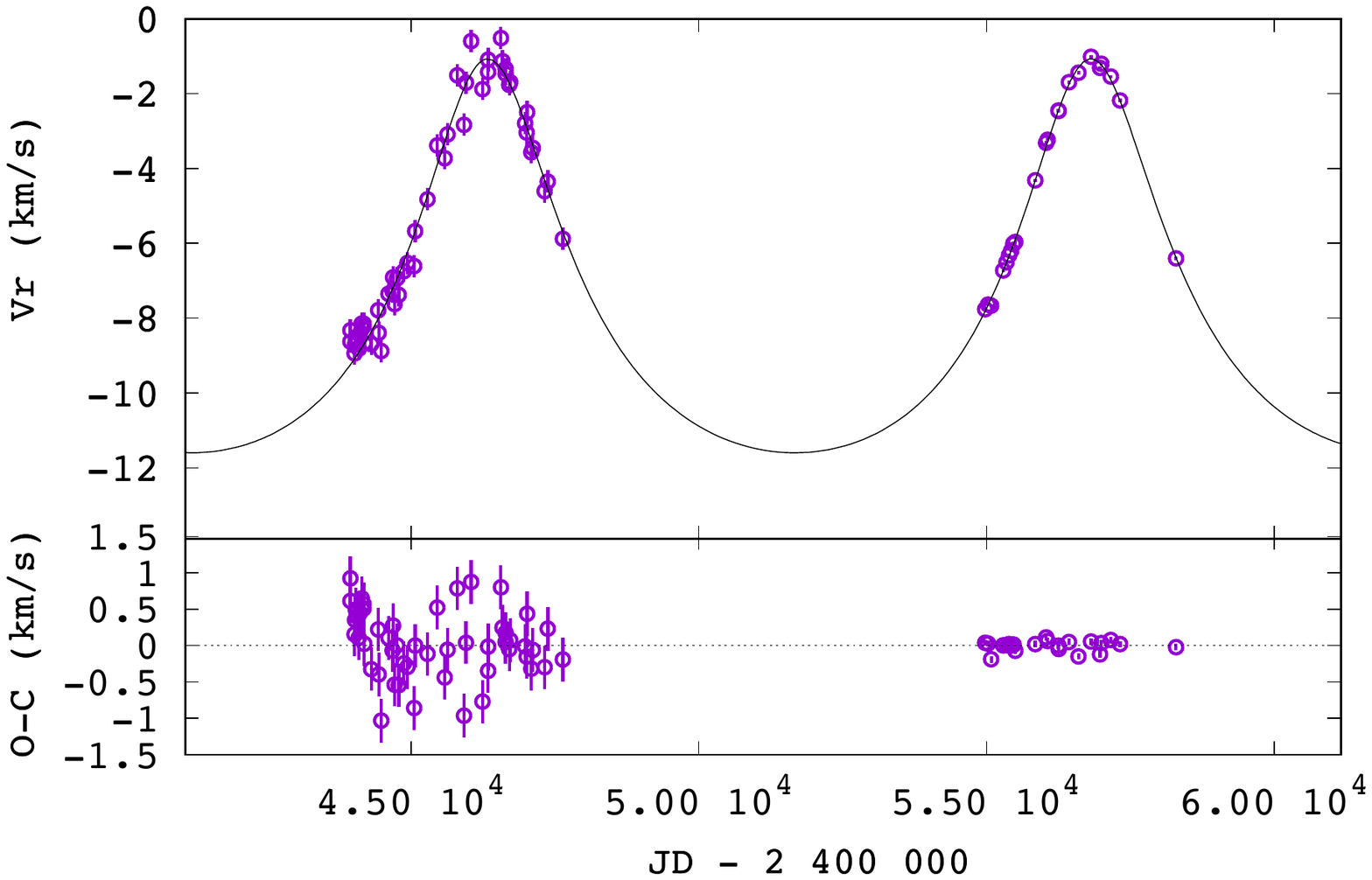}
\caption{\label{Fig:Orbit_199394} 
Two possible orbital solutions for the mild barium star HD~199394. Top panel:  Preliminary solution with $P = 14.3$~yr and $e = 0.11$; Bottom panel:  Another solution with $P =$28.7~y, and $e = $0.36 is also possible, although less likely given its associated mass function of $0.128\pm0.007$~\Msun, as compared to $0.030\pm0.001$~\Msun\ for the 14~yr orbit. Older data are from CORAVEL, newer from HERMES. Lower panels: O-C residuals. An offset of +0.5~\kms\ has been applied to the CORAVEL data.}
\end{figure}

\begin{figure}
\includegraphics[width=9cm]{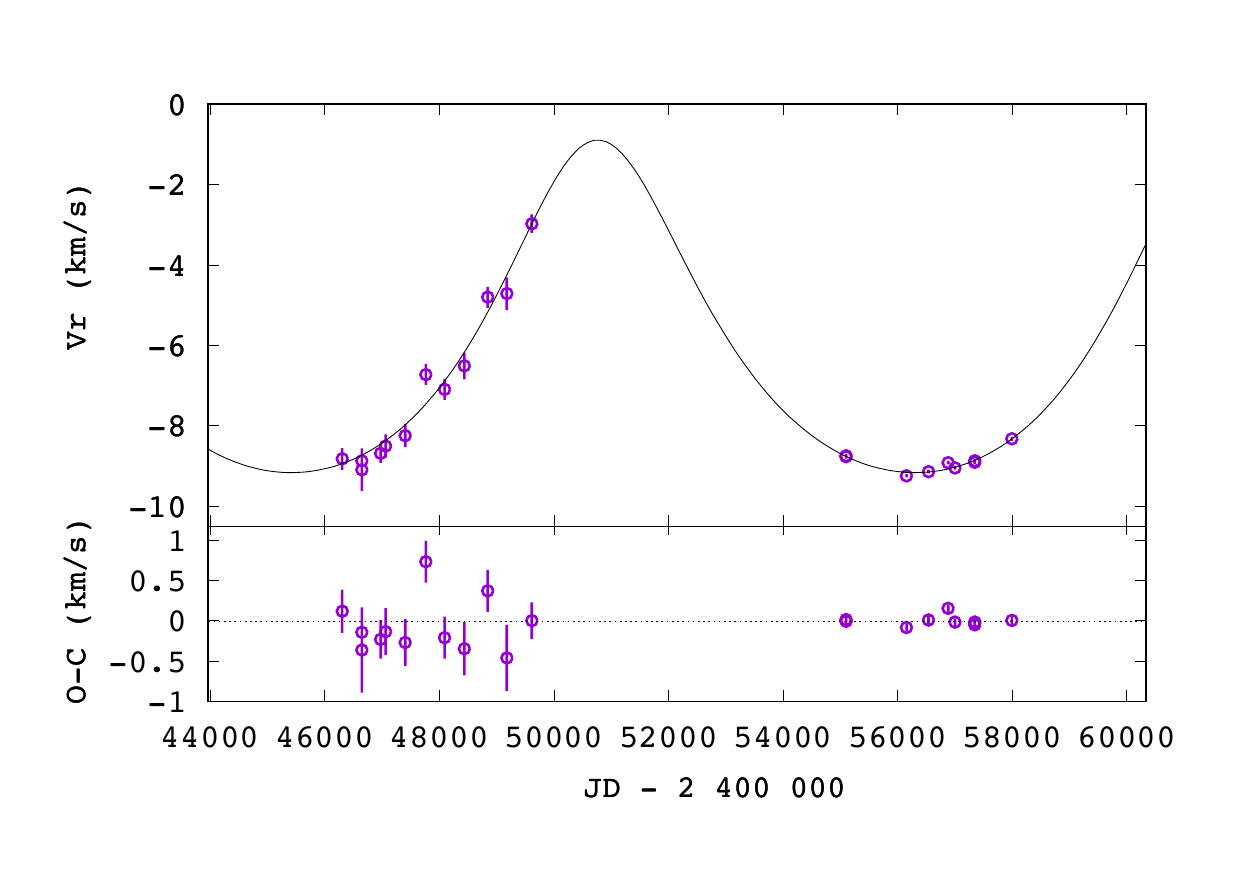}
\caption{\label{Fig:Orbit_211954} 
Upper panel: Radial velocities of the strong barium star HD~211954 and the
associated orbit. Older data are from CORAVEL, newer from HERMES. Lower panel: O-C residuals.}
\end{figure}

\begin{figure}
\includegraphics[width=9cm]{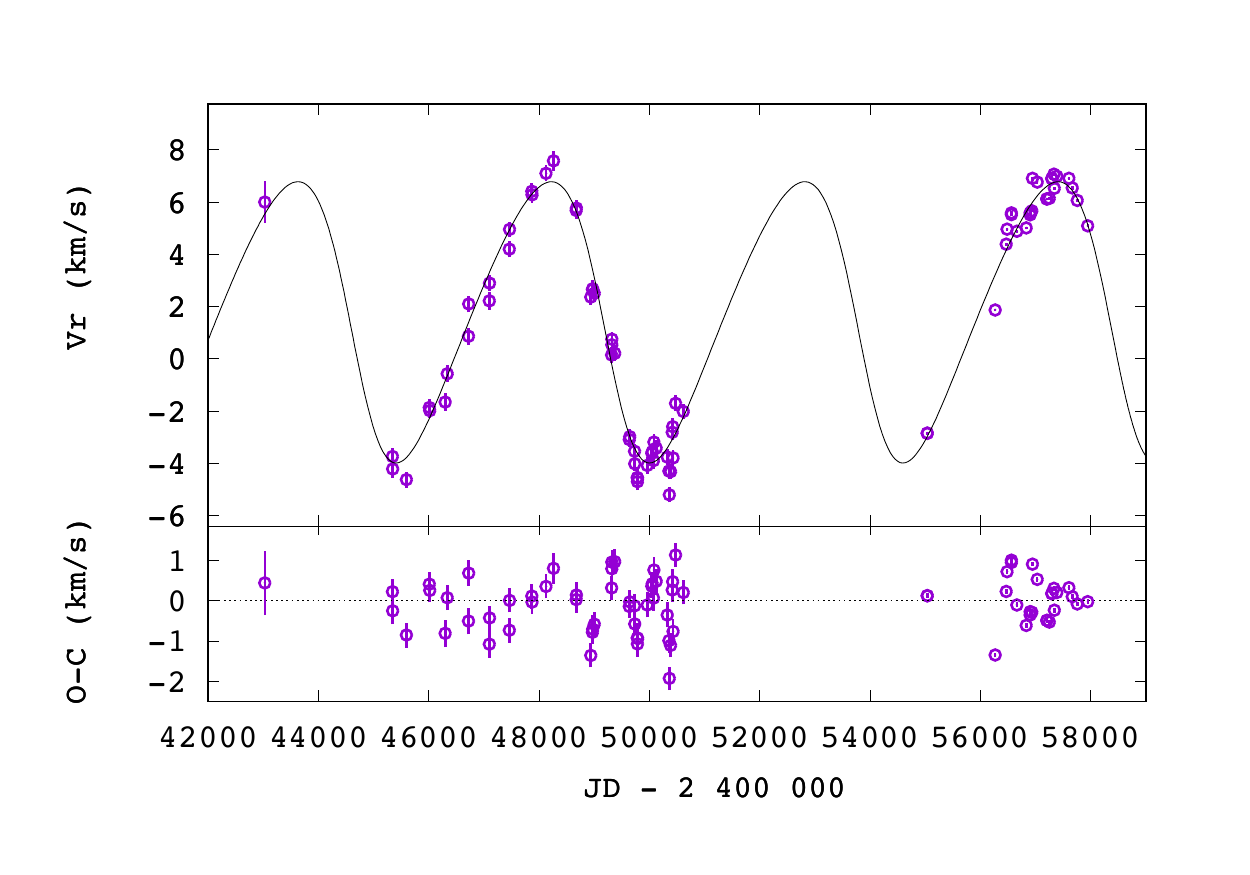}
\caption{\label{Fig:Orbit_HD7351} 
Upper panel: Radial velocities of the symbiotic S star HD~7351 = HR~363 and the
associated orbit. Older data are from CORAVEL, newer from HERMES. Lower panel: O-C residuals.}
\end{figure}

\begin{figure}
\includegraphics[width=9cm]{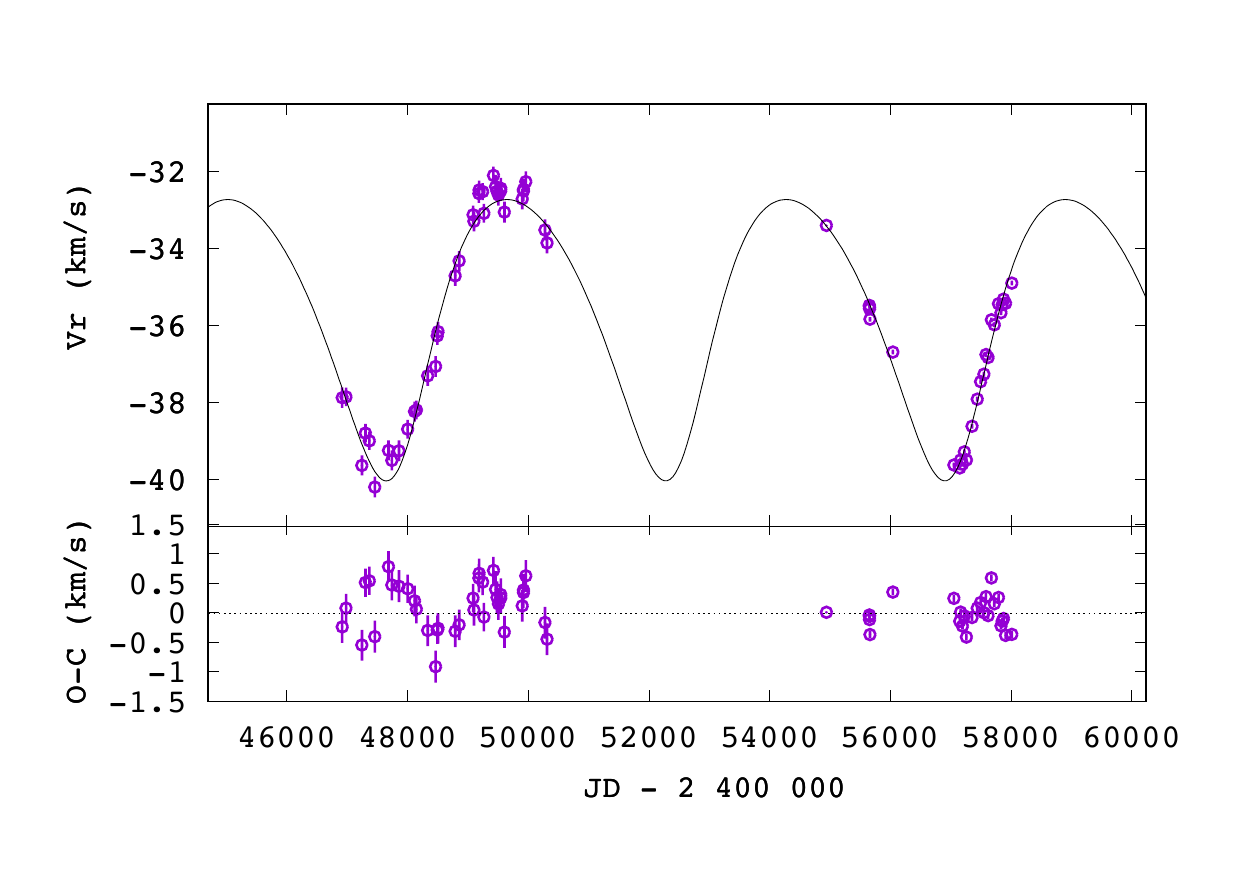}
\caption{\label{Fig:Orbit_170970} 
Upper panel: Radial velocities of the S star HD~170970 and the
associated orbit. Older data are from CORAVEL, newer from HERMES. Lower panel: O-C residuals.}
\end{figure}

\begin{figure}
\includegraphics[width=9cm]{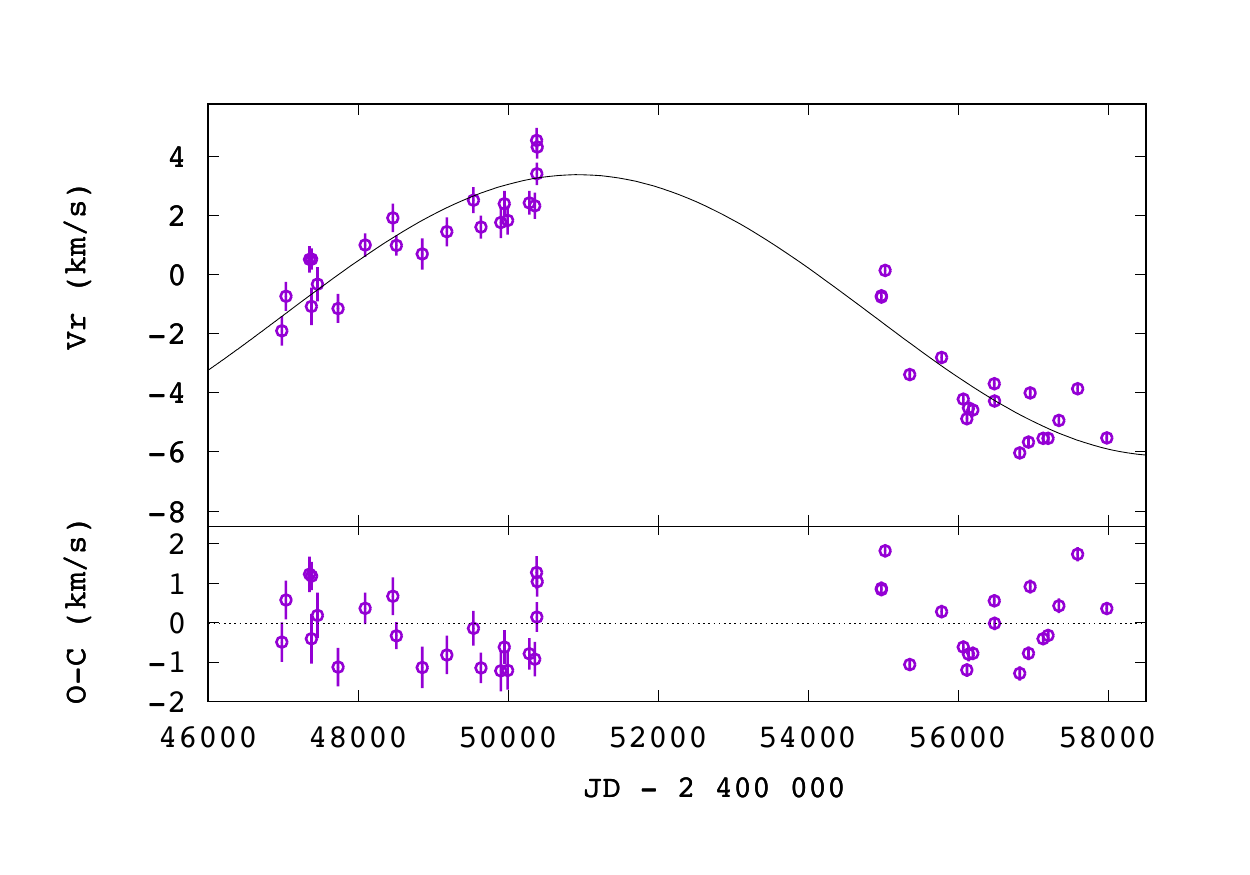}
\caption{\label{Fig:184185} 
Upper panel: Radial velocities of the S star HD 184185 and a preliminary orbit with $P = 43$~yr and $e = 0$.
Older data are from CORAVEL, newer from HERMES. Lower panel: O-C residuals.}
\end{figure}

\begin{figure}
\includegraphics[width=9cm]{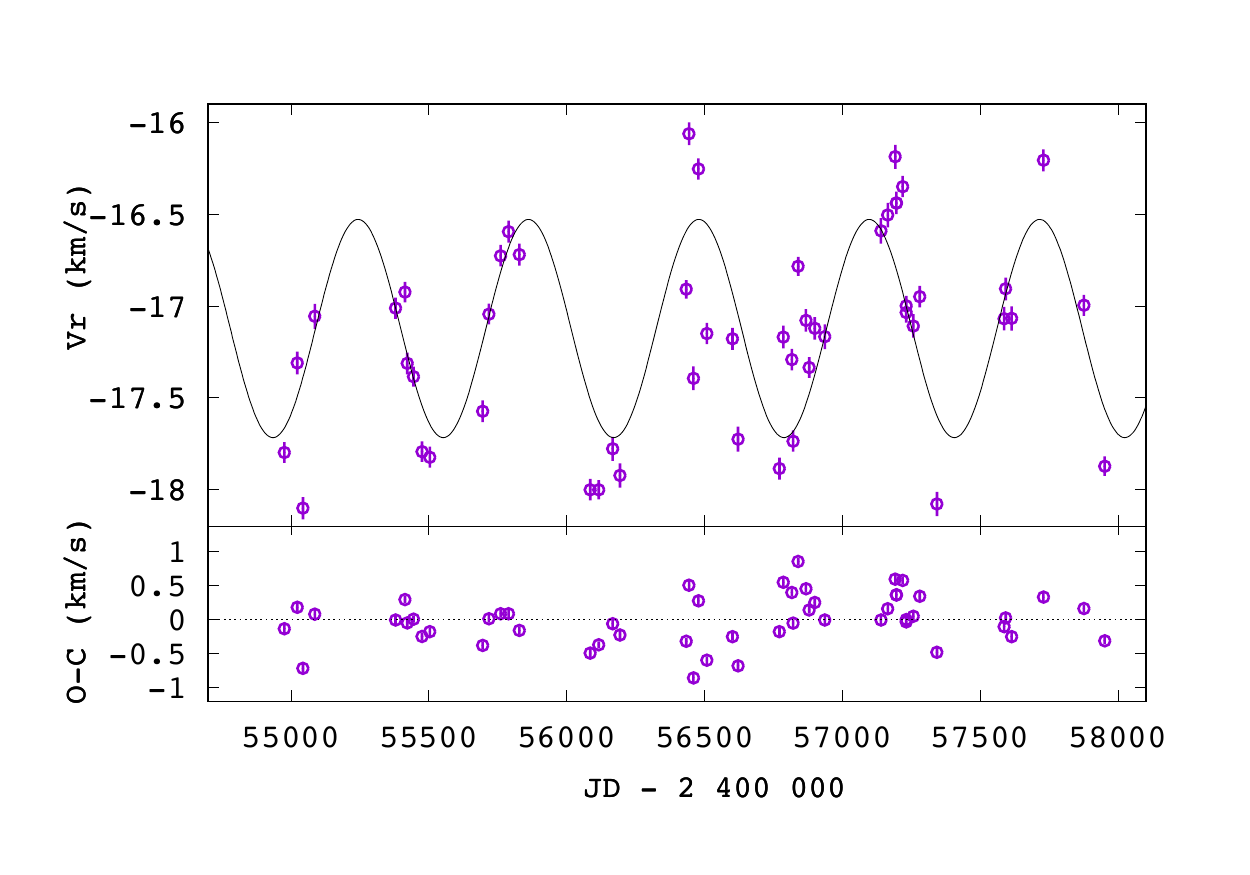}
\caption{\label{Fig:Orbit_189581} 
Upper panel: Radial velocities of the S star HD~189581 and the
associated orbit. Lower panel: O-C residuals.}
\end{figure}

\begin{figure}
\includegraphics[width=9cm]{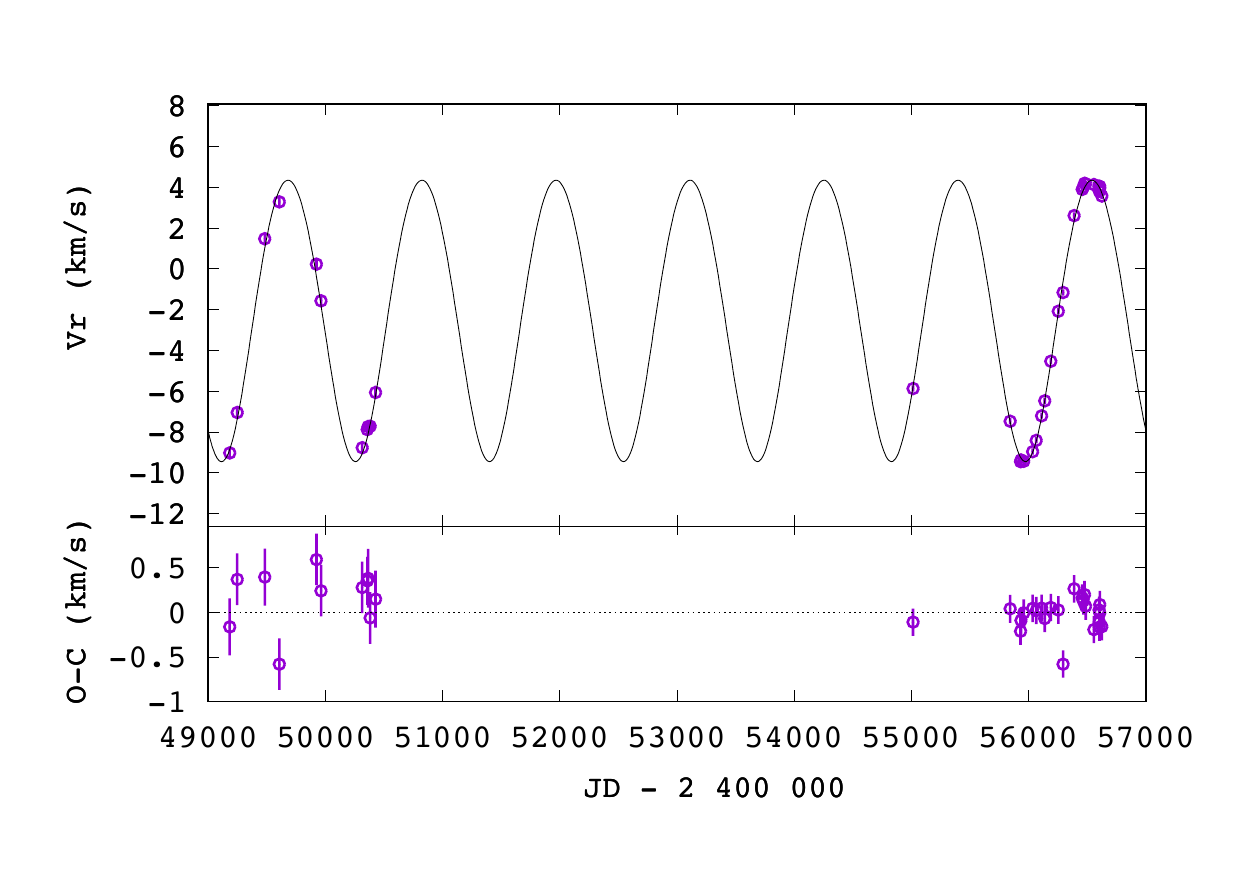}
\caption{\label{Fig:Orbit_215336} 
Upper panel: Radial velocities of the S star HD~215336 and the
associated orbit. Older data are from CORAVEL, newer from HERMES. Lower panel: O-C residuals.}
\end{figure}

\begin{figure}
\vspace{-3cm}
\includegraphics[width=9cm]{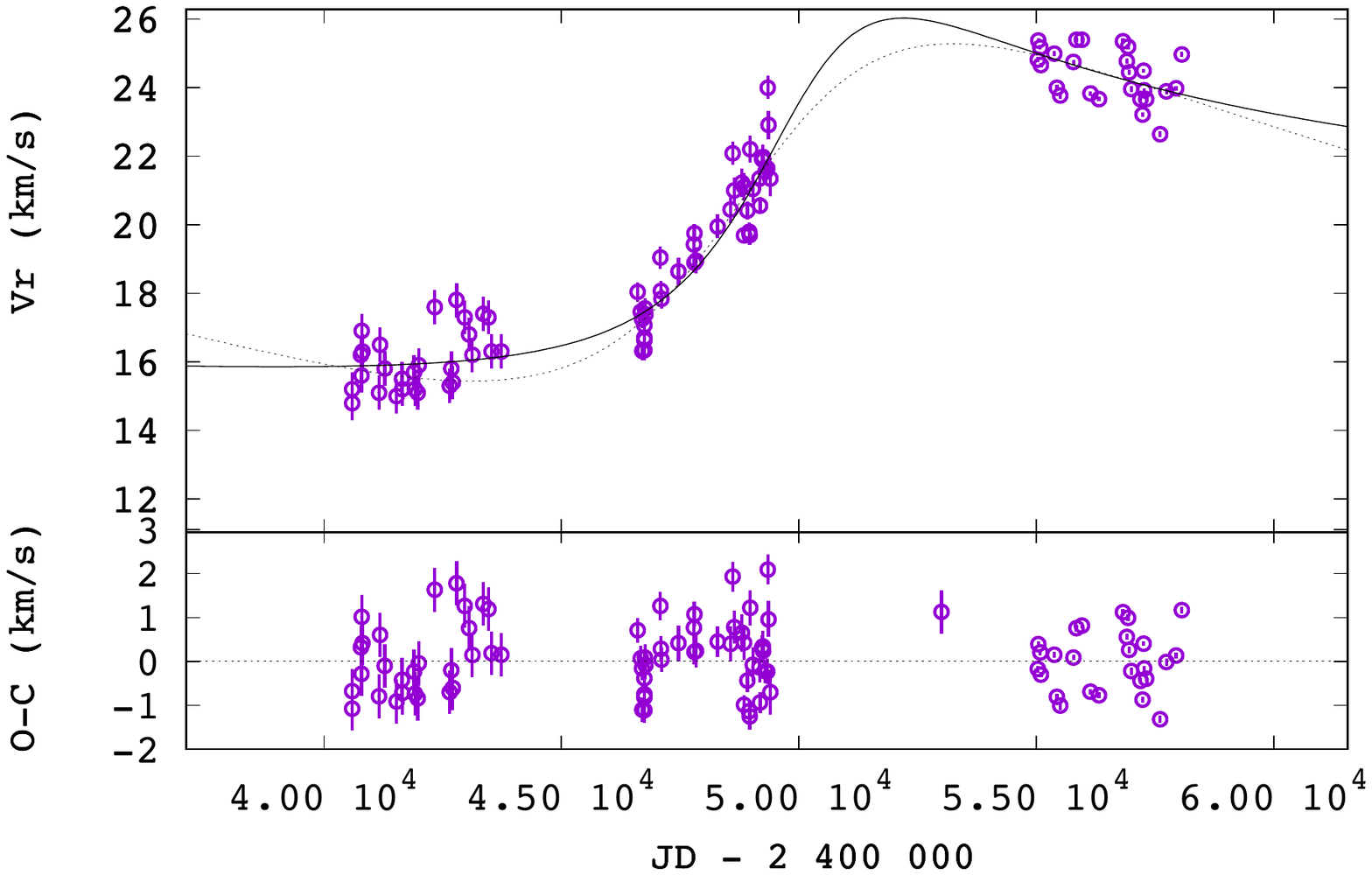}
\vspace{-3.5cm}
\caption{\label{Fig:Orbit_57Peg} 
Upper panel: Radial velocities of the S star HD 218634 (57~Peg) and preliminary orbits with $P = 532$~yr and $e = 0.8$ (solid line), or   $P = 106$~yr and $e = 0.4$ (dashed line). Older CORAVEL data are from R. Griffin (priv. comm.), 
newer from HERMES. Lower panel: O-C residuals.}
\end{figure}

\begin{figure}
\includegraphics[width=9cm]{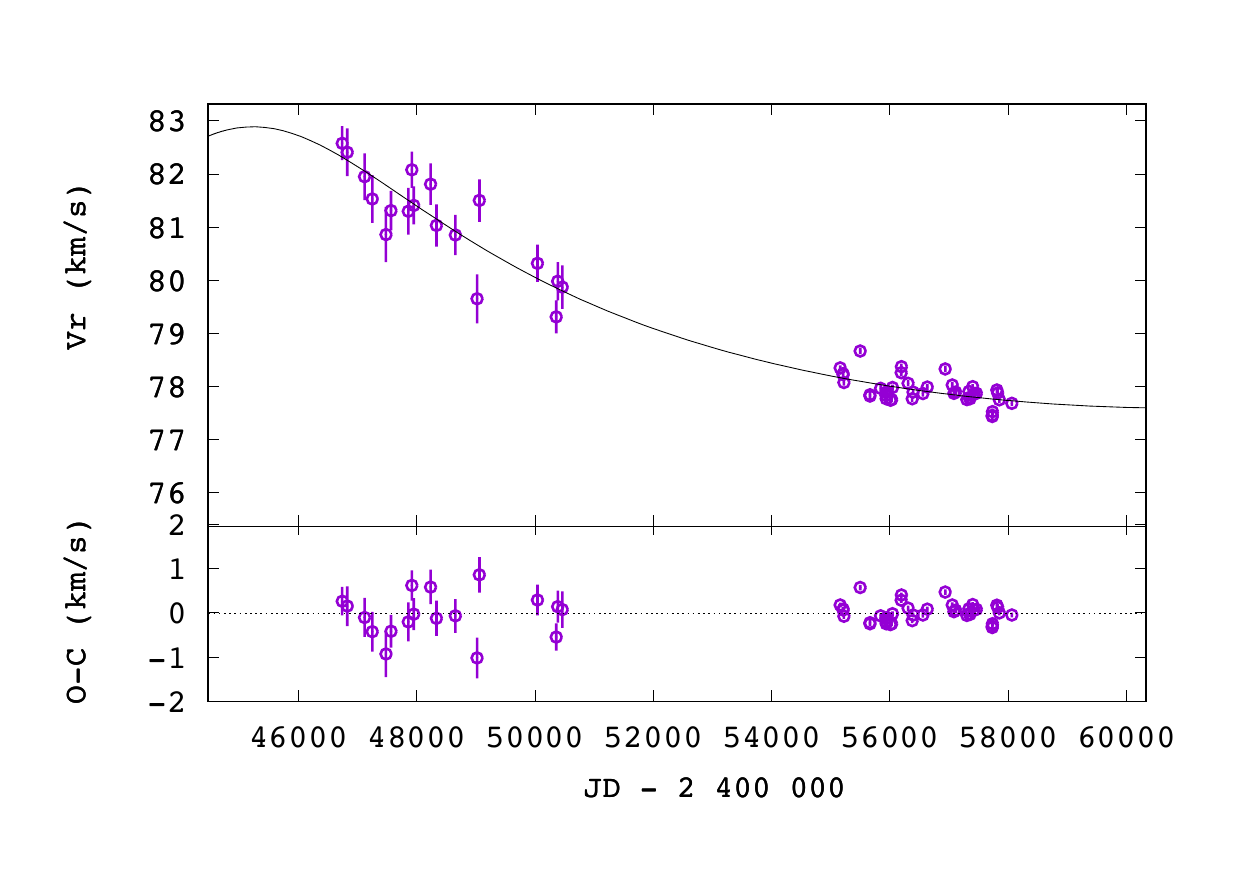}
\caption{\label{Fig:288833} 
Upper panel: Radial velocities of the S star HDE 288833 and a preliminary orbit with $P = 78$~yr and $e = 0.35$! 
Older data are from CORAVEL, newer from HERMES. Lower panel: O-C residuals.}
\end{figure}

\begin{figure}
\includegraphics[width=9cm]{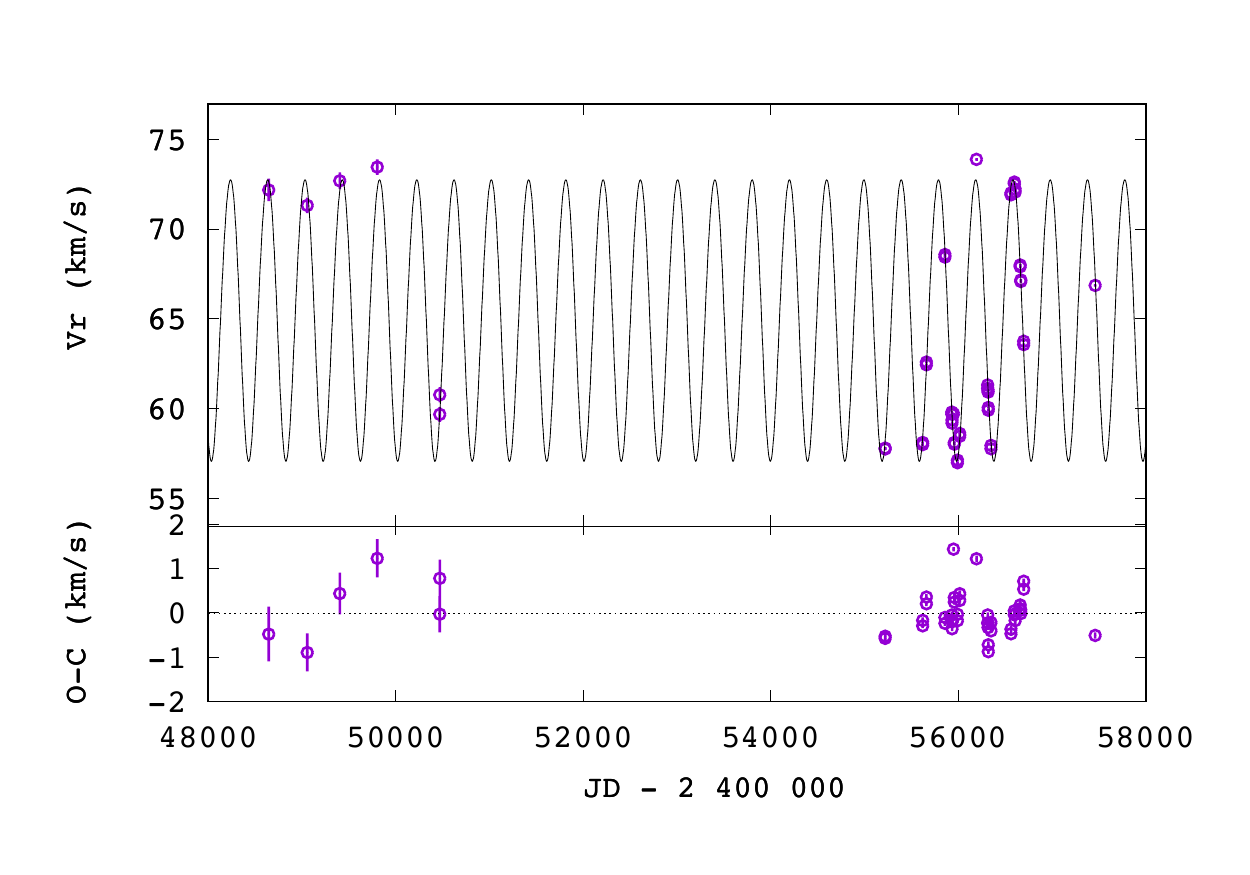}
\caption{\label{Fig:-28.3719} 
Upper panel: Radial velocities of the S star CD -28$^\circ$3719 (= Hen 4-18) and the
associated orbit. Older data are from CORAVEL, newer from HERMES. Lower panel: O-C residuals.}
\end{figure}

\begin{figure}
\includegraphics[width=9cm]{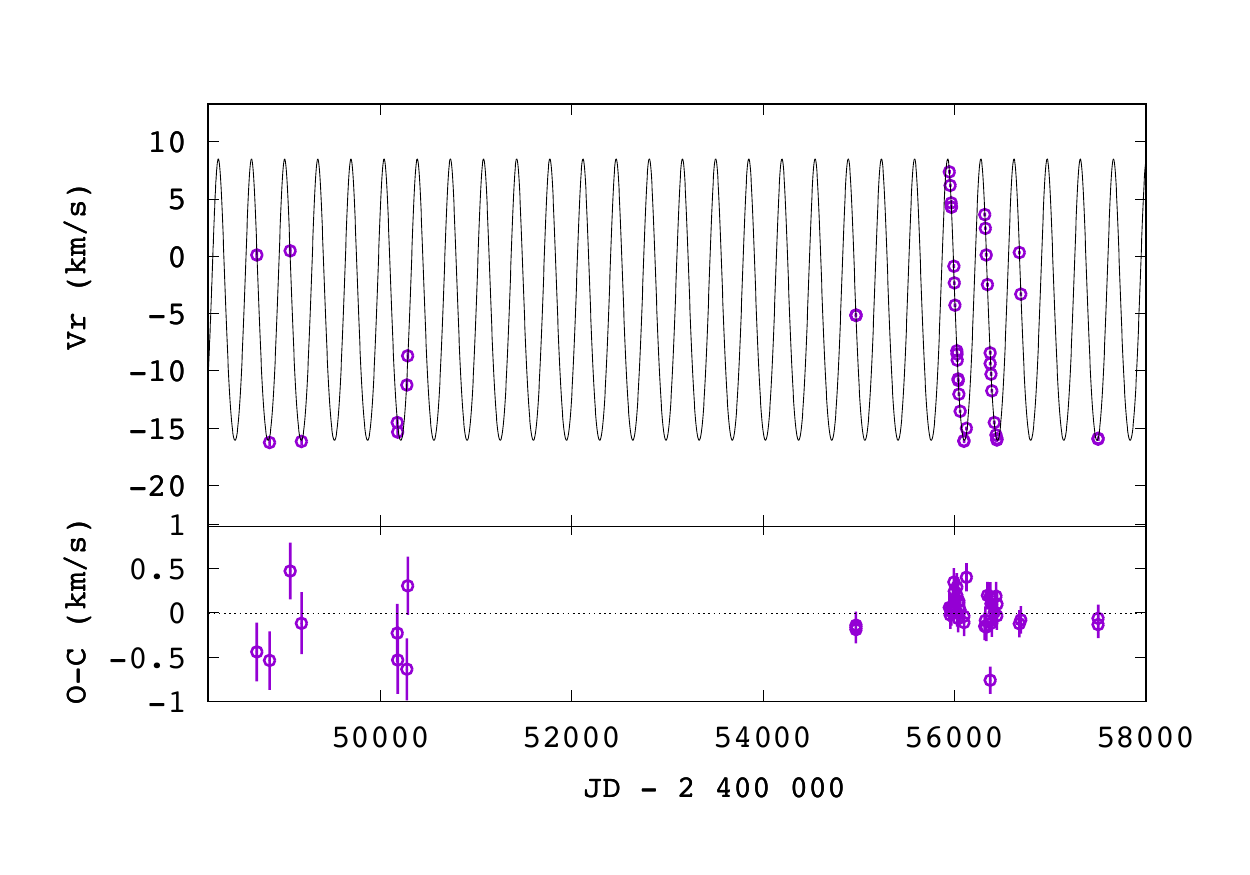}
\caption{\label{Fig:-25.10393} 
Upper panel: Radial velocities of the S star CD -25$^\circ$10393 (= Hen 4-147) and the
associated orbit. Older data are from CORAVEL, newer from HERMES. Lower panel: O-C residuals.}
\end{figure}

\begin{figure}
\includegraphics[width=9cm]{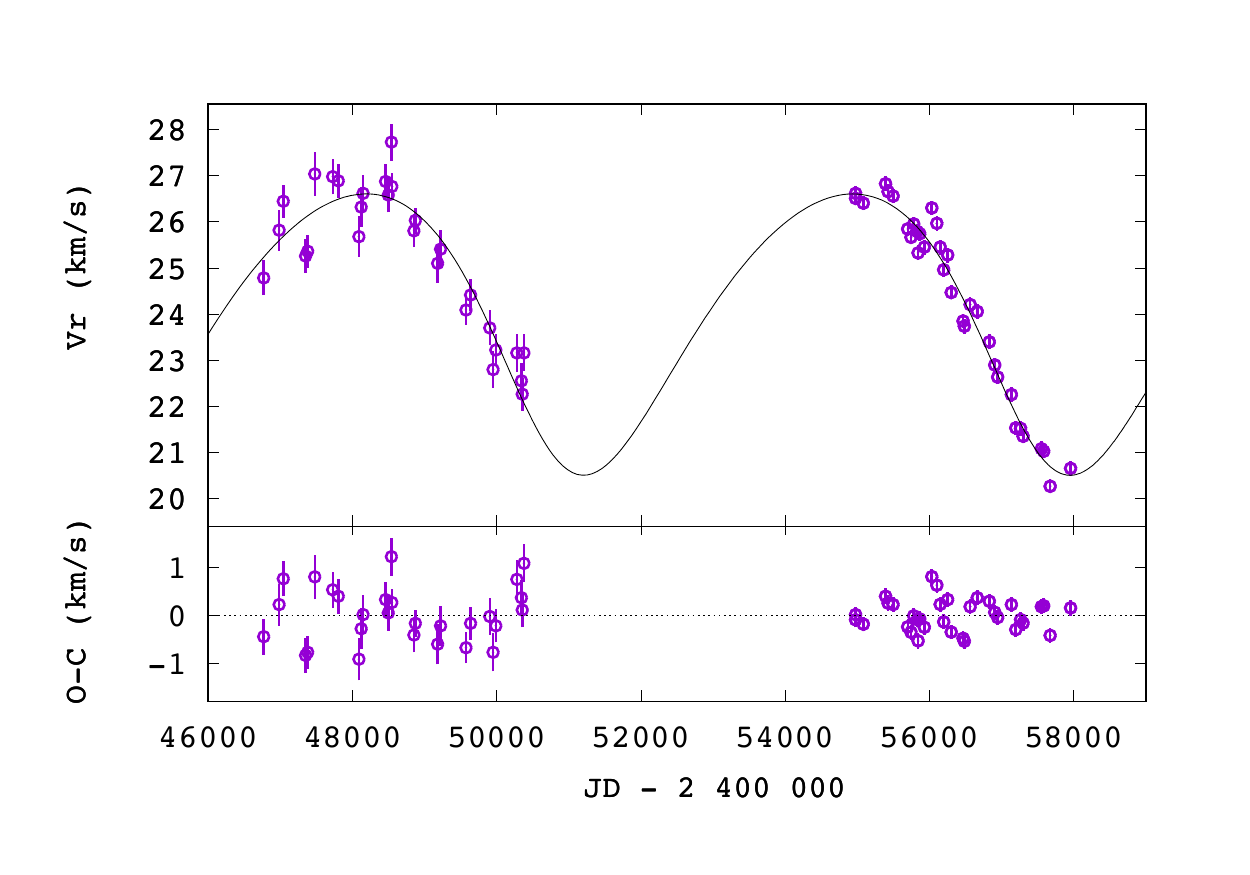}
\caption{\label{Fig:31.4391} 
Upper panel: Radial velocities of the S star BD~$+31^{\circ}$4391 and the associated orbit.
Older data are from CORAVEL, newer from HERMES. Lower panel: O-C residuals.}
\end{figure}

\begin{figure}
\vspace{-3.5cm}
\includegraphics[width=9cm]{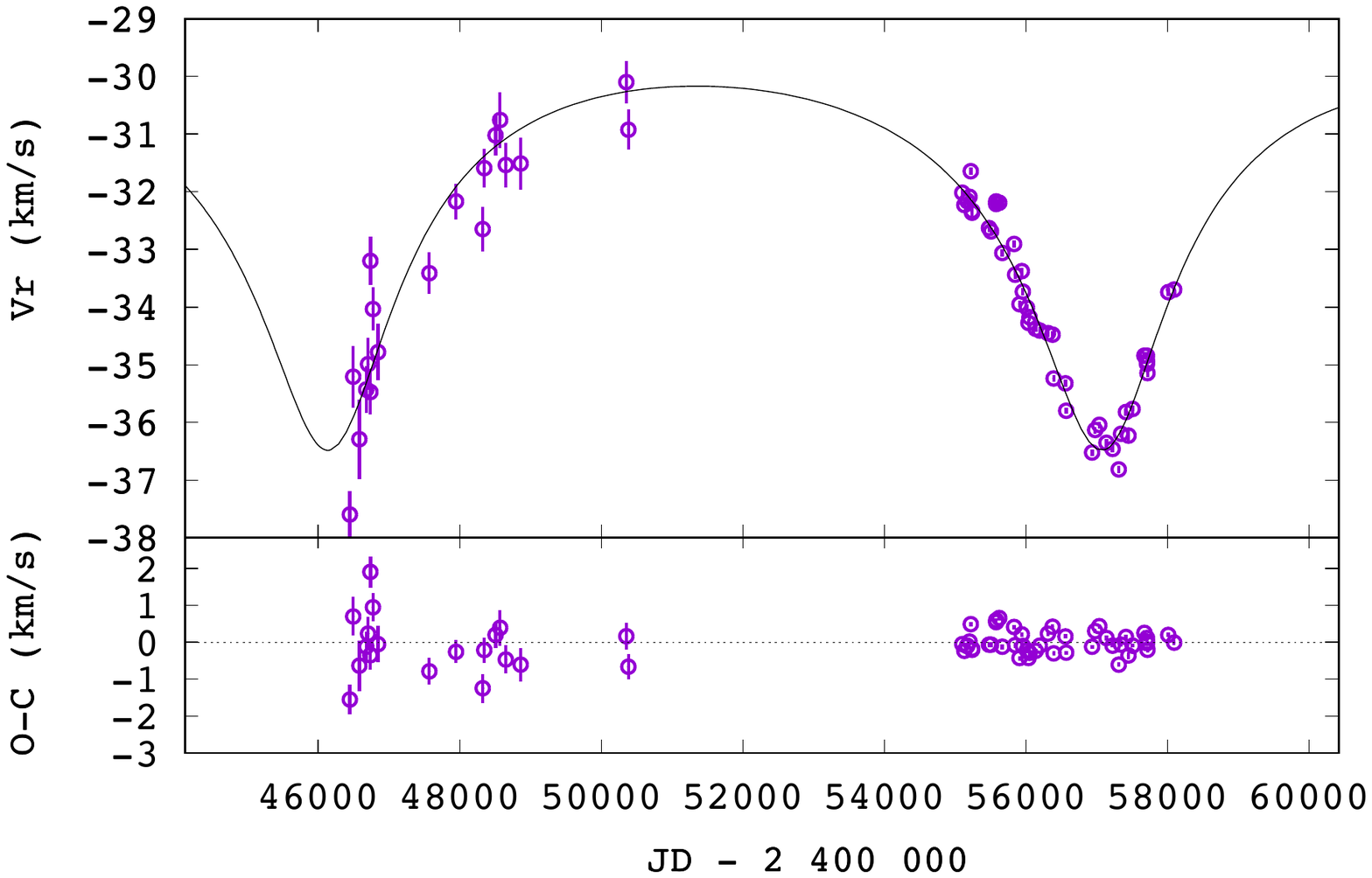}
\vspace{-3.5cm}\\
\caption{\label{Fig:79.156} 
Upper panel: Radial velocities of the S star BD~+79$^{\circ}$156  and the
associated orbit.
Older data are from CORAVEL, newer from HERMES. Lower panel: O-C residuals.}
\end{figure}

\begin{figure}
\includegraphics[width=9cm]{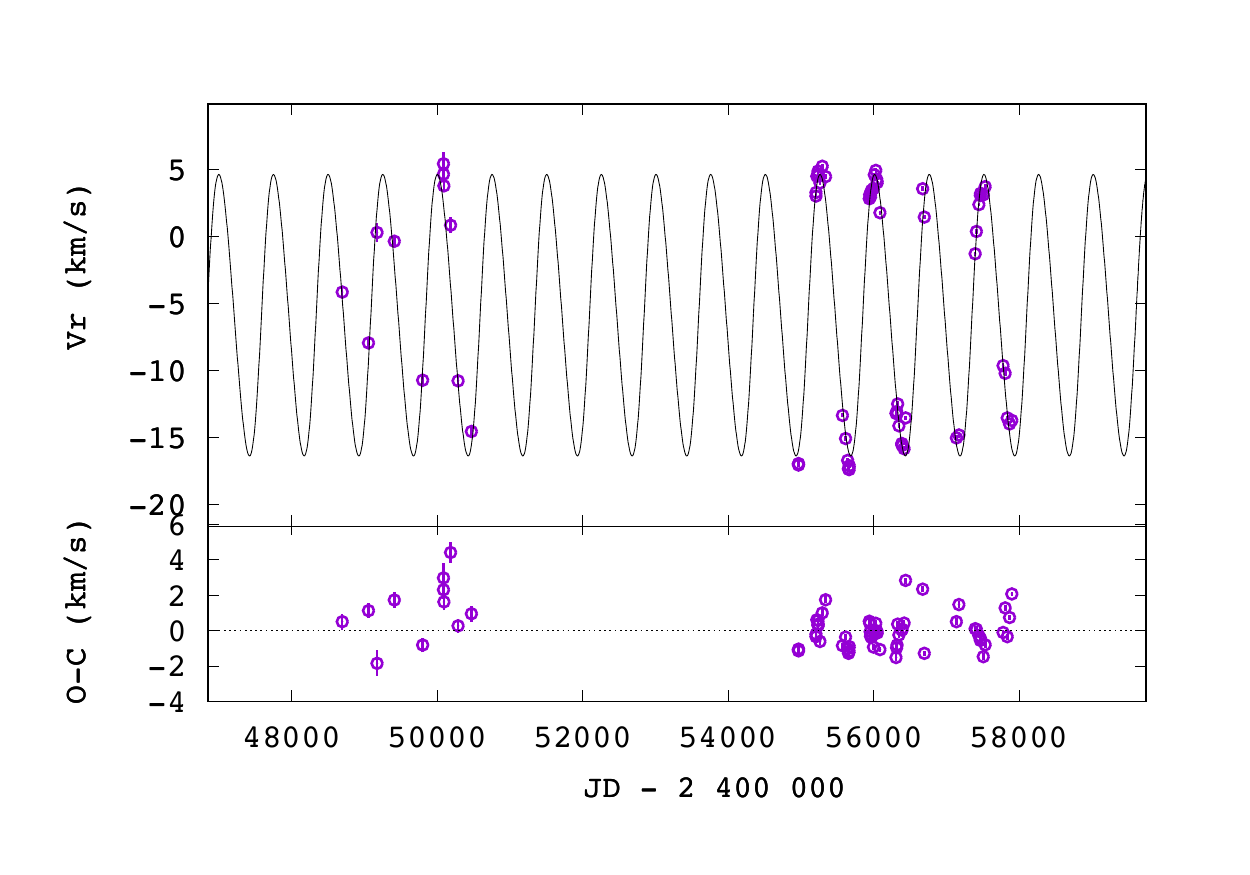}
\caption{\label{Fig:V420Hya} 
Upper panel: Radial velocities of the S star V420 Hya and the
associated orbit. Older data are from CORAVEL, newer from HERMES. Lower panel: O-C residuals.}
\end{figure}

\begin{figure}
\includegraphics[width=9cm]{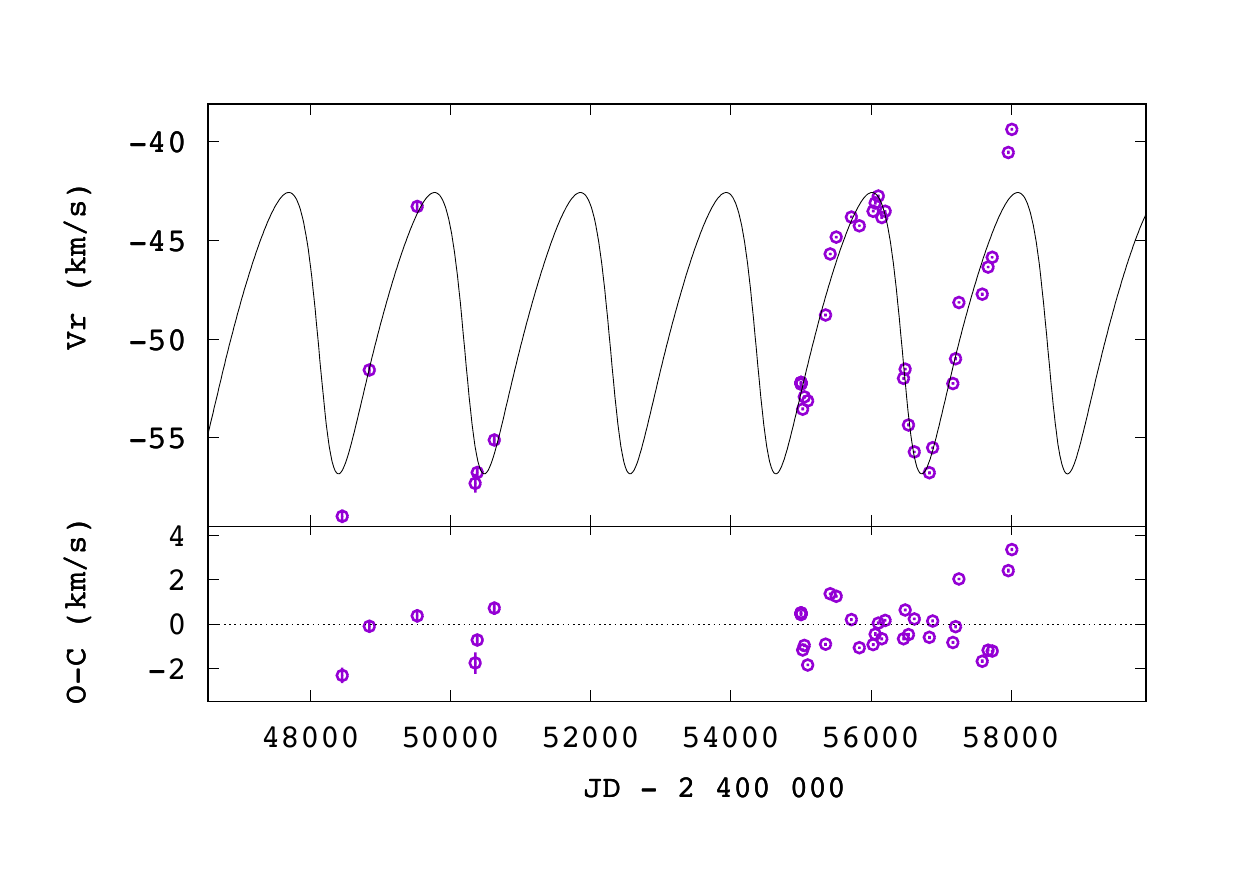}
\caption{\label{Fig:Orbit_ERDel} 
Upper panel: Radial velocities of the S star ER~Del and the
associated orbit. Older data are from CORAVEL, newer from HERMES. Lower panel: O-C residuals.}
\end{figure}
\clearpage
\newpage
\section{Fe line list}

Table~\ref{Tab:Felinelist} presents the Fe lines used to derive the metallicities of barium stars. 

\begin{table}
\caption[]{\label{Tab:Felinelist}
The Fe lines used to derive the metallicities of barium stars, along with their excitation potential and oscillator strength. 
}
\begin{tabular}{llrl}
\hline\\
$\lambda$ & $\chi_{\rm low}$ & $\log gf$ \\
(\AA) & (eV)\\
\hline\\ 
5217.919  &3.640 &-1.719  &Fe I\\
5223.183  &3.635 &-1.783  &Fe I\\
5231.395  &3.573 &-2.951  &Fe I\\
5232.940  &2.940 &-0.076  &Fe I\\
5236.202  &4.186 &-1.497  &Fe I\\
5243.776  &4.256 &-1.050  &Fe I\\
5272.268  &5.033 &-1.038  &Fe I\\
5285.127  &4.434 &-1.540  &Fe I\\
5302.300  &3.283 &-0.720  &Fe I\\
5321.108  &4.434 &-1.089  &Fe I\\
5322.041  &2.279 &-2.802  &Fe I\\
5324.179  &3.211 &-0.103  &Fe I\\
5326.142  &3.573 &-2.071  &Fe I\\
5339.929  &3.266 &-0.684  &Fe I\\
5364.871  &4.445 & 0.228  &Fe I\\
5365.399  &3.573 &-1.020  &Fe I\\
5379.574  &3.694 &-1.514  &Fe I\\
5398.279  &4.445 &-0.630  &Fe I\\
5405.775  &0.990 &-1.858  &Fe I\\
5406.775  &4.371 &-1.620  &Fe I\\
5410.910  &4.473 & 0.339  &Fe I\\
5412.784  &4.434 &-1.716  &Fe I\\
5417.033  &4.415 &-1.580  &Fe I\\
5434.524  &1.011 &-2.119  &Fe I\\
5436.295  &4.386 &-1.440  &Fe I\\
5445.042  &4.386 &-0.020  &Fe I\\
5501.465  &0.958 &-3.046  &Fe I\\
5506.779  &0.990 &-2.793  &Fe I\\
5567.391  &2.608 &-2.617  &Fe I\\
5568.810  &3.635 &-2.850  &Fe I\\
5569.618  &3.417 &-0.486  &Fe I\\
5572.842  &3.396 &-0.275  &Fe I\\
5573.102  &4.191 &-1.317  &Fe I\\
5576.089  &3.430 &-0.900  &Fe I\\
5586.756  &3.368 &-0.120  &Fe I\\
5587.574  &4.143 &-1.750  &Fe I\\
5811.914  &4.143 &-2.330  &Fe I\\
5852.219  &4.548 &-1.230  &Fe I\\
5853.148  &1.485 &-5.180  &Fe I\\
5853.683  &4.191 &-2.590  &Fe I\\
5855.076  &4.608 &-1.478  &Fe I\\
5856.088  &4.294 &-1.327  &Fe I\\
5857.802  &5.033 &-1.767  &Fe I\\
5858.778  &4.220 &-2.160  &Fe I\\
5859.586  &4.549 &-0.419  &Fe I\\
5927.789  &4.652 &-0.990  &Fe I\\
5929.677  &4.548 &-1.310  &Fe I\\
5930.180  &4.652 &-0.230  &Fe I\\
5934.655  &3.928 &-1.070  &Fe I\\
5958.333  &2.176 &-4.160  &Fe I\\
5425.257  &3.199 &-3.220  &Fe II\\
5432.967  &3.267 &-3.527  &Fe II\\
5534.847  &3.245 &-2.865  &Fe II\\
5991.376  &3.153 &-3.647  &Fe II\\
6238.392  &3.889 &-2.600  &Fe II\\
6247.557  &3.892 &-2.435  &Fe II\\
6416.919  &3.892 &-2.877  &Fe II\\
6432.680  &2.891 &-3.570  &Fe II\\
6456.383  &3.903 &-2.185  &Fe II\\
\hline\\
\end{tabular}
\end{table}

\end{document}